\documentclass[useAMS,usenatbib]{mn2e}
\usepackage{graphicx}
\usepackage{color}
\usepackage{soul} 
\usepackage{url}
\usepackage{txfonts}
\usepackage{longtable}

\definecolor{myred}{rgb}{0.7,0.0,0.2}

\definecolor{myblue}{rgb}{0.0,0.2,0.7}

\definecolor{mygreen}{rgb}{0.2,0.7,0.0}

\title[Symbiotic stars in M31]{First detection and characterization of symbiotic stars in M31}
\author[J. Miko{\l}ajewska et al.]{Joanna
Miko{\l}ajewska$^{1}$\thanks{E-mail: mikolaj@camk.edu.pl}, Nelson Caldwell$^{2}$, Michael M. Shara$^{3}$
\\
$^{1}$N. Copernicus Astronomical Center, Bartycka 18, PL 00--716 Warsaw, Poland\\
$^{2}$Harvard-Smithsonian Center for Astrophysics, Cambridge, MA 02138, USA\\
$^{3}$Department of Astrophysics, American Museum of Natural History, Central Park West at 79th Street, New York, NY 10024, USA\\
}

\begin{document}

\date{Accepted  Received }

\pagerange{\pageref{firstpage}--\pageref{lastpage}} \pubyear{}

\maketitle

\label{firstpage}

\begin{abstract}
Symbiotic binaries are putative progenitors of type Ia supernovae. The census of Galactic symbiotic binaries is so incomplete that we cannot reliably estimate the total population of these stars, and use it to check whether that number is consistent with the observed type Ia supernova rate in spiral galaxies. We have thus begun a survey of the nearest counterpart of our own Galaxy, namely M31, where a relatively complete census of symbiotic stars is achievable. We report the first detections and spectrographic characterizations of 35 symbiotic binaries in M31, and compare these stars with the symbiotic population in the Milky Way. These newly detected M31 symbiotic binaries are remarkably similar to galactic symbiotics, though we are clearly only sampling (in this feasibility study) the most luminous symbiotics in M31. We have also found, in M31, the symbiotic star (M31SyS\,J004233.17+412720.7) with the highest ionization level known amongst all symbiotics.  An optical outburst of the M31 symbiotic star M31SyS\,J004322.50+413940.9 was probably a nova-like outburst, the first symbiotic outburst detected outside the Milky Way and Magellanic Clouds. 

 \end{abstract}

\begin{keywords}
surveys -- binaries: symbiotic -- planetary nebulae: general -- M31  \end{keywords}


\section{Introduction}

\subsection{Motivation}

While not a single progenitor of a SNIa has been observed before the explosion, there is a consensus 
that these supernovae result from the thermonuclear disruption of Carbon-Oxygen white dwarfs (CO-WDs) reaching 
the Chandrasekhar mass (\citealt{Paczynski1985}; \citealt{Webbink1984};  \citealt{Iben1984}). 
This must be due to either 1) rapid mass accretion onto a WD from a non-degenerate companion (single-degenerate or SD model); 
or 2) mass transfer between and/or merger of two WDs (double-degenerate or DD model). 
Each of the proposed scenarios has its strengths and weaknesses \citep[][and references therein]{distefano2013}.
Two particularly well-observed SNIa (in M101 \citealt{Li2011}; and in the LMC \citealt{schaef2012}) 
almost certainly were of the DD type. A third SN Ia with hydrogen in its late spectrum was likely a symbiotic star of the SD type \citep{Dilday2012}. 

Some SD binaries that are currently symbiotic stars (hereafter SySt)  must inevitably evolve into DD systems (e.g. \citealt{DiStefano2010}; \citealt{Mik2013}). 
Symbiotic stars are thus potential SD supernovae {\bf and} the progenitors of DD supernovae.   

SySt are amongst the longest orbital period interacting binaries. The components are an evolved cool giant and an accreting, hot, 
luminous companion (usually a WD) surrounded by a dense ionized nebula. 
Depending on the nature of the cool giant, two main classes of symbiotic stars have been defined. 
The S-types (stellar) are normal M giants with orbital periods of the order of a few years \citep[see e.g.][and references therein]{Mik2012}. 
The D-types (dusty) contain Mira variable primaries surrounded by warm dust \citep{Whitelock1987}; orbital periods are decades or longer (e.g. \citealt{Schmid2002}; \citealt{Grom2009}). 
In addition to ionized and neutral regions, accretion/excretion discs, interacting winds, dust-forming regions, and bipolar outflows with jets are observed 
(see e.g. \citealt[][2012]{Mik2007}, for recent reviews). Some SySt must give rise to nova explosions (\citealt{Yaron2005}; \citealt{Shara2010}). 
Thus, in addition to their importance as SNIa progenitors, symbiotic stars offer insight into all interacting binaries that include evolved giants and accreting WDs during any phase of their evolution.

\begin{figure*}
\resizebox{\hsize}{!}{\includegraphics{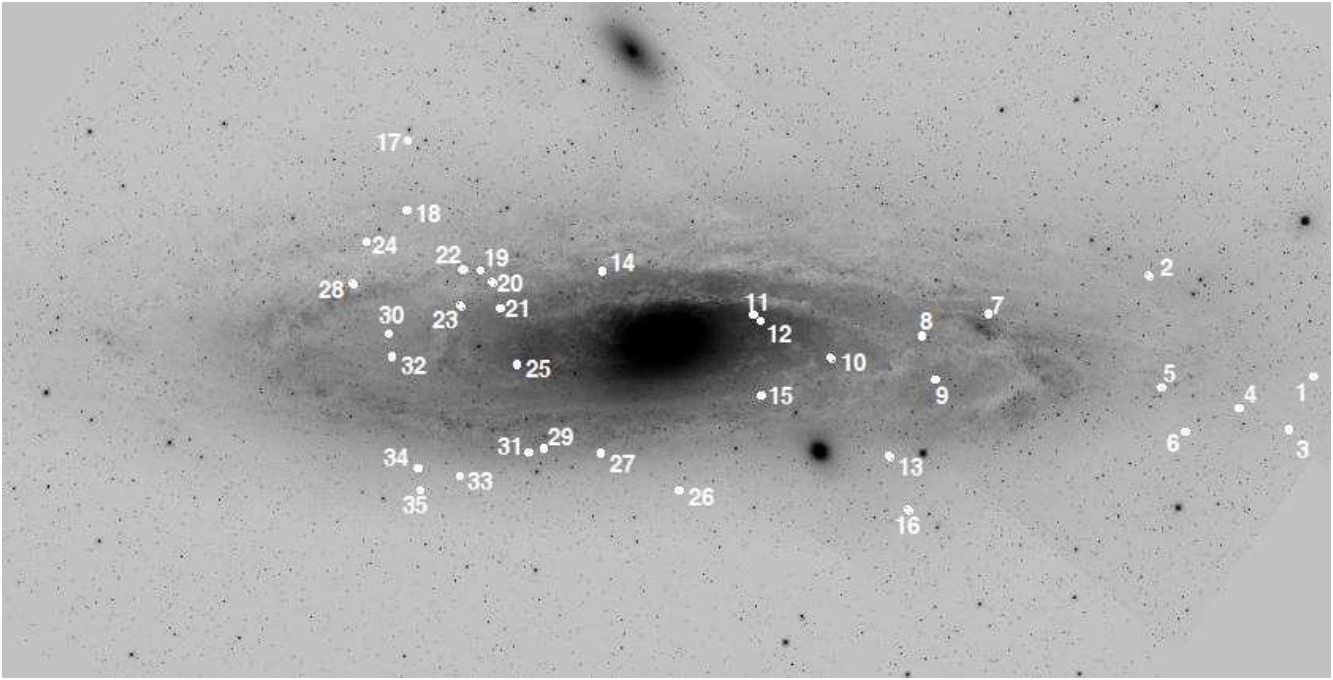}}
\caption{The distribution of M31 SySt and possible SySt, identified by their ordinal numbers in Table\,\ref{Tid}, and overlaid on the  DSS optical image of M31. The field of view is $2.\degr66\times1.\degr27$.}\label{m31}
\end{figure*}

Only $\sim$300 SySt are known today -- mostly in the Milky Way, where our sample is very incomplete. 
This incompleteness is due to the lack of deep, all-sky surveys at wavelengths
where
SySt distinguish themselves from all other stars. As an example, there is no
publicly-available,
deep H$\alpha$ survey at most moderate and high Galactic latitudes; nor is there a
deep Brackett-gamma
imaging survey of the Galactic Plane.

Why should we hunt for more, particularly in external galaxies? 
The most important reasons to acquire large, complete samples of extragalactic SySt are
\begin{enumerate}
\item to enable a direct comparison of symbiotic total numbers in galaxies with the numbers expected in these galaxies if SySt are SNIa progenitors;
\item to enable the spatial distributions of SySt in galaxies to be delineated, relative to the spiral arms and massive-star formation regions in each galaxy, for comparison with SNIa in spiral galaxies; 
\item to eventually enable the characterization of a SySt {\bf before} its eruption as an SNIa, a major confirmation of a prediction of binary stellar evolution theory.
\end{enumerate}

Current estimates of the population of Milky Way SySt vary from 3,000 \citep{Allen1984}, to 30,000 \citep{Kenyonetal1993} up to 400,000 (\citealt{MunRen1992}; \citealt{Magrini2003}). 
However, fewer than 300 SySt have been detected in the Milky Way, and a few dozen more in the closest galaxies 
(e.g. \citealt{Bel2000};  \citealt[][and references therein]{Corradi2012}; \citealt[][2014;]{MMU2013} \citealt{MM2014}; \citealt{Goncalves2008}; \citealt{kniazev2009};  \citealt{oliveira2013}). 

\citet{Hakobyan2011} demonstrated that a significant fraction of SNIa in spiral galaxies preferentially occur in young, star-forming regions. 
If SySt are associated with SNIa, then they, too, should show some preference for the spiral arms and HII regions in spiral galaxies. 
All existing surveys for SySt are so badly incomplete that we can say nothing about their global populations in any other galaxy. 
We are equally ignorant of the global spatial distribution of SySt in the Milky Way. This paper is a demonstration of feasibility of acquiring a complete census of the SySt in the Local Group of galaxies, 
and a first step in tracing the locations of SySt in M31. 

If SySt (or some subset of all SySt) are really the progenitors of SNIa, then eventually one of these objects will erupt in a Local Group galaxy (including the Milky Way). 
The larger the sample of spectrographically confirmed SySt (in any galaxy) that astronomers will have, the greater the possibility that one of these SySt will be the progenitor. 
It is thus clear that a sample of {\bf at least} 10\,000 SySt is required to have a reasonable chance of observing one erupt during an astronomer's lifetime.  
This would be an enormous success for stellar evolution theory.  

\subsection{Why M31?}

With these motivations in mind, we have begun a concerted search for extragalactic SySt.  
The powerful H$\alpha$ emission of SySt is as distinctive in halo and extragalactic stars as in their local counterparts. SySt are simultaneously very blue (in $U$--$B$ colours) and very red (in $V$--$I$ colours), 
due to the simultaneous presence of a very hot white dwarf and and cool red giant. 
Thus a large survey for extragalactic SySt should be based on a survey with (at least) $UBVI$ and H$\alpha$ magnitudes for 
a large number of extragalactic stars. 

At first glance the logical galaxies to start with are the LMC and SMC. They certainly are the closest major Local Group members, and a few SySt have been detected in each. 
The strongest drawbacks to these galaxies are their apparent sizes and lack of suitable databases. About 150 square degrees would have to be surveyed to fully cover the LMC and SMC. 
No suitably deep, well-resolved, publicly available $UBVI$H$\alpha$ atlases or catalogs of the LMC and SMC exist. 

Fortunately there {\bf is} a large, deep, uniform, well-calibrated and documented, publicly-available $UBVI$H$\alpha$ (as well as \mbox{[O\,{\sc iii}]} and \mbox{[S\,{\sc ii}]}) survey 
and catalog of nine other Local Group galaxies that is ideal for our purposes. \citet{Massey2006} used 42 nights of the KPNO and CTIO 4 meter telescopes, 
Mosaic cameras and the Lowell Observatory 1.1 meter telescope to exhaustively image, calibrate and catalog the galaxies M31, M33, IC10, NGC 6822, WLM, Sextans A, Sextans B, Pegasus and Phoenix. 
Careful astrometry was used to ensure stellar positions accurate to 0.2 arcsec. Painstaking photometric calibrations of each CCD chip assured 
3 per cent photometry at $U \sim B \sim V \sim R \sim I \sim 22$, and as faint as H$\alpha \sim 21$. 
This corresponds well to the magnitudes expected for SySt in these galaxies. It also ensures colours accurate to better than 5 per cent. 
Massey et al focused (successfully) on a search for Luminous Blue Variables in their nine galaxies. The LGGS catalog is freely available from the NOAO archives. 
It contains the positions, magnitudes and $UBVRI$ colours of 518\,000 stars in M31 and M33, and 88\,000 stars in the other galaxies. 
One of us (NC) has already collected spectra of stars with strong H$\alpha$ emission in M31 detected in LGGS, and it is the results of this spectrographic reconnaissance that we report in this paper.

In Sect.\,\ref{obs} we describe the data and their reductions. The coordinates and spectra of 35 new M31 SySt are presented in Sect.\,\ref{analysis}. 
We characterize these stars, and contrast and compare them with SySt in our own Galaxy in Sect.\,\ref{discussion}. A brief summary of our results is found in Sect.\,\ref{conclusions}.

\begin{table*}
 \centering
  \caption{List of new and possible symbiotic stars with accurate coordinates, their other cross identifiers, and the reference to the number of figure presenting the spectrum}\label{Tid}
  \begin{tabular}{@{}rllllclc@{}}
  \hline
No. &  M31SyS\,J & RA(2000) & DEC(2000) & LGGS\,J & [AMB2011] & CrossIDs & Fig. \\
\hline
1 &  003846.16$+$400717.0 & 00:38:46.159 & 40:07:16.995 &  & && A1\\
2 &  003857.17$+$403132.2 & 00:38:57.166 & 40:31:32.216 & 003857.19+403132.2 & ~106&  & A1 \\
3 & 003923.79$+$400543.4 & 00:39:23.792 & 40:05:43.378 & 003923.80$+$400543.5  & & & A1\\
4 & 003932.01$+$401224.7 & 00:39:32.010& 40:12:24.708 &  & &SDSS9\,J003931.98$+$401224.8 & A1\\
5 & 003951.85$+$402153.8 & 00:39:51.847& 40:21:53.773 &  & ~407 & & A1 \\
6 & 004005.96$+$401604.3 & 00:40:05.964& 40:16:04.325 & & &2MASS\,J00400597$+$401604 & A1\\
7 &  004021.10$+$404501.1& 00:40:21.096& 40:45:01.085 & 004021.11$+$404501.2 & ~575& & A1\\
8 & 004059.40$+$405003.1 & 00:40:59.400& 40:50:03.127 &  &1022 & & A1\\
9 & 004117.11$+$404524.1 & 00:41:17.112& 40:45:24.115 & 004117.09$+$404524.2 & 1172& & A2\\
10 &  004147.71$+$405737.1 & 00:41:47.712& 40:57:37.082 & & 1418& [PAC]61099 & A2\\
            &  & & & & &M31N 1993-05 & \\
11 & 004155.61$+$410846.7 & 00:41:55.608& 41:08:46.685 & 004155.60$+$410846.7 & 1455 &2MASS\,J00415562$+$4108487 & A2\\
12 &  004156.21$+$410735.0 & 00:41:56.208& 41:07:35.040 & & 1458  & & A2\\
13 & 004216.70$+$404415.7 & 00:42:16.704& 40:44:15.725 & 004216.72$+$404415.8 &1586 &2MASS\,J00421671$+$4044158 & A2\\
            &  & & & & & [PAC]57435& \\
            &  & & & & & NBVF020684&\\
14 & 004233.17$+$412720.7 & 00:42:33.167& 41:27:20.712 & & 1681 & 2MASS\,J00423313$+$4127207 & \ref{pn566}\\
15 & 004235.59$+$410148.0 & 00:42:35.592& 41:01:47.996 & &1693 & [PAC]93855 & A2 \\
16 & 004237.49$+$403813.2 & 00:42:37.488 &40:38:13.204 & 004237.49$+$403813.2 & 1696 & & A2\\
17 & 004241.96$+$415656.4& 00:42:41.962 & 41:56:56.405 & & & & A2\cr
18 & 004319.98$+$415137.5 & 00:43:19.976& 41:51:37.540 & & &NBW9012546 & A3\\
19 & 004322.50$+$413940.9 &00:43:22.495&41:39:40.875 &004322.50$+$413940.9 & 2115 &2MASS\,J00432249$+$4139409 & \ref{pn739}\\
            &  & & & & & BATC\_004322.51$+$413941.1& \\
              && & & &  & M31V\,J00432248$+$4139408 &\\
              && & & & & [PAC]47174&\\
20   & 004323.68$+$413733.6& 00:43:23.682& 41:37:33.557 & & 2122 & WISE\,J004323.68$+$413733.6 & \ref{pn744}\\
             & && & & &M31\,J00432365$+$4137335 &\\
              & & && & &[PAC]47312 &\\
21 &  004334.79$+$413447.9 & 00:43:34.794& 41:34:47.869 & 004334.78$+$413447.9 & 2270  & M31V\,J00433475$+$4134477& A3 \\
              & & & && & [PAC]46291 &\\
22 & 004335.01$+$414358.2& 0:43:35.010 & 41:43:58.216 & & PN275 &  & A3\cr
23 & 004349.54$+$413855.9 & 00:43:49.543& 41:38:55.886 & & &M31\,J00434953$+$413855 & \ref{qso44}\\
24 & 004353.59$+$415323.3& 00:43:53.592& 41:53:23.284 & & 2478 &NBW9056509 & A3\\
               & & & & & & M31V\,J00435354$+$4153232&\\
25 & 004358.20$+$412850.9 & 00:43:58.200& 41:28:50.882 & &2528  &M31V\,J00435816$+$4128508& A3\\
26 & 004359.52$+$410253.5& 00:43:59.517& 41:02:53.515 & & 2541  && A3\\
27 & 004411.71$+$411336.0 & 00:44:11.713 & 41:13:35.994 & 004411.71$+$411336.1 & 2703 & BATC\,J004411.66$+$411335.9 & 2\\
              & & & & & & D31\,J004411.7$+$4111335.9 &\\
28 & 004421.89$+$415125.6 & 00:44:21.891& 41:51:25.648 & 004421.90$+$415125.7 & 2808 & M31V\,J00442188\_4151255 & A3\\
29 &004432.09$+$411940.8 & 00:44:32.094 & 41:19:40.769 & 004432.06$+$411940.6 & 2996 & NBVE012366 & A3\\
     && & & & & 2MASS\,J00443214$+$4119408 &\\
30 & 004433.74$+$414402.8 & 00:44:33.741& 41:44:02.789 & 004433.74$+$414402.7 & 3020 &2MASS\,J00443374$+$4144032 & A4\\
        &  & & & & &  M31\,J00443372$+$4144026& \\
        &  & & & & &  D31\,J004433.7$+$414403.0&\\
31 & 004440.50$+$412052.2 & 00:44:40.495 & 41:20:52.194 & & PN\,339 & & A4\cr
32 & 004445.03$+$414156.5 & 00:44:45.030& 41:41:56.473 & 004445.02$+$414156.5 & & M31\,J00444500$+$4141563& A4\\
33 &  004521.55$+$412557.3 & 00:45:21.554 & 41:25:57.339 & &  & M31\,J00452155$+$4125571 & A4\cr
34 & 004534.07$+$413049.0 & 00:45:34.069& 41:30:48.958 & 004534.04$+$413049.0 & 3609  & M31V\,J00453405$+$4130487 & A4\\
35 & 004545.41$+$412851.6 & 00:45:45.408 & 41:28:51.596 & 004545.34$+$412851.7 & 3699 & & A4\\
\hline
\end{tabular}
\begin{list}{}{}
\item {[AMB2011]} XXXX - number in the catalog of HII regions in M31 (table 2; \citealt{azimlu})
\item {[AMB2011]} PNXXX - number in the catalog of PNe in M31 (table 3; \citealt{azimlu})
\item {[PAC]}... - POINT-AGAPE Catalogue identifier  \citep{An2004}
\item NW... - indentifier in The Guide Star Catalog Version 2.3.2 (GSC2.3; STScI, 2006)
\item BATC... - BATC Data Release One (Zhou et al. 1995--2005)
\item M31... - \citet{VRJ2006}
\item D31... - DIRECT \citep{DIRECT} 
\end{list}
\end{table*}

\begin{table*}
 \centering
  \caption{Photometric data of new and likely new SySt}\label{Tphot}
  \begin{tabular}{@{}lrrrrrrrrrrrrrr@{}}
  \hline
 M31SyS\,J  & \multicolumn{5}{c}{LGGS}  & \multicolumn{6}{c}{PHAT}  & \multicolumn{3}{c}{2MASS}\\
   & $V$  & $B$--$V$ & $U$--$B$  & $V$--$I$ & $I$ & $m_{\rm 275}$ & $m_{\rm 336}$ & $m_{\rm 475}$ & $m_{\rm 814}$ & $m_{\rm 110}$ & $m_{\rm 160}$ & $J$ & $H$ & $K$ \\
\hline
003846.16$+$400717.0&   & & & & 20.77 \cr
003857.17$+$403132.2   & 20.85 & 0.31&-1.23 & 0.98 \cr
003923.79$+$400543.4  & 21.00 & 0.32 & -0.66 & 1.35 \cr
004156.21$+$410735.0 &   & &  & & 20.37\cr 
003951.85$+$402153.8 &   & &  & & 20.59\cr 
004005.96$+$401604.3 &  &  &  &  & 19.76& &  &  & & & &  18.3~ & 16.6~ & 16.75 \cr
 004021.10$+$404501.1 & 23.03 & 0.98 &-1.41 &2.61 \cr
004059.40$+$405003.1&  &  & & & $\ga$21 \cr
004117.11$+$404524.1 & 22.78 &-0.06 &-1.47&1.90 \cr
004147.71$+$405737.1 & & & & & 20.47 \cr
004155.61$+$410846.7& 21.26 & 0.38&-0.84& 1.40 & & & & & & & &16.5 & 16.01& 15.1 \\
004156.21$+$410735.0 &   & &  & & 20.54\cr 
004216.70$+$404415.7 & 20.70 & 0.60&-0.32& 1.31 & & & & & & & & 18.05 & 17.9 & 16.8 \\
004233.17$+$412720.7  & &  &  &  & 19.77& &  &  & & &  & 16.82 & 15.88 & 15.50 \\
004235.59$+$410148.0&   & &  & & 20.48\cr 
004237.49$+$403813.2 & 22.35 & 0.61 &-0.44 & 1.05 \cr
004319.98$+$415137.5&   & &  & & 19.94\cr 
004322.50$+$413940.9  & 20.35 & 0.84 &-0.50 & 1.25 && & & & & & & 17.39 & 16.17 & 15.91 \\
004323.68$+$413733.6&   & &  & & 20.91\cr 
004334.79$+$413447.9  & 20.91 & 0.39 &-0.67 & 1.35 &
& 21.452& 20.910 & 20.804 &19.452  & 18.440&17.480 \cr
004349.54$+$413855.9  & & & &  & $\ga$ 21 &21.410 & 21.330 & 23.066 & 20.487 & 18.915& 17.879\cr
004353.59$+$415323.3&   & &  & & 20.62\cr 
004358.20$+$412850.9& & & & & $\ga$21 &26.408 & 23.973 & 23.954 & 21.056 & 19.373 & 18.269\cr
004359.52$+$410253.5&   & &  & & 20.55\cr 
004411.71$+$411336.0  & 20.81& 0.77&-0.59&1.68 & &
21.449 & 21.248 & 22.014 & 19.458& 18.676 & 17.975 \cr
004421.89$+$415125.6 &21.27& -0.02& -0.84& 1.36&
 & 27.562 & & 22.366 &20.327 & 18.571 & 17.588 \cr
004432.09$+$411940.8 & 21.15 & 0.28 &-0.08 & 0.44& & & & &  &  &  & 17.2 & 15.9 & 15.46\cr	
004433.74$+$414402.8&21.57&0.36&-1.00& 2.37 & &
23.066& 21.848& 23.028& 19.842 &17.924& 16.745 & 16.97 & 15.98 & 15.52 \cr
004440.50$+$412052.2 & & & & & & 22.786  & 21.526 & 22.110  &  22.212  &  20.664  & 19.028  \cr     
004445.03$+$414156.5& 22.35&-0.33&-0.31&1.95 &
 &21.404 &21.256 & 22.448 & 21.401 & 18.086 & 17.301 \cr
004521.55$+$412557.3 & & & & & &  23.911 &  22.432 & 23.130 & 23.219 & 21.206 & 19.629 \cr
004534.07$+$413049.0& 20.89& 0.50&-0.82& 1.27 &
 & 21.548 & 20.793 & 21.145 & 19.699 & 18.996 & 18.149 \cr
004545.41$+$412851.6 & 20.05& 0.63&-0.71&1.03 & 19.09\cr
\hline
\end{tabular}
\end{table*}

\section{Observations and data reduction}\label{obs}

The symbiotic stars discussed in this paper were found as "by-products" of a spectrographic survey of M31 aimed at detecting and characterizing HII regions and planetary nebulae (PNe)(Sanders et al (2012)). HII region candidates were chosen because they displayed strong H$\alpha$ emission in images of the publicly-available LGGS \citep{LGGSHa}. Planetary nebula candidates were chosen from the planetary nebula catalog of \citet{Merrett06}, as well as from strong and unresolved  \mbox{[O\,{\sc iii}]} emission objects in the LGGS images. A second set of spectra was collected in 2011 (using the catalog of \citealt{azimlu}), which also contained objects selected on the basis of H$\alpha$ emission. In all, the spectra of about 700 objects which displayed H$\alpha$ or \mbox{[O\,{\sc iii}]} emission were collected. About 95\% of the candidates did turn out to be HII regions and PNe, as expected.

Upon visual inspection of all the spectra, we found that a small
number of objects were neither PNe nor HII regions. While all display a strong H$\alpha$
line, they have little or no emission in the familiar forbidden lines
of \mbox{[N\,{\sc ii}]}, \mbox{[S\,{\sc ii}]}, and \mbox{[O\,{\sc ii}]}. 
Instead, they show lines of high ionization potential.
These 35 objects, listed in Table\,\ref{Tid}, are the subject of this paper. We emphasize that none of these candidates were selected for spectroscopy with the expectation that they might be SySt. Any statistical analyses of the SySt in M31, based on this sample alone, should take this into consideration.

The spectra themselves were obtained with the Hectospec multi-fiber positioner and spectrograph on the 6.5m MMT telescope \citep{fabricant05}, 
as described in \citet{caldwell2009}.  The Hectospec 270 gpm grating was used and provided
spectral coverage from roughly $3700-9200${\AA} at a resolution of $\sim5${\AA}.  Some spectra did not cover  \mbox{[O\,{\sc ii}]}$\lambda 3727$, 
because of the design of the spectrograph, and the small blueshift of M31. The observations were made
in the period from 2004--2007 and 2011,
and were reduced in the uniform manner outlined in \citet{caldwell2009}. The frames were first de-biased and flat fielded.  
Individual spectra were then  extracted and wavelength calibrated.  Sky subtraction is achieved with Hectospec by
averaging spectra from "blank sky" fibers from the same exposures or by offsetting the telescope by a few arcseconds. 
Standard star spectra obtained intermittently were used for flux calibration and instrumental response.  These
relative flux corrections were carefully applied to ensure that the relative line flux ratios would be accurate.
The total exposures were between 1800 and 4800s.

16 of our targets have LGGS photometry \citep{Massey2006} and their $UBVI$ are collected in Table\,\ref{Tphot}. The uncertainty on individual measurements is less than 0.03 mag for objects brighter than $\sim$ 22 mag, and 0.1--0.3 for the fainter ones.
Since \citet{Massey2006} required
a detection in all filters for inclusion in their catalog, very red stars
are left out. So, additional $I$ photometry from the LGGS was obtained for most of the candidates that were not in Massey et al.'s catalog.
These $I$ magnitudes (Table\,\ref{Tphot}) were
obtained by measuring aperture photometry of stars in relevant mosaicked images,
cross-correlating those stars with ones actually in the Massey catalog, deriving a transformation,
and then applying that transformation to the measurements of the missing stars.  
The estimated errors are around 0.3 mag for these. Three objects, M31SyS\,J004059.40$+$405003.1, M31SyS\,J004358.20$+$412850.9, and M31SyS\,J004349.54$+$413855.9,
were too faint to be measured using this technique, though they do show up in the $I$ band images.

The PHAT survey \citep{Dalcanton2012} has imaged the NE half of the disk
of M31 in 6 filters from the UV to the near IR, thus approximately
half of our SySt sample lies within that survey. Details of the point-spread
function fitting photometry are contained in \citet{Williams2014}.
What we have done is search the PHAT photometric catalogs for coordinate
matches with our SySt candidates. Since the fiber aperture is 1.5\,arcsec in diameter,
we allowed a tolerance of 0.75\,arcsec in the coordinate matching. Of the 10--20
stars in the catalog within that coordinate tolerance, we chose the star with the brightest
F814W magnitude to be the SySt. Typically, the brightest star was more than 1
magnitude brighter than the next brightest, so it seems safe to assume we have
chosen the correct star. Of the 11 SySts with PHAT photometry, only one
proved to be problematic, with several stars of similar magnitude within the fiber
aperture. We do not report values for this case.

Table\,\ref{Tphot} also gives the 2MASS $JHK$ photometry for some of our objects. The 
locations of all of our targets in M31 are displayed in Fig.\,\ref{m31}.

\section{Identification and classification of S\lowercase{y}S\lowercase{t} in M31}\label{analysis}

\begin{figure*}
\centerline{\includegraphics[width=150mm]{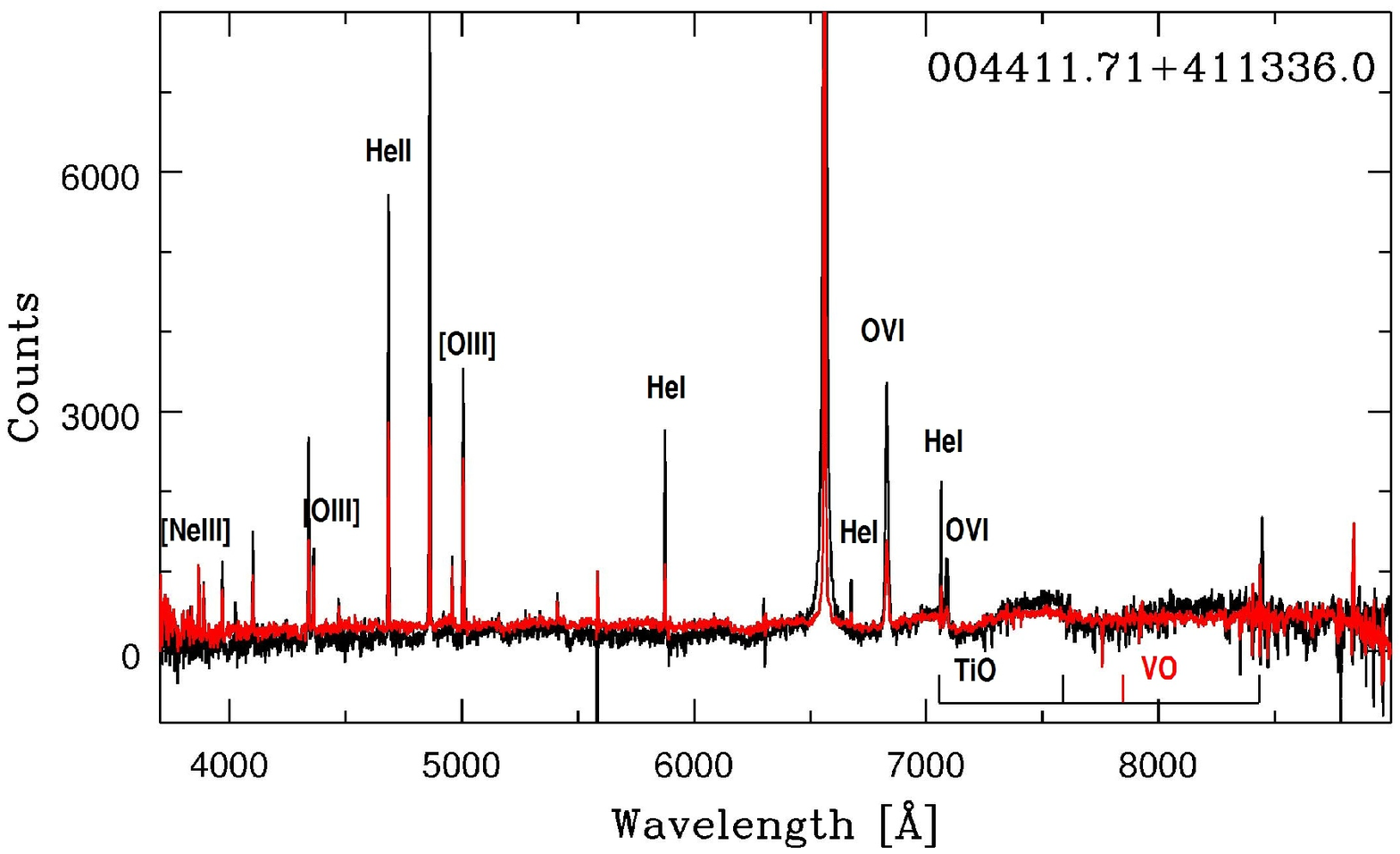}}
\caption{The spectrum of M31SyS\,J004411.71$+$411336.0 which presents all the defining features of a classical S-type SySt. The black and red colours correspond to the spectra taken in 2011 and 2006, respectively. Note the possible change in relative emission line ratios.}\label{942}
\vspace{0.5cm}
\centerline{\includegraphics[width=150mm]{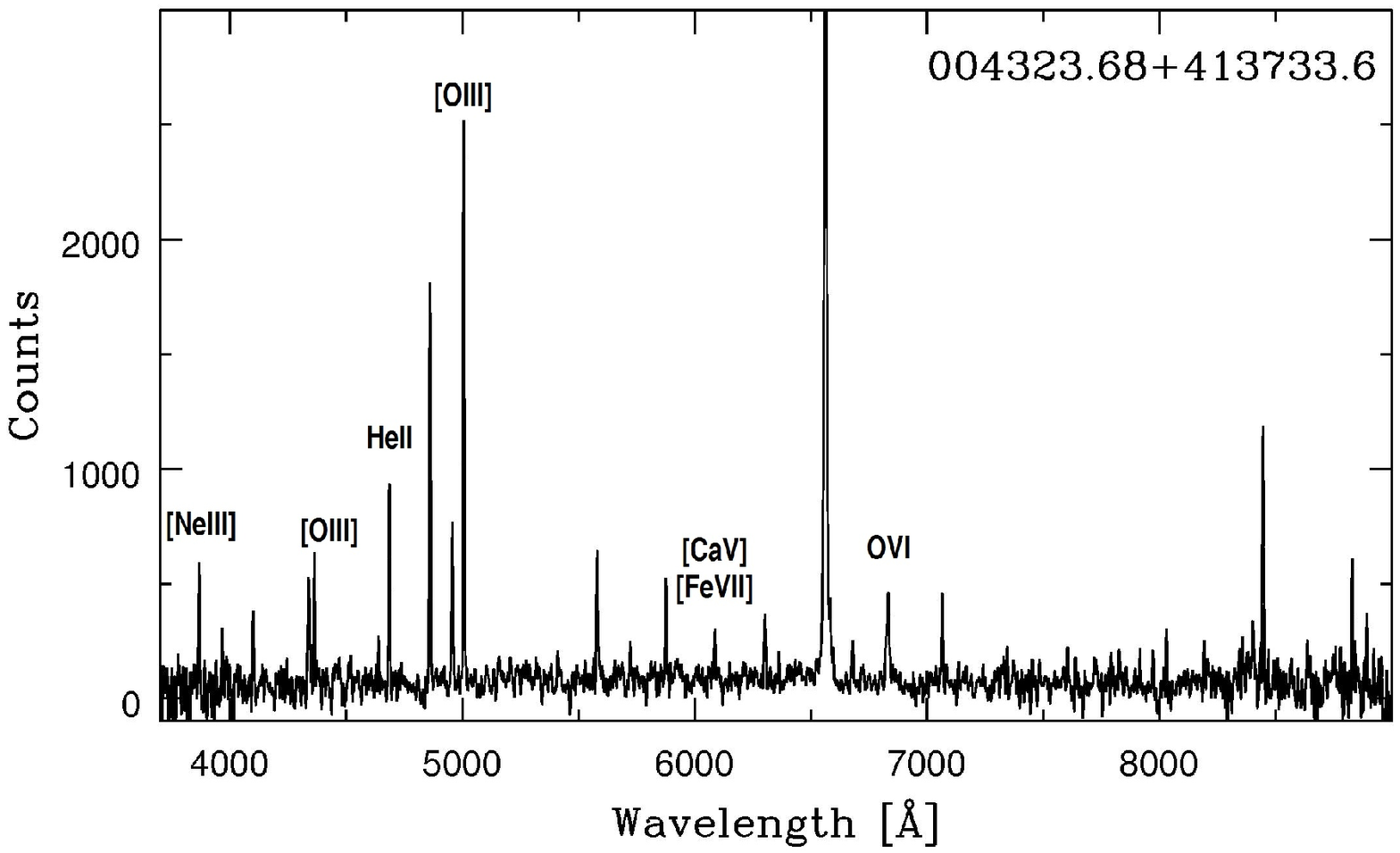}}
\caption{The spectrum of M31SyS\,J004323.68$+$413733.6. The symbiotic nature of this object is demonstrated by the presence of the broad Raman scattered \mbox{O\,\sc{vi}} lines whereas the continuum is too weak to reveal any M giant features.}\label{pn744}
\end{figure*}

\begin{figure*}
\centerline{\includegraphics[width=150mm]{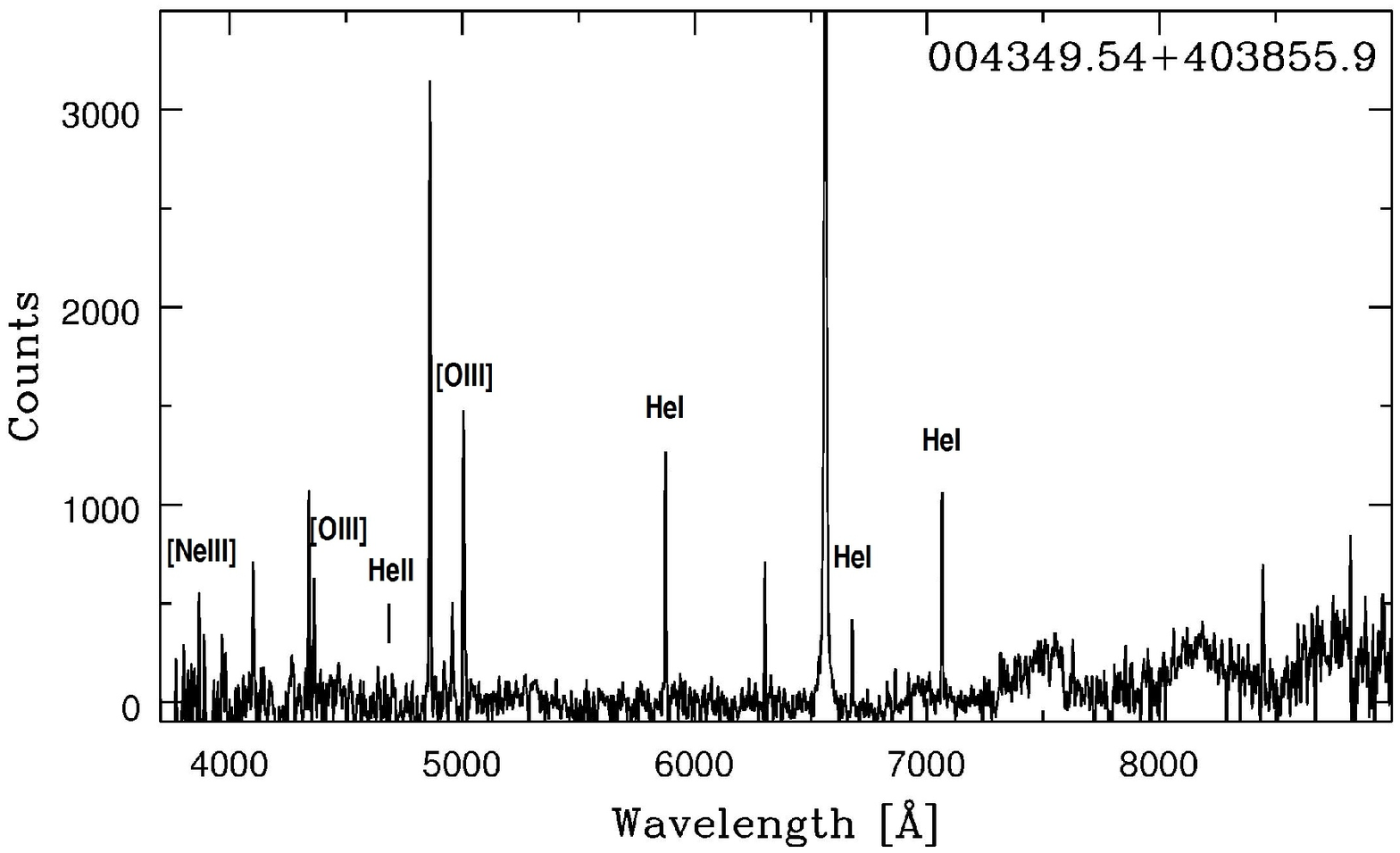}}
\caption{The spectrum of M31SyS\,J004349.54$+$413855.9 which is typical of a SySt with a lower ionization degree as indicated by the lack of measureable \mbox{He\,\sc{ii}}$\lambda\,4686$ line.}\label{qso44}
\label{pn942}
\end{figure*}

The optical spectra of SySt are characterized by the presence of absorption features and continua appropriate for a late-type M giant, 
and strong nebular emission lines of Balmer \mbox{H\,\sc{i}}, \mbox{He\,\sc{ii}}, \mbox{He\,\sc{i}}, \mbox{He\,\sc{ii}} 
and forbidden lines of  \mbox{[O\,\sc{iii}]}, \mbox{[Ne\,\sc{iii}]}, \mbox{[Ne\,\sc{v}]}, and \mbox{[Fe\,\sc{vii}]}. 
Optical spectra of large samples of SySt were catalogued by \citet{Allen1984} and \citet{mz2002}.
The spectra of many SySt also show a broad emission feature at  about $\lambda6830\,\AA$, and a factor of $\sim$ 4 weaker 
but similar feature at about $\lambda7088\,\AA$. \citet{allen1980} found that $\sim 50$ per cent SySt exhibit this band. 
He also found that the $\lambda6830\,\AA$ feature is only observed in SySt. \citet{schmid1989} first  identified the $\lambda\lambda6830,7088$ lines 
as Raman scattering of the \mbox{O\,\sc{vi}} resonance lines $\lambda\lambda1032,1038$ by neutral hydrogen. 
The proposed  production mechanism of these features confirms their intrinsic connection with the interaction between the hot component (copious in \mbox{O\,\sc{vi}} photons) 
and the cool component (with its strong stellar wind) of the SySt. 
Thus the presence of the Raman scattered lines provides a very strong criterion for the identification and classification of SySt.

The composition of SySt is also reflected by their broad-band colours.  They appear blue in the UV, where the nebular line and continuum emission dominates the spectral energy distribution (SED), and red in the optical and near IR, because of the emerging red giant spectrum.


Table\,\ref{Tid} gives the names 
and coordinates of 31 bona-fide and 4 possible SySt in M31 that were classified as such systems following the criteria in \citet{MAS1997} and \citet{Bel2000}. 
According to the IAU recommendations, the name was built from the object position and considering the observational angular resolution. The resulting format is M31\,SyS JHHMMSSss$+$DDMMSSs where the acronym M31\,SyS indicates that they are also SySt. The finder charts are available via dedicated web page created by one of us (NC)\footnote{http://www.cfa.harvard.edu/oir/eg/m31clusters/symb/symb.html}. 
These adopted criteria include the following:

\begin{description}
\item (a) Presence of absorption features of a late-type giant: in practice this means TiO, and sometimes VO  bands.
\item (b) Presence of strong  \mbox{H\,\sc{i}}, \mbox{He\,\sc{i}} emission lines, together with emission lines of ions with ionization potential (IP) of at least 30 eV (e.g. \mbox{[O\,\sc{iii}]}).
\item (c) Presence of the Raman scattered  \mbox{O\,\sc{vi}} $\lambda6830$ even if no features of the cool star are found.
\end{description}

Examples of the spectra presenting all the defining features of SySt are shown in Figs.\,2--\ref{qso44} whereas the spectra for all remaining SySt presented in this study are shown in the Appendix A (available online). The last column of Table\,\ref{Tid} also gives the reference number of the figure in which the spectrum is presented.

Unfortunately, since our spectra come from a fiber system,  only a few of them had
a 'local' background subtracted. Typically the background was far away (10--20 arcminutes).
Thus, some spectra in the disk and bulge show features due to their surroundings.
Good examples are M31SyS\,J004156.21+410735.0 and M31SyS\,J004155.61+410846.7, which shows bulge absorption features as well as the stars' emission
lines. The absorption features can be mistaken for a single red giant star, complicating the SySt classification. 

The symbiotic nature of these objects is also confirmed by their brightnesses in $I$ and red $V$--$I$ colours. In particular, they all are a few magnitudes brighter than the $I$ mag predicted by their H$\alpha$ fluxes and by assuming that their red/near IR emission is of nebular origin; whereas the presence of a bright, cool giant star is consistent with their observed red magnitudes.

The hot components of SySt often show strong optical outbursts characterized by the appearance of bright  A- or F-type continuua, as well as strong  \mbox{H\,\sc{i}} 
as well as \mbox{O\,\sc{i}}, \mbox{[O\,\sc{i}]}, \mbox{Fe\,\sc{ii}}, \mbox{[Fe\,\sc{ii}]},
and other low-excitation emission lines. The stronger emission lines present more or less complex P Cyg structure, while the red giant bands are only visible in the red/infrared spectral range. In our M31 SySt sample, this could be the case of M31SyS\,J004322.50+413940.9 (Fig.\,\ref{pn739}, and Sect.\,\ref{individual}).
In the present study we classify some objects showing only a low-excitation nebular spectrum in addition to TiO bands in their red spectra as possible SySt.

Table\,\ref{Tsp} summarizes the spectroscopic characteristics of our sample, in particular, the selected emission line flux ratios, the highest IP emission lines visible in the spectrum, 
and where possible the spectral classification of the red giant.

\begin{figure}
\centerline{\includegraphics[width=0.86\columnwidth]{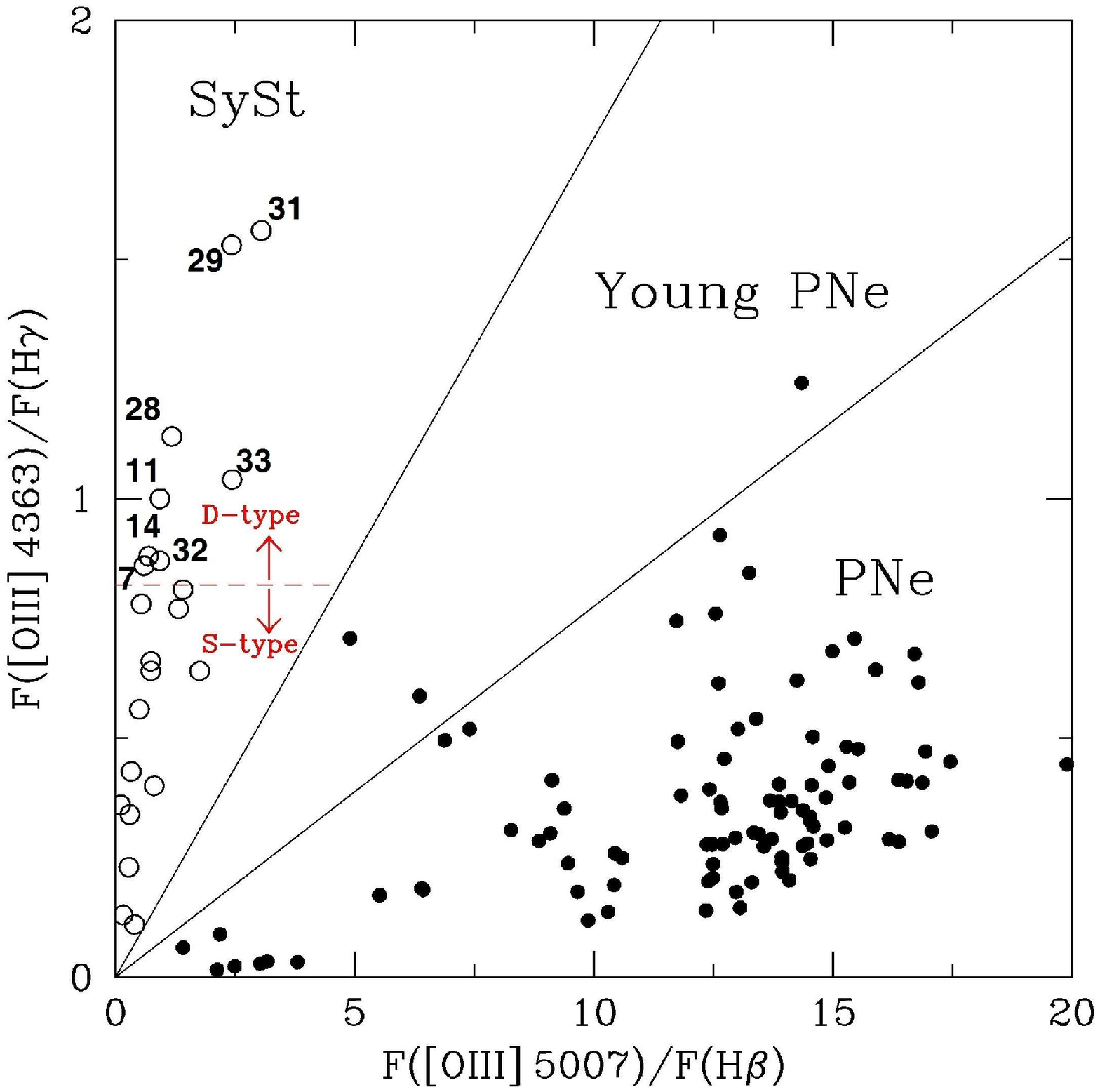}}
\caption{The M31 SySt together with the M31 PNe \citep{sanders2012} in the [O\,{\sc iii}] diagram. The two straight lines separate the regions occupied by SySt, young PNe and PNe \citep{gmmc1995}. The dashed line represent the division line between the S- and D-types. Possible D-type candidates are identified by their ordinal numbers in Table\,\ref{Tid}.}
\label{OIII}
\end{figure}

\begin{figure}
\centerline{\includegraphics[width=0.9\columnwidth]{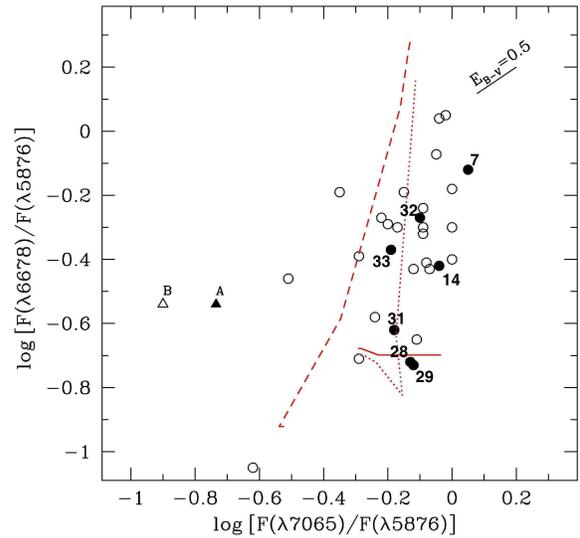}}
\caption{The M31 SySt in the He\,{\sc i} diagram. The filled circles represent those SySt with \mbox{[O\,\sc{iii}]}/H$\beta \ga 0.83$. The solid, dotted, and dashed lines represent the models for $T_{\rm e}=10^4$\,K and $n_{\rm e}=10^6$, $10^{10}$, and $10^{12}\,  \rm cm^{-3}$, repectively  (table 5 of \citealt{proga1994}).}
\label{HeI}
\end{figure}

Unfortunately, the above criteria are insufficient to detect D-type SySt unless they emit the Raman scattered  \mbox{O\,\sc{vi}}\,$\lambda6830$ (which is the case for only $\sim$50\,per cent  of SySt). 
Nebulae around these D-type SySt look very much like genuine PNe because their Mira-type giants are, in most cases, embedded in warm dust shells, and the cool giant signatures are invisible in the optical spectra.  
The difference between the D-type SySt and PNe is, in principle, only detectable via near IR colours including the $JHKL$ bands (e.g. \citealt{Whitelock1987};
\citealt{Schmeja2001}). 
The presence of the cool giant in an increasing fraction of these D-type SySt has been confirmed by detection of Mira-type pulsations in their $JHKL$ light curves (\citealt{Whitelock1987}; \citealt{Grometal2009}). 

\citet{gmmc1995} demonstrated that a diagnostic diagram based only on the \mbox{[O\,{\sc iii}]} lines at 5007 and 4363\,\AA \ effectively separates SySt from planetary nebulae. It also distinguishes S- and D-type SySt from each other. 
We plotted the new M31 SySt together with the PNe in M31 \citep{sanders2012} in this diagram in Fig.\,\ref{OIII}. The new M31  SySt all lie exactly in the region occupied by SySt, whereas the M31 PNe fall into the region of PNe defined by \citet{gmmc1995}. 
Similarly, \citet{proga1994} show that the \mbox{He\,{\sc i}}\,$\lambda\lambda$5876, 6678 line ratio distinguishes between S- and D-type SySt 
in that $F(\lambda6678)/F(\lambda5876) \ga 0.5$ for S-types and $F(\lambda6678)/F(\lambda5876) \sim 0.25$ for D-types. The \mbox{He\,{\sc i}} diagram for our new SySt is presented in Fig.\,\ref{HeI}. 
Based on the locations of these objects in these two diagrams we estimate their IR types (Table\,\ref{Tpar}). Those SySt with \mbox{[O\,\sc{iii}]}/H$\beta \ga 0.83$, i.e. located in Fig.\,\ref{OIII} above the division line between the S- and D-types, are plotted as filled circles. We classified as D-type only 4 of these objects: M31SyS J004421.89+415125.6 (but see discussion of this object in Sect.\,\ref{individual}), M31SyS J004432.09+411940.8, and M31SyS J004440.50+412052.2, and possibly M31SyS J004445.03+414156.5. The position of  M31SyS\,J004021.10+404501.1 in Fig.\,\ref{HeI} indicates that it is almost certainly S-type. 
This classification is, however, only provisional and should be confirmed by near IR observations. 
In the case of some objects, located in both diagrams near the dividing line between S- and D-types, the classification is very uncertain or impossible.  

The absorption features of a cool component are present in most of the sample. Their spectral types (listed in Table\,\ref{Tsp}) were mostly identified based on the depths of the heads of strong TiO bands at $\lambda\lambda$7054, 7589, and 8432, and in some cases the VO band at $\lambda$7845. 
The TiO bands seem to be present for some objects, but their under-exposed continua or spectra, contaminated by features of the surroundings due to poor background subtraction, make the spectral type estimate impossible. We note this by classifying their red components as 'M'.

The majority of our new SySt sample show  the \mbox{He\,{\sc ii}}\,$\lambda4686$ emission line  (26 out of 31), and many of them also show the Raman-scattered \mbox{O\,{\sc vi}} lines
(17 out of 31) and \mbox{[Fe\,{\sc vii}]}\,$\lambda\lambda5721$, 6089 (10 out of 31). The highest ionization emission line detected is coronal \mbox{[Fe\,{\sc x}]}\,$\lambda6375$ in M31SyS\,J004233.17+412720.7 and M31SyS\,J003846.16+400717.0, strongly suggesting that these SySt may be supersoft x-ray sources (see Sect.\, \ref{individual}).

\begin{table*}
 \centering
  \caption{Emission line ratios (H$\beta$=100), and the ion with the highest ionization potential observed in the spectra of the new M31 SySt. In addition, the last column gives the observed H$_\alpha$ fluxes taken from \citet{azimlu} catalog.}\label{Tsp}
  \begin{tabular}{@{}lrrrrrrrrrrrll@{}}
\hline
 M31SyS\,J  & H$\alpha$ & H$\gamma$ & H$\delta$ & \mbox{He\,\sc{ii}} & \multicolumn{2}{c}{[O\,{\sc iii}]} & [Fe\,{\sc vii}] & O\,{\sc vi} & \multicolumn{3}{c}{He\,{\sc i}}  & Ion  & F(H$\alpha$)$^3$ \cr
         &    &                  &                 &          $\lambda$4686&  $\lambda$4363& $\lambda$5007 & $\lambda$6086 & $\lambda$6830 & $\lambda$5876 & $\lambda$6678 & $\lambda$7065 \cr
\hline
\multicolumn{13}{c}{Symbiotic stars}  \cr
\hline
003846.16$+$400717.0& 710 & 27 & 13 &101 &  &  &  & 47 & 19 & 9 &  & Fe$^{+9}$&  \\
003857.17$+$403132.2& 874 & 18 & 5 & & 2 & 4 &  &  & 16 & 6 & 12 & C$^{+4}$ & 4.35\\
003932.01$+$401224.7 & 742 & 25 & 8 & 44 & 9 & 11 &4 & 43 & 27 & 13 & 22 & O$^{+5}$ &  \\
003951.85$+$402153.8 & 562 & 27 & 12 & 15 &  & 64 &  &  & 34 & 17 & 23 & He$^{+2}$ &1.08 \\
004005.96$+$401604.3 & 736 & 25 & 11& 52 &  & 11 &  & 20 & 22 & 24 & 20 & O$^{+5}$ & \\
 004021.10$+$404501.1 & 887 & 49 & 30 & 43 & 42 & 60 &  &  & 30 & 23 & 34 & He$^{+2}$ & 4.49\\
004059.40$+$405003.1 & 638 & 35 & 24 & 66 & 14 & 81 & 27 & 52 & 6 &  &  & O$^{+5}$ & 1.77 \\
004117.11$+$404524.1 & 983 & 36 & 24 & 50 & 23 & 74 &  & 28 & 18 & 9 & 18 & O$^{+5}$  & 4.99 \\
004147.71$+$405737.1 & 601 & 39 & 27 & 25 &  & 40 &  &  & 52 & 21 & 26 & He$^{+2}$  & 0.71 \\
004155.61$+$410846.7 & 566 & 54 &  & & 54 & 93 &  &  & 48 & 22 &  & O$^{+2}$  & 0.81 \\
004156.21$+$410735.0& 825 & 56 & & 110 & 44 & 54 & 13 & 43 & 47 & 9 & 24 & O$^{+5}$  & 3.22 \\
004216.70$+$404415.7 & 454 & 45 & 36 & 8 &  & 27 &  &  & 23 & 8 & 7 & He$^{+2}$& 1.99  \\
004233.17$+$412720.7 & 1064 & 30 & 26 & 103 & 26 & 69 & 28 & 35 & 43 & 16 & 39 &  Fe$^{+9}$ & 2.40 \\
004235.59$+$410148.0 & 841 &  &  & 80 &  & 144 &  &  & 65 & 33 & 41 & He$^{+2}$& 0.69 \\
004241.96$+$415656.4&452&46& 32 & &32&176&&&19&5&11&Ne$^{+2}$ & \cr
004319.98$+$415137.5 & 527 & 30 & 18 & 41 & 15 &  &  & 34 & 30 & 16 & 18 & O$^{+5}$  & \\
004323.68$+$413733.6 & 1115 & 27 & 16 & 50 & 21 & 132 & 17 & 51 & 28 & 14 & 23 & O$^{+5}$ & 4.29\\ 
004334.79$+$413447.9 & 716 & 29 & 18 &  & 10 & 30 &  &  & 31 & 20 & 14 & Ne$^{+2}$  & 2.12\\
004335.01$+$414358.2 & 429 & 42 & 32 & & 34 & 104 & & & 32 & 27 & 29 & Ne$^{+2}$ & 0.88 \\
004349.54$+$413855.9 & 572 & 31 &  22 & & 17 & 50 &  &5 & 38 & 14 & 32 & Ne$^{+2}$  &\\
004353.59$+$415323.3 & 528 & 38 & 23 & 47 &4 & 16 & 20 &  & 17 & 11 & 12 & Fe$^{+6}$ & 2.56 \\
004358.20$+$412850.9 & 403 & 55 & 19 & 11 &  & 12 &  &  & 23 & 26 & 22 & He$^{+2}$ & 0.85 \\
004359.52$+$410253.5 & 553 & 47 & 34 & 116 &  &  &  & 46 & 12 &  & 10 & O$^{+5}$  & 1.05\\
004411.71$+$411336.0$^1$ & 1133 & 26 & 14 & 57 & 11 & 33 & 1 & 74 & 23 & 9 & 19 & O$^{+5}$ & 8.61\\
004411.71$+$411336.0$^2$ & 1017 & 45 & 29 & 84 & 30 & 74 &  & 85 & 27 & 6 & 21 & O$^{+5}$ & \\
004421.89$+$415125.6 & 576 & 37 & 24 & 12 & 42 & 118 &  &  & 53 & 10 & 39 & He$^{+2}$ & 6.24 \\
004432.09$+$411940.8&543&34& 18 &84&52&243&21&23&16&3&12&O$^{+5}$  & 1.88\\
004433.74$+$414402.8 & 639 & 35 & 18 & 31 & 8 & 28 & 7 & 2 & 28 & 16 & 23 &  O$^{+5}$  & 4.60\\
004440.50$+$412052.2&370&41& 18 &24&64&305&&&21&5&14&He$^{+2}$ & 1.00\cr
004445.03$+$414156.5 & 1103 & 28 & 23 & 50 & 24 & 93 &  & 77 & 24 & 13 & 19 & O$^{+5}$  & \\
004521.55$+$412557.3 &623&34& 17 & 55&37&244&21&12&14&6&9&O$^{+5}$ &  \\
004534.07$+$413049.0 & 454 & 43 & 31 & 17 &  & 40 &  &  & 24 & 16 & 24 & He$^{+2}$ & 1.73 \cr

\hline
\multicolumn{13}{c}{Possible symbiotic stars}  \cr
\hline
003923.79$+$400543.4 & 733 &  &  &  & &  &  &  & 34 &3 & 8 & He$^{+}$  & \\
004237.49$+$403813.2 & 932 & 37 & 20 &  &  &  &  &  & 10 &4 & 10 & He$^{+}$ & 2.19 \\
004322.50$+$413940.9 & \multicolumn{2}{c}{P Cyg profile}& & &  &  &  &  &  &  &    & H$^{+}$  & 2.00 \\
004545.41$+$412851.6 & 495 & 52 & 32 &  &  &  &  &  &  &  &  & Fe$^{+2}$ & 2.97\\
\hline
\end{tabular}
\begin{list}{}{}
\item {$^1$} observed in November 2006
\item {$^2$} observed in  October 2011
\item {$^3$} in units of $10^{-15}\, \rm erg\,s^{-1}\,cm^{-2}$
\end{list}
\end{table*}

\section{Characterization of the new S\lowercase{y}S\lowercase{t} in M31}\label{discussion}

We searched for matches to all new and possible SySt in all catalogs available in the CDS database via VizieR\footnote{http://vizier.u-strasbg.fr/viz-bin/VizieR}. 
Table\,\ref{Tid} lists all matches within 0.5 arcsec radius. 
24 objects are listed by \citet{azimlu} in their the catalogs of \mbox{H\,{\sc ii}} regions and PNe. They also provide estimates for the observed H$\alpha$ fluxes (see Table\,\ref{Tsp}), the M31 interstellar extinction, and the extinction-corrected total H$\alpha$ luminosity, $L(\rm H\alpha)$ (see also Sect.\,\ref{parameters}). 

Six objects coincide at the 0.2--0.5 arcsec level with the long period variables found by POINT-AGAPE survey of M31 \citep{An2004}.
These coincidences seem to be real. The periods quoted in the POINT-AGAPE Catalogue (PAC) for these SySt, 
$P \sim 300$--1000 days, fall in the range covered by the orbital periods of known SySt \citep{Mik2012}. These are often classified as semiregular long-period variables because of their complex photometric behaviours (both components can be intrinsic variables). 
Several SySt also have matches in the catalogues of \citet{VRJ2006}.
We briefly discuss these and some other noteworthy individual objects in Sect.\,\ref{individual}.

\subsection{Physical parameters}\label{parameters}

The physical parameters of the SySt sample are given in Table\,\ref{Tpar}. 
To estimate the hot component temperature, $T_{\rm h}$, we employed the simple formula of the form $T_{\rm h}$[1000\,K]=$IP_{\rm max}$[eV], where $IP_{\rm max}$ is the highest observed ionization potential \citep{MN94}. The accuracy of this method is $\sim$10 per cent provided that the highest ionization stage is indeed observed. Since the highest ionization stages easily observed in the optical spectra of SySt are those of Fe$^{+6}$ and O$^{+5}$, this method fails for $T_{\rm h} \ga 115\,000$\,K. The method also fails for S-type SySt with small and dense nebulae, and with $54\,000\, {\rm K} \ga T_{\rm h} \la 115\,000$\,K because of the lack of strong permitted lines corresponding to the highest $IP$ in this temperature range whereas  the forbidden emission lines are either faint or absent due to high nebular density. Therefore we set the upper limit for $T_{\rm h}$ by the \mbox{He\,{\sc ii}}4686/H$\beta$ ratio, and assuming case B recombination and cosmic He/H. 

The interstellar reddening values, $E(B-V)$, were either adopted from \citet{azimlu} or estimated from the \mbox{H\,{\sc i}} and \mbox{He\,{\sc i}} recombination lines. Unfortunately, in S-type SySt as well as some D-types, there are significant departures from case B conditions due to self absorption effects, and the reddening-free Balmer decrements are very steep with H$\alpha$/H$\beta \sim 5$--10 (e.g. \citealt{proga1996}). Similarly, the \mbox{He\,{\sc i}} line ratios are affected by both optical depth and collisions, especially in S-type SySt (e.g. \citealt{proga1994}, and references therein). Thus the  $E(B-V)$'s  in Table\,\ref{Tpar} were estimated using results of \citet{netzer1975} (\mbox{H\,{\sc i}}) and \citet{proga1994} (\mbox{He\,{\sc i}}). Although their accuracy is not better than $\sim 0.1$, the differences between our estimates and the values quoted by \citet{azimlu} in all but one case did not exceed 0.1. The absolute magnitudes were calculated assuming the distance to M31, \mbox{ $m$--$M$}=24.47, corresponding to $d$=780\,kpc \citep{VRJ2006}, and extinction-corrected.
Whenever possible, the extinction-corrected total H$\alpha$ luminosity, $L(\rm H\alpha)$, is also given in Table\,\ref{Tpar}.

\begin{table*}
 \centering
  \caption{Physical parameters of new SySt in M31}\label{Tpar}
  \begin{tabular}{@{}lcccrcccccc@{}}
\hline
M31SyS\,J & IR Type & Sp Type & E(B-V) & Ion & $T_{\rm h}\,[10^3$\,K] & $M_{\rm I}$ & $M_{\rm 814}$ & $M_{\rm U}$ & $M_{\rm 336}$ & L(H$\alpha$)$^2$\\
\hline
\multicolumn{11}{c}{Symbiotic stars}  \cr
\hline
003846.16$+$400717.0& S & M4 & 0.2 & Fe$^{+9}$ & 235$\div$320 &-4.1 &  &  &  & \\
003857.17$+$403132.2& S & M4 & 0$^1$ & C$^{+4}$ & 64   &-4.6 &  &-4.5 &  & 3.18/83\\
003932.01$+$401224.7 & S & M5 & 0.3 & O$^{+5}$  & 114$\div$160 &-4.6 &  &  &  & \\
003951.85$+$402153.8 & S & M3 & 0.17$^1$ & He$^{+2}$ & 54$\div$100 &-4.2 &  &  &  & 1.18/30\\
004005.96$+$401604.3 & S & M5/6 & 0.2 & O$^{+5}$ & 114$\div$170 &-5.1 &  &  &  &\\ 
 004021.10$+$404501.1 & S & M  & 0.20$^1$ & He$^{+2}$ & 54$\div$160 &-4.4 &  &-2.9 &  & 5.22/136\\
004059.40$+$405003.1 &  S & M  & 0.21$^1$ & O$^{+5}$ & 114$\div$200 & $\ga$-3.9 &  &  &  & 2.10/57\\
004117.11$+$404524.1 & S &  & 0.2 & O$^{+5}$ & 114$\div$170 &-4.0 &  &-4.1 &  & 5.82/143\\
004147.71$+$405737.1 &  S  & M& 0.3 & He$^{+2}$ & 54$\div$125 &-4.5 &  &  &  & 1.04/27\\
004155.61$+$410846.7 &   & M  & 0.09$^1$ & O$^{+2}$ & 35   &-4.8 &  &-4.1 &  & 0.74/19\\
004156.21$+$410735.0& & M   & 0.02$^1$ & O$^{+5}$ & 114$\div$360 &-4.0 &  &  &  & 2.56/66\\
004216.70$+$404415.7 & S & M  & 0$^1$ & He$^{+2}$ & 54$\div$90 &-5.1 &  &-3.5 &  & 1.46/38\\
004233.17$+$412720.7 &  S & M4 & 0.22$^1$ &  Fe$^{+9}$ & 235$\div$320 &-5.1 &  &  &  & 2.95/77\\
004235.59$+$410148.0 & S & M  & 0.53$^1$ & He$^{+2}$ & 54$\div$240 &-5.0 &  &  &  & 1.75/46\\
004241.96$+$415656.4 & D: &  &  & Ne$^{+2}$ & 41  &  &  &  &  & \\
004319.98$+$415137.5 & S & M4 & 0 & O$^{+5}$ & 114$\div$155 &-4.5 &  &  &  & \\
004323.68$+$413733.6 & S &  & 0.24$^1$ & O$^{+5}$ & 114$\div$170 &-4.0 &  &  &  & 5.48/143\\
004334.79$+$413447.9 & S & M  & 0.18$^1$ & Ne$^{+2}$ & 41  &-5.3 &-5.4 &-4.7 &-4.5 & 2.35/61\\
004335.01$+$414358.2 & S & M4/5& 0.2 & Ne$^{+2}$ & 41 & & & & & 1.03/27 \cr
004349.54$+$413855.9 & S & M5/6 & 0.2 & Ne$^{+2}$ & 41   & $\ga$-3.8  &-4.4 &  &-4.1 & \\
004353.59$+$415323.3 & S & M6 & 0.11$^1$ & Fe$^{+6}$ & 100$\div$160 &-4.1 &  &  &  & 2.43/64\\
004358.20$+$412850.9 & S & M  & 0.26$^1$ & He$^{+2}$ & 54$\div$100 & $\ga$-3.9 &-3.9 &  &-1.8 & 1.15/30\\
004359.52$+$410253.5 &  S & M4/5 & 0.1 & O$^{+5}$ & 114$\div$300 &-4.1 &  &  &  & 0.97/25\\
004411.71$+$411336.0 &  S  & M4/6 & 0.11$^1$ & O$^{+5}$ & 114$\div$260 &-5.5 &-5.2 &-4.0 &-3.8 & 8.26/216\\
004421.89$+$415125.6 & D   & M &0.3 & He$^{+2}$ & 54$\div$100 &-5.2 &  &-5.6 &  & 9.5/243\\
004432.09$+$411940.8 & D &  & 0.23$^1$ & O$^{+5}$ & 114$\div$260 &-3.3 &-3.6 &-4.2 &  & 2.73/71\\
004433.74$+$414402.8 & S & M4 & 0 & Fe$^{+6}$ & 100$\div$145 &-5.3 &-4.6 &-3.5 &-2.6 & 3.37/88\\
004440.50$+$412052.2 & D &  & 0.21$^1$ & He$^{+2}$ & 54$\div$125 &  &-2.6 &  &-4.0 & 1.21/32\\
004445.03$+$414156.5 & S & M5 & 0.4 & O$^{+5}$ & 114$\div$170 &-4.9 &-3.8 &-4.7 &-5.2 & \\
004521.55$+$412557.3 &  D &  & 0.5 & O$^{+5}$ & 114$\div$180 &  &-2.1 &  &-4.5 & \\
004534.07$+$413049.0 & S &  & 0.1 & He$^{+2}$ & 54$\div$110 &-5.1 &-5.0 &-4.4 &-4.2 & 1.60/42\\
\hline
\multicolumn{11}{c}{Possible symbiotic stars}  \cr
\hline
003923.79$+$400543.4 &   & M & 0 & He$^{+}$ & 25   &-4.8 &  &-3.8 &  & \\
004237.49$+$403813.2 &   & M  & 0.5 & He$^{+}$ & 25   &-4.1 &  &-4.4 &  & 5.10/132\\
004322.50$+$413940.9 &   & M & 0.40$^1$ & H$^{+}$ & 14   &-6.1 &  &-5.7 &  & 3.79/99\\
004545.41$+$412851.6 &   & M3/5 & 0 & Fe$^{+2}$   & 16  &-5.4 &  &-4.5 &  & 2.18/57\\
\hline
\end{tabular}
\begin{list}{}{}
\item {$^1$} adopted from \citet{azimlu}
\item {$^2$} in units of $10^{35}\,\rm erg\,s^{-1}$/L$\sun$ 
\end{list}
\end{table*}

\begin{figure}
\resizebox{\hsize}{!}{\includegraphics{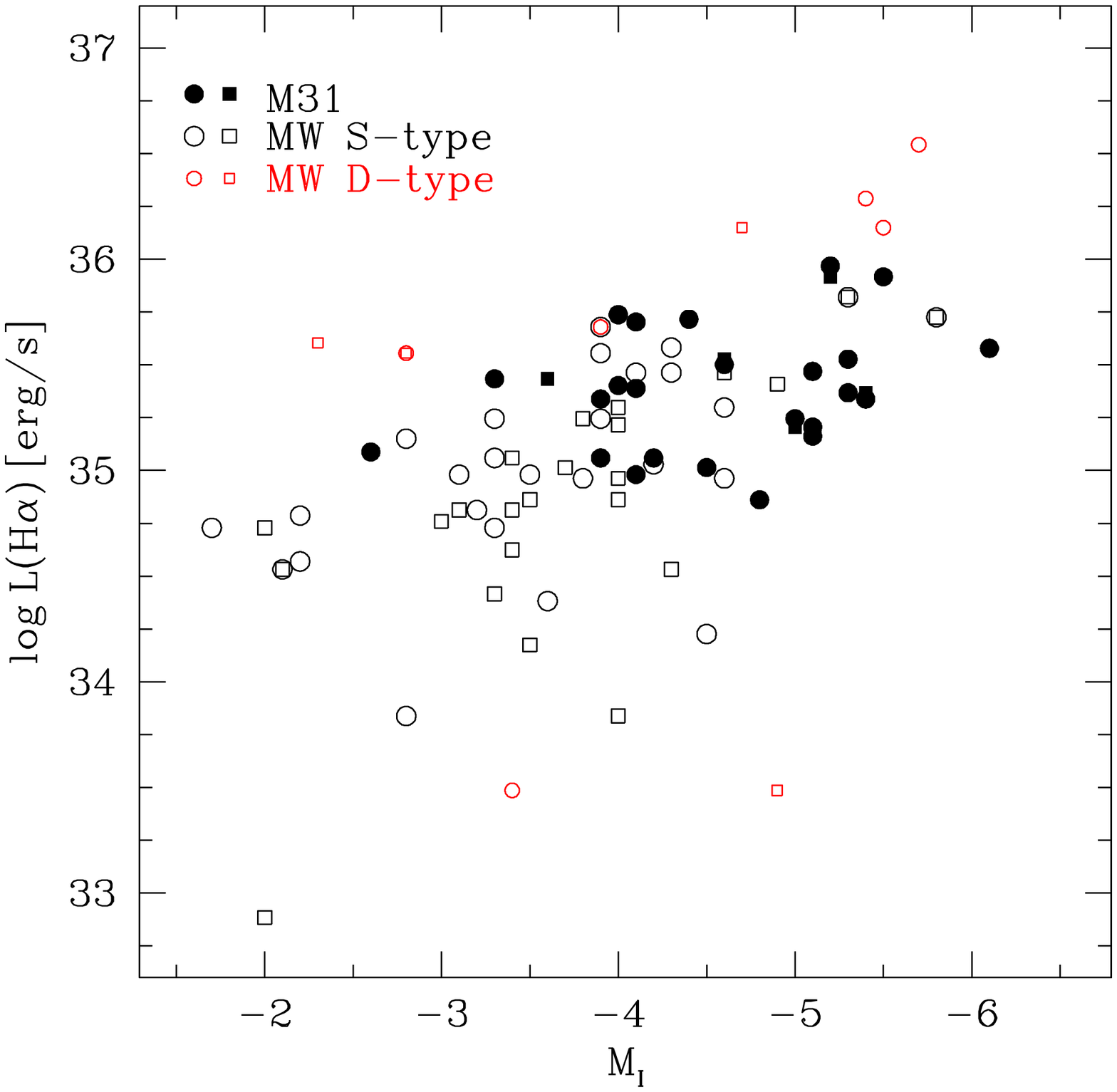}}
\caption{The H$\alpha$ total luminosity vs. $M_{\rm I}$ for the M31 (filled symbols) and those in Milky Way (open symbols).}\label{corr}
\end{figure}

The H$\alpha$ total luminosity, and the absolute $U$ magnitude ($M_{\rm U}$) characterize the nebular emission which is related to the hot component luminosity and temperature.  Thus it is not surprising that there is some positive correlation between these two quantities.
The absolute $I$ magnitude ($M_{\rm I}$) characterizes the cool star in S-type SySt. In the case of dust obscured symbiotic Miras (D-types), their $I$ magnitudes may not be representative of their intrinsic brightnesses, and in the most extreme cases, they can be dominated by the nebular emission.
Although there is not sufficient information about the near-IR $JHKL$ colours for our M31 SySt, their PHAT $m_{\rm 814}-m_{\rm 160}$ colours, whenever available, seem to be consistent with cool stellar photospheres rather than warm dust emission. Therefore the correlation between $L(\rm H\alpha)$ and $M_{\rm I}$ 
(Fig.\,\ref{corr}) indicates that there may be some relation between the hot component luminosity and that of the cool component. Such an effect is also observed in Galactic SySt (Fig.\,\ref{corr}). 
Moreover \citet{MAS1997} found a relation between the hot component luminosity and the spectral class of the cool giant in the sense that the brightest hot components are the companions of the coolest and the most luminous M giants.

\subsection{Comparison of the M31 SySt with those in our own Galaxy}

In the following section, we compare the M31 sample with the SySt in the Milky Way. 
To calculate the  H$\alpha$ total luminosity, $L(\rm H\alpha)$, and the absolute $U$ and $I$ magnitudes, $M_{\rm U}$ and $M_{\rm I}$ for the galactic SySt, we collected the available relevant information. In particular, the 
the H$\alpha$ fluxes for various samples of SySt were published by: \citet{Blair1983},  \citet{MAS1997}, \citet{Gutierrez1999}, and for individual SySt by: \citet{Kenyon1987},
\citet{Kenyon1991}, \citet{Kenyon1993}, \citet{Brandi2005}, 
\citet{Mik1992}, \citet{Schmid1990}. Most of these papers also contain information about the  interstellar reddening, $E(B-V)$, and distances. Whenever different estimates for the same object were available, the $E(B-V)$'s were critically revised by one of us (JM). 
The $UBVI$ photometry was taken from \citet{munari1992}, and also from the DENIS catalogue available in the CDS database via VizieR.
Although the DENIS $I$ filter and the Cousins $I$ filter, used by \citet{munari1992} and \citet{Massey2006}, differ slightly in their pivot wavelengths, the resulting magnitudes differ by no more than 0.1--0.2, which is less than the intrinsic variability of the majority of symbiotic red giants (e.g. \citealt{grom2013}, \citealt{angeloni2013}).
For these galactic SySt we used the H$\alpha$ fluxes, and $U$ mag observed during quiescence. 
During optical outbursts of any type (Z And-type, high states of accretion-powered SySt, symbiotic novae; e.g. \citealt{Mik2003}) 
the continuum is cooler, and $UBV$ magnitudes much higher.
 
\begin{figure}
\resizebox{\hsize}{!}{\includegraphics{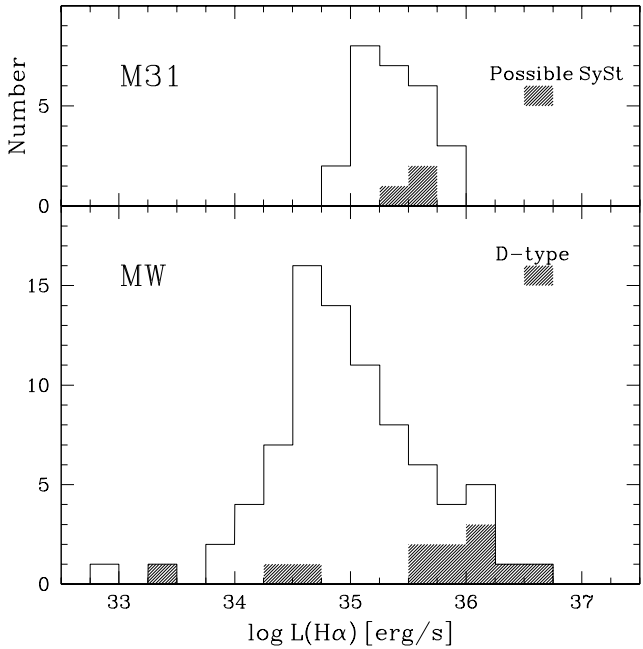}}
\caption{The H$\alpha$ total luminosity, $L(\rm H\alpha)$, distribution. The values for the M31 SySt are from \citet{azimlu} catalog; those for the galactic SySt - see text}\label{LHalpha}
\end{figure}

\begin{figure}
\resizebox{\hsize}{!}{\includegraphics{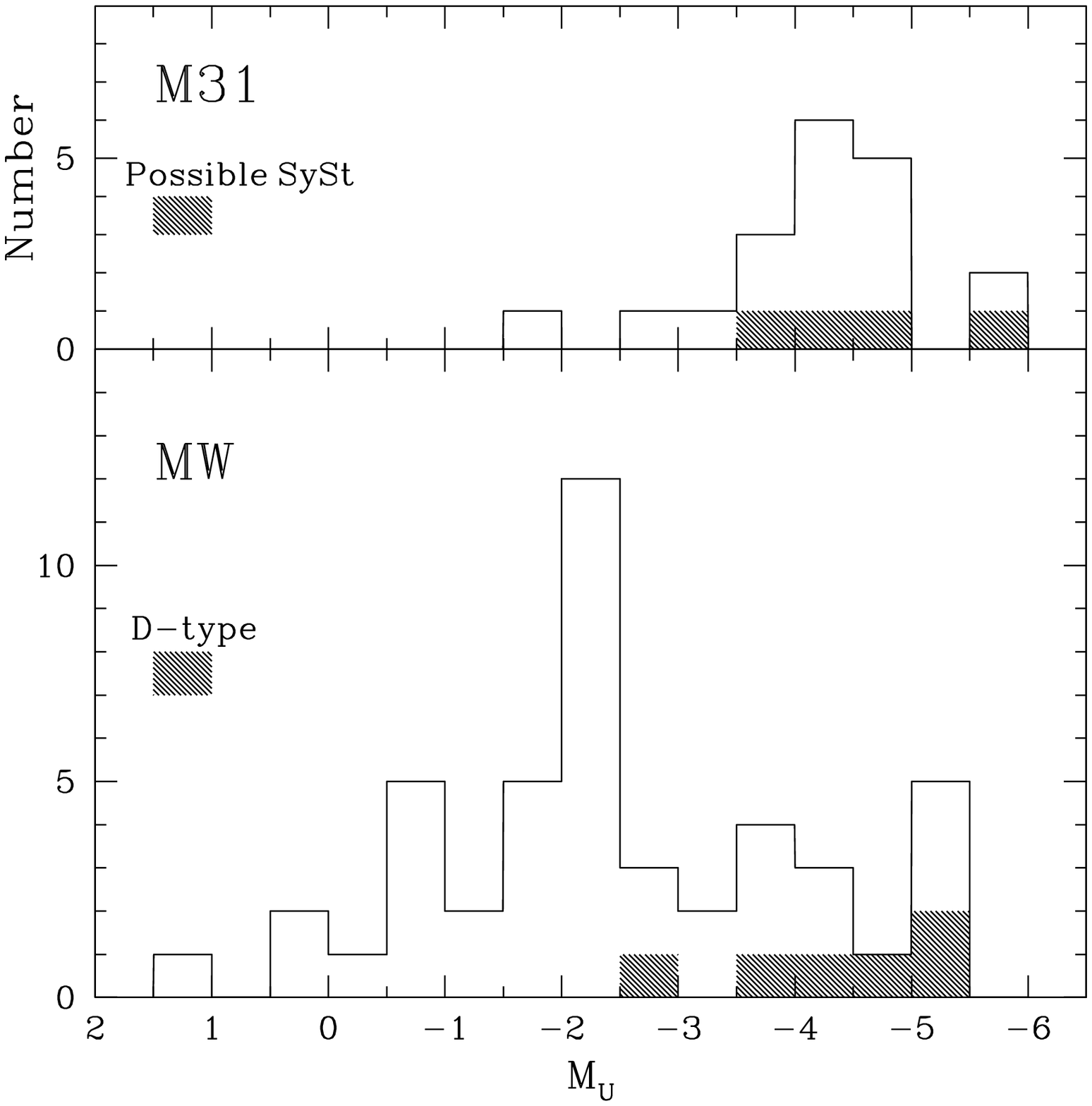}}
\caption{The absolute magnitude,$M_U$, distribution.}\label{MU}
\end{figure}

\begin{figure}
\resizebox{\hsize}{!}{\includegraphics{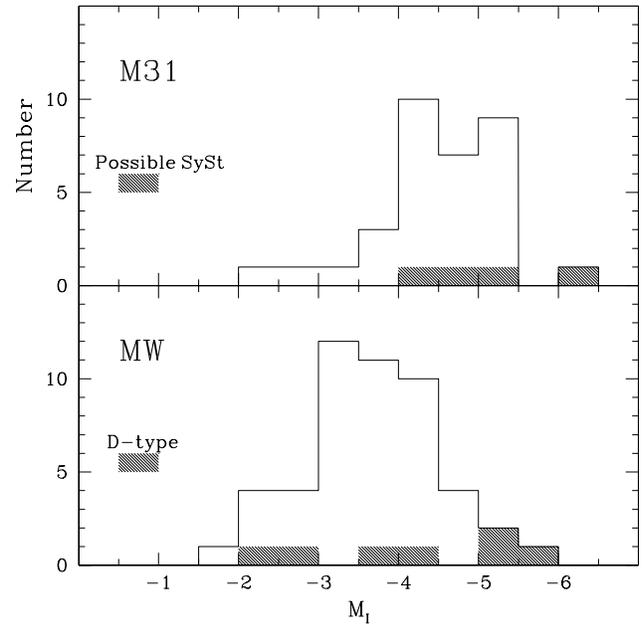}}
\caption{The absolute magnitude,$M_I$, distribution.}\label{MI}
\end{figure}

Figs.\,\ref{LHalpha}--\ref{MI} compare the distribution of the H$\alpha$ total luminosity and the absolute magnitudes,  $M_{\rm U}$ and $M_{\rm I}$,  of the M31 and galactic SySt. 
Some of the galactic D-type SySt show surprisingly low $M_{\rm I}$. This is because their Mira components are obscured by optically thick dust, and their $I$ magnitudes do not represent their intrinsic brightnesses. 
These plots show that we are clearly sampling similar populations, but
that we are incomplete in M31 for all but the most luminous SySt.
With a targeted campaign, we should be able to detect objects down to $\log\,L(\rm H\alpha)=34.3$ (the limit of the Azimlu et al. (2011) catalog), well below the peak in the MW, and thus
allow a rigorous comparison between the populations in the two galaxies.

Finally, \citet{frank2009} claimed that the intrinsic luminosity function in \mbox{[O\,\sc{iii}]} $\lambda$\,5007 line of the galactic symbiotic nebulae (SyStNLF) have very similar cutoff luminosity and general shape to those of the planetary nebulae. Unfortunately, we cannot estimate the \mbox{[O\,\sc{iii}]} luminosity for our M31 sample because the fiber spectra do not have absolute flux calibration. Nevertheless, the relative  \mbox{[O\,\sc{iii}]}\,5007/H$\alpha$ fluxes for majority of our M31 SySt are lower than $\sim 0.1$, and they never exceed 1. This fact combined with the total H$\alpha$ luminosity indicates that the  \mbox{[O\,\sc{iii}]} luminosity is much lower than the PN "universal" bright cutoff. In particular, the most luminous D-type SySt in our sample have $L$(\mbox{[O\,\sc{iii}]}) ~66, 39, and 32 L\sun, respectively, which is at least an order of magnitude lower than the value of the cutoff, $L$(\mbox{[O\,\sc{iii}]}) $\simeq$ 600\,L\sun, reported by \citet{frank2009}. 

\subsection{Notes on some individual objects}\label{individual}

\subsubsection{M31SyS\,J004147.71$+$405737.1}

\citet{Shafter} reported an optical nova, M31N 1993-05, $\sim 1.5$ arcsec away from M31SyS\,J004147.71+405737.1. However, their nova position's accuracy is, at best, $\sim 1$ arcsec, and the putative nova and our SySt may well be the same object.
M31N 1993-05, with its $m_{\rm H\alpha}=18.9$, was one of faintest nova candidates in their survey. Taking into account their H$\alpha$ filter width, $\Delta\,\lambda$=70\,\AA, and that $m_{\rm H\alpha}=0$ for $F(\rm H\alpha)$=$1.77 \times 10^{-7} \rm erg\,cm^{-2}\,s^{-1}$, we estimated $F(\rm H\alpha)$=$4.9 \times 10^{-15} \rm erg\,cm^{-2}\,s^{-1}$  for the nova. This is only $7\times$ higher than $F(\rm H\alpha)$=$7.1 \times 10^{-16} \rm erg\,cm^{-2}\,s^{-1}$ measured for M31SyS\,J004147.71$+$405737.1 between 2000 August and 2002 September by \citet{azimlu}. 
If the coincidence between M31SyS\,J004147.71$+$405737.1 and M31N 1993--05 is real, then it was 
a Z And-type symbiotic outburst rather than a real thermonuclear nova.
There is also a red variable,  [PAC]\,61099, only  0.16 arcsec away from our position of M31SyS\,J004147.71$+$405737.1, and the reported $P \sim 900^{\rm d}$  \citep{An2004} is consistent with the orbital periods observed for galactic S-type SySt, including those with the Z And-type outburst activity \citep{Mik2012}.

\subsubsection{M31SyS\,J004216.70$+$404415.7}

This object has a very close 2MASS\,J00421671$+$4044158 counterpart ($\la 0.2$ arcsec). The 2MASS $JHK$ combined with the LGGS $I$ are then consistent with an un-reddened M giant. In addition, there is a red variable, [PAC] \,57435 $\la 0.3$ arsec), with $P \sim 720^{\rm d}$, which seems to be the same object. The near IR colours and the period are then consistent with an S-type SySt.

\subsubsection{M31SyS\,J004233.17$+$412720.7}

\begin{figure*}
\centerline{\includegraphics[width=150mm]{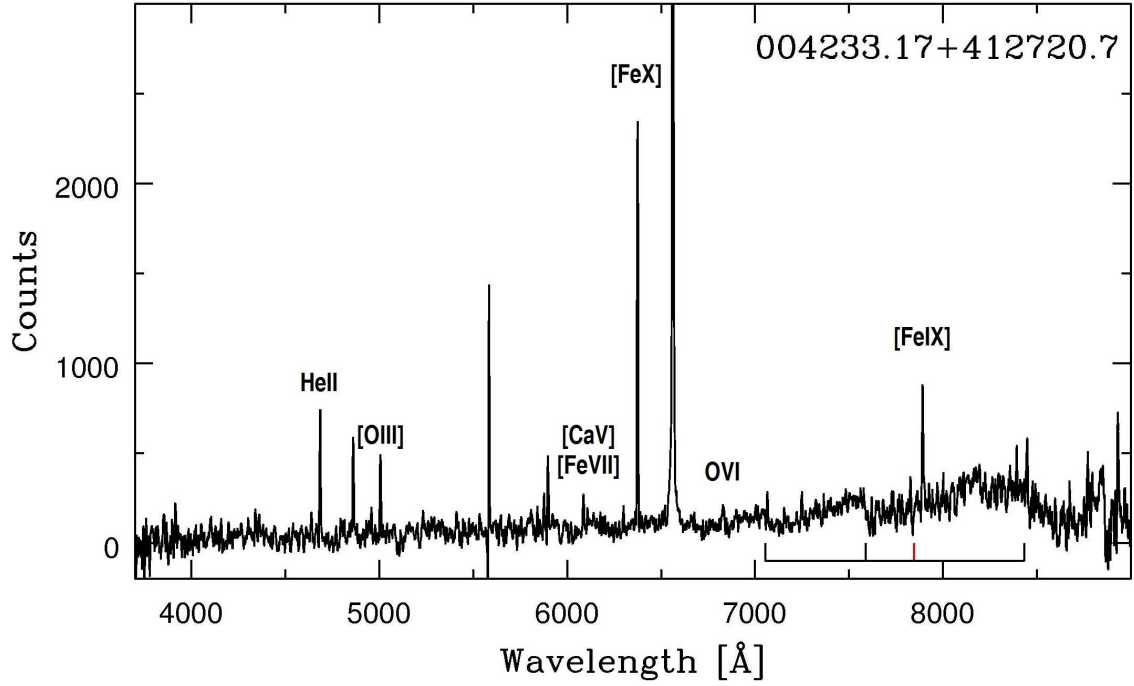}}
\caption{The spectrum of M31SyS\,J004233.17$+$412720.7, the SySt with the highest degree of ionization amongst all known SySt. The \mbox{[Fe\,{\sc x}]}\,$\lambda6374$ and \mbox{[Fe\,{\sc ix}]}\,$\lambda7892$ emission lines, and other high IP lines are indentified as well as the strongest molecular bands from the M giant.}\label{pn566}
\end{figure*}

The most exceptional property of this object is the presence of very strong coronal \mbox{[Fe\,{\sc x}]}\,$\lambda6374$, and \mbox{[Fe\,{\sc ix}]}\,$\lambda7892$ emission lines, that combined with the fairly strong Raman scattered \mbox{O\,{\sc vi}}, and \mbox{[Fe\,{\sc vii}]} emission lines (Fig.\,\ref{pn566}) places M31SyS\,J004233.17$+$412720.7 amongst those SySt with the highest degree of ionization. 
The \mbox{[Fe\,{\sc x}]} is  observed in some classical novae (e.g. \citealt{mclau1953}), and symbiotic recurrent novae \citep{Williams1991} during their nebular phase, but it is extremely rare in classical SySt. In principle, the presence of this line is well-documented only in SMC 3 which is also the most luminous supersoft X-ray source  (SSXS) associated with a SySt (e.g. \citealt{Jordan1996}, \citealt{Orio2007}), however, it is not as strong as in M31SyS\,J004233.17$+$412720.7 ( \mbox{[Fe\,{\sc x}]}/H$\beta \sim 0.5$ in SMC3, and $\sim 3$ in M31SyS\,J004233.17$+$412720.7).
A weak \mbox{[Fe\,{\sc x}]} was also reported in V2116Oph/GX1$+$4, the  famous SySt with a neutron star accompanied by an M6 giant \citep{David1977}, and recently also in the C-rich SySt 2MASS\,J16003761--4835228 \citep{MM2014}.

Among our new M31 SySt, this emission was detected in only one other object -- M31SyS\,J003846.16$+$400717.0 (Fig.\,A1).
There is no X-ray counterpart to M31SyS\,J004233.17$+$412720.7 and any other of our new SySt.
The nearest Chandra source, [VG2007] 257, is about 15 arsec away from M31SyS\,J004233.17$+$412720.7. This does not mean that the SySt does not produce any X-ray emission, as such emission could be absorbed locally (by the red giant wind), and/or by the interstellar medium. 
There is a bright red giant,  2MASS\,J00423313$+$4127207, $\sim 0.4$ arcsec away from the position of the SySt. Assuming that it is the cool component of M31SyS\,J004233.17$+$412720.7, the 2MASS $JHK$ colours are consistent with a moderately reddened, $E(B-V) \sim 0.4$, M4 giant. However, the $I-K \sim 4.3$ would then be too red for such a star. One cannot, however, exclude that the spectral type is in fact later, and/or the star is variable.

\subsubsection{M31SyS\,J004235.59$+$410148.0}

There is a red variable, {PAC]\,93855, with $P \sim 560^{\rm d}$, $\sim$ 0.5 arcsec away. The field is rather crowded, with a few other objects with similar $I$ mag, and this may be an accidental coincidence. 

\subsubsection{M31SyS\,J004322.50$+$413940.9}

The spectrum of this object, presented in Fig.\,\ref{pn739}, shows a very rich, low-excitation emission line spectrum, including \mbox{H\,{\sc i}}, numerous \mbox{Fe\,{\sc ii}} and  \mbox{[Fe\,{\sc ii}]}, and \mbox{Na\,{\sc i}}\,D, and very strong \mbox{Ca\,{\sc ii}}\,T lines.
The \mbox{H\,{\sc i}} and other permitted lines show P Cyg profiles with absorption component blue-shifted by $\sim 350\,\rm km\,s^{-1}$ with respect to the emission peak. In addition, the red giant presence is indicated by the TiO bands, and the rising continuum in the red spectral range.
Such a spectrum is typical for SySt during their strong optical outbursts. 
In particular, the similarity between the emission line spectrum of M31SyS\,J004322.50$+$413940.9 and that of RX Pup during its optical outbursts (e.g. \citealt{Swings1976}; \citealt{Mik1999}, and references therein) is remarkable. 
The $UBV$ colours and absolute magnitudes, as well as total H$\alpha$ luminosity of these two objects are also very similar: $M_{\rm U}\sim -5.2$ to --5.7, and $\sim -5.7$ and  $L(\rm H\alpha) \sim 180\, \rm L_{\sun}$, and $\sim 100\, \rm L_{\sun}$, in RX Pup and M31SyS\,J004322.50+413940.9, respectively.  

The coincidences between the position of the SySt and 2MASS  J00432249+4139409 (0.06 arcsec), M31V J00432248+4139408 (0.13 arcsec), BATC\_004322.51$+$413941.1 (0.26 arcsec) and [PAC]\,47174 (0.27 arcsec) are all real because there is no other bright
object near enough to be the SySt.
The $JHK$ colours are consistent with a moderately reddened, $E(B$--$V)\sim 0.5$, M4 giant. The presence of a red giant is also evident from the BATC spectral energy distribution. The $\sim$900-day period reported by \citet{An2004} very likely represents orbital variation. The shorter period $P \sim 80^{\rm d}$ \citep{VRJ2006} may be due to low-amplitude semiregular (SR) pulsation of the red giant, which is very common in Galactic and MC SySt (\citealt{grom2013}; \citealt{angeloni2013}).  

\begin{figure*}
\centerline{\includegraphics[width=150mm]{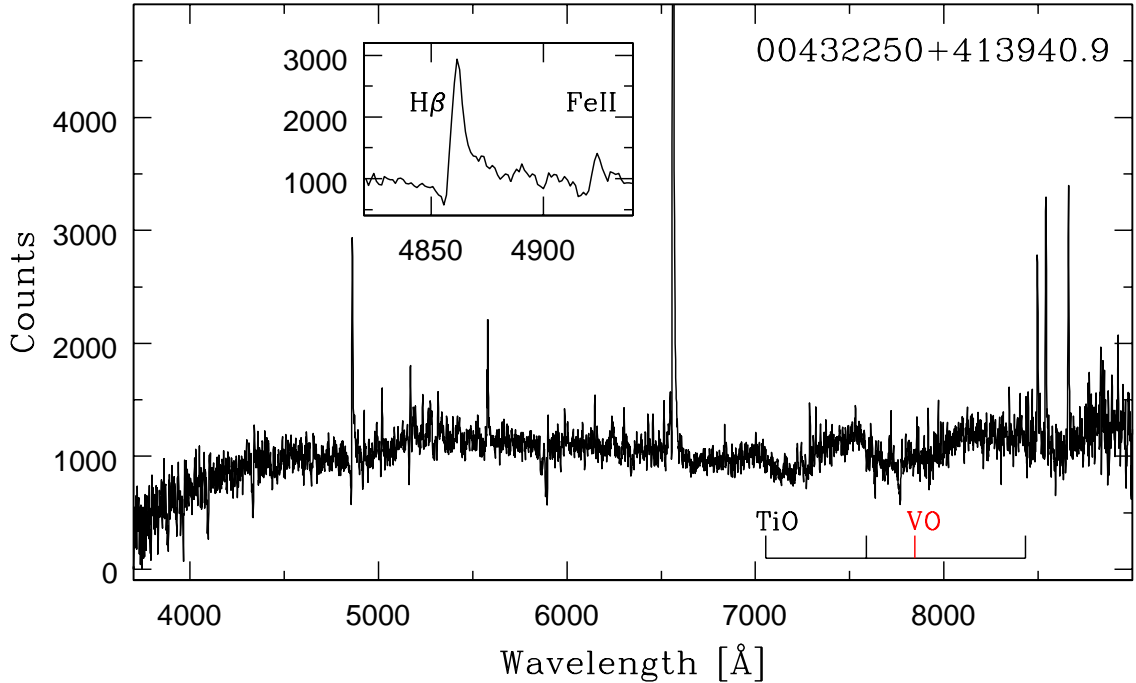}}
\caption{The spectrum of M31SyS\,J004322.50$+$413940.9, a possible SySt caught during an optical outburst.}\label{pn739}
\end{figure*}

\subsubsection{M31SyS\,J004323.68$+$413733.6}

The spectrum of this object (Fig.\,\ref{pn744}) has a very weak continuum, and its symbiotic nature is indicated only by the \mbox{O\,\sc{vi}} Raman scattered lines, and other high ionization emission lines. However, M31\,J00432365$+$4137335 is only $\la0.2$ arcsec away, with $V=22.5$ and $B=23.1$ \citep{VRJ2006}. It was also detected by WISE, and in addition there is the nearby red variable, [PAC]\,47312, with $P \sim 1000^{\rm d}$. The WISE magnitudes and colours are consistent with those of a red giant star.

\subsubsection{M31SyS\,J004334.79$+$413447.9}

\citet{VRJ2006} report a variable star, M31V J00433475$+$4134477, with  $V$=20.95 and $B$=21.46 very similar to LGGS magnitudes of the SySt (Table\,\ref{Tphot}), and with  $P\sim83^{\rm d}$  (the light curves could be interpreted as due to a SR variable) which they identified as the LGGS object. There is also  a red variable, [PAC]\,46291, with $P \sim 260^{\rm d}$, which could be the same object. The HST F814W image\footnote{http://www.cfa.harvard.edu/oir/eg/m31clusters/symb/symb.html} does not show any other bright and red object so close to M31SyS\,J004334.79$+$413447.9 (all these objects are within $\la 0.4$ arcsec).

\subsubsection{M31SyS\,J004411.71$+$411336.0}

This object was observed in November 2006, and October 2011. The two spectra, shown in Fig.\,2, do not show significant differences except for some change in relative emission line intensity ratios (e.g. \mbox{He\,\sc{ii}}4686/H$\beta$, \mbox{[O\,\sc{iii}]}/H$\beta$). These changes can reflect some orbitally related changes or be due to  intrinsic variability of the hot component. 
A comparison of the PHAT magnitudes (table\,\ref{Tphot}) with the blue and near IR magnitudes from BATC DR1 suggests that M31SyS\,J004411.71$+$411336.0 was $\sim 1$ mag brigher in the blue optical range during the BAT observations whereas its near IR brightness was more or less the same. 

\subsubsection{M31SyS\,J004421.89$+$415125.6}

\citet{VRJ2006} report a variable star, M31V J00442188+4151255, coinciding with the position of the SySt within $\sim 0.1$ arcsec, and with $V$=20.97, $B$=21.35, and low-amplitude variability with $P\sim80^{\rm d}$ which is most likely due to SR pulsations of the red giant. This would then indicate that M31SyS\,J004421.89$+$415125.6 is not a D-type SySt as suggested by its position in the \mbox{[O\,\sc{iii}]} diagram.

\subsubsection{M31SyS\,J004433.74$+$414402.8}

\citet{Bonanos2003} report a variable red star ($V$=20.82; $B$=21.29, $I$=19.07), D31\,J004433.7+414403.0, which they tentatively classify as Type II Cepheid. 
However, its red colour (V-I =1.75) and positional coincidence with the red giant 2MASS J00443374+4144032 suggest that it may be M31SyS J004433.74+414402.8, especially since there is no other nearby red object in the HST F814 image bright enough to be a match.
The $34^{\rm d}$-periodicity would then reflect a SR pulsation of the red component.

\subsubsection{M31SyS\,J004534.07$+$413049.0}     

\citet{VRJ2006} report a variable star, M31V\,J00453405+4130487, with similar $BV$ magnitudes and colour to those of the SySt, and  $P \sim 80^{\rm d}$  mostly likely due to SR variability of the red giant component.


\section{Conclusions}\label{conclusions}

We presented and discussed the first SySt detected in M31. 

The main conclusions are as follows:
\begin{enumerate}
\item A total of 31 SySt were confirmed. We tentatively identified 24 of them as S-types and 5 as D-types, however this classification must be confirmed by near-IR photometry. 4  additional objects were classified as possible symbiotic stars.
\item A comparison of these new SySt with the galactic SySt indicates that we are sampling similar binary star populations, however, we are incomplete in M31 for all but the most luminous SySt.
\item M31SyS\,J004233.17$+$412720.7 was found to be the SySt with the highest ionization level among all known SySt. 
\item Although M31SyS\,J004322.50$+$413940.9 was classified as a possible SySt, it is very likely that in 2006, when the Hectospec spectrum was taken, it was at an optical outburst maximum. Further observations are necessary to confirm its symbiotic nature.  
\end{enumerate}

More detailed spectral analysis of these new SySt will be the subject of a follow-up paper II.


\section{Acknowledgments}

This study has been supported in part by the Polish NCN grant
DEC-2013/10/M/ST9/00086.
We gratefully acknowledge Nathan Sanders, for help
with the emission line tables, and the fine support at the MMT Observatory, and
The Local Group Galaxy Survey conducted at NOAO by Phil Massey and
collaborators.
MS thanks Allen Shafter for helpful details about  M31N 1993--05.
We also thank the referee, R. Corradi, for useful comments.
This research has made use of the VizieR catalogue access tool, operated at CDS, Strasbourg, France.


\newpage

\appendix

\section[]{Hectospec spectra of the M31 S\lowercase{y}S\lowercase{t} and possible S\lowercase{y}S\lowercase{t}}

\begin{figure*}
\centerline{\includegraphics[width=\columnwidth]{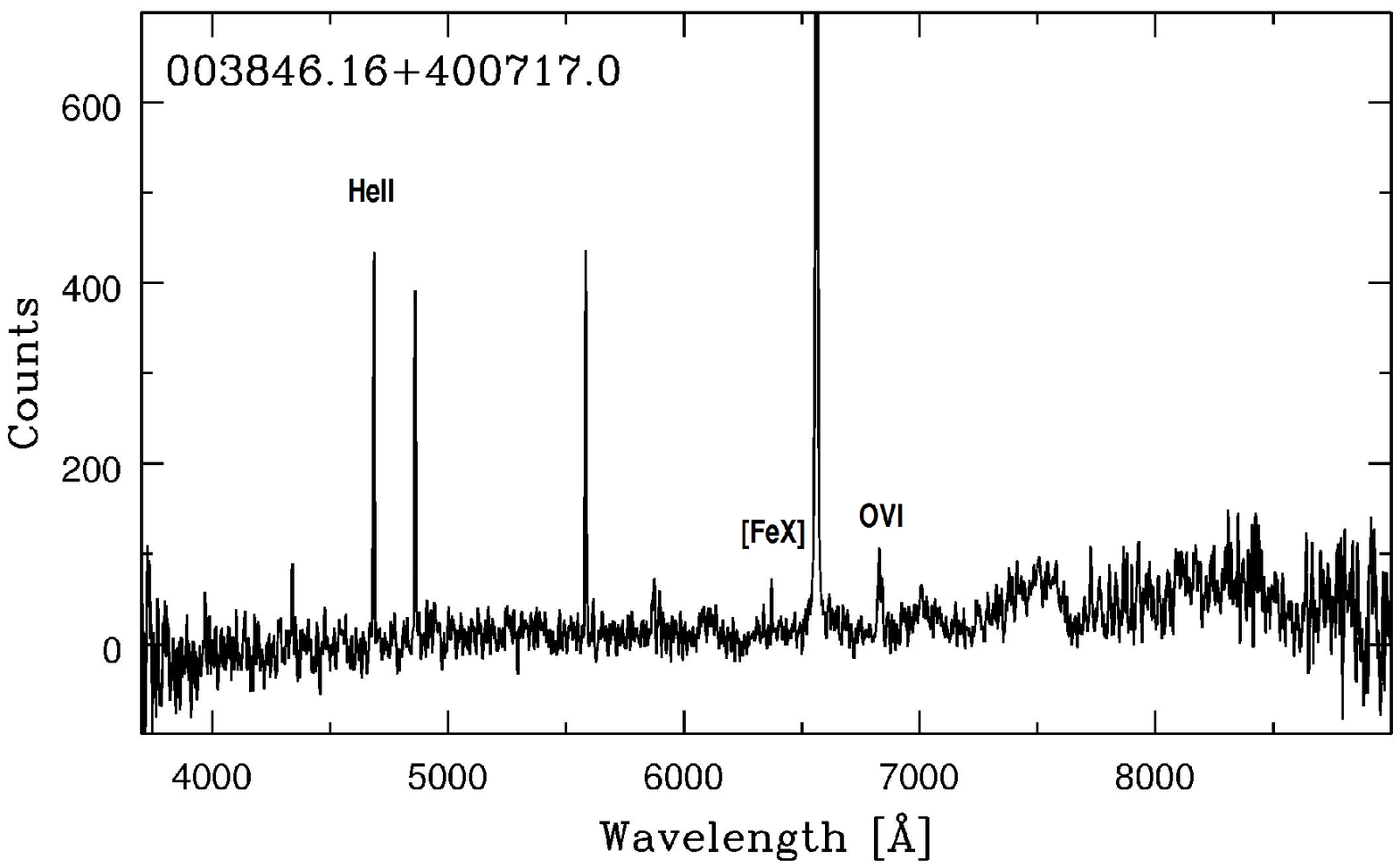}
\includegraphics[width=\columnwidth]{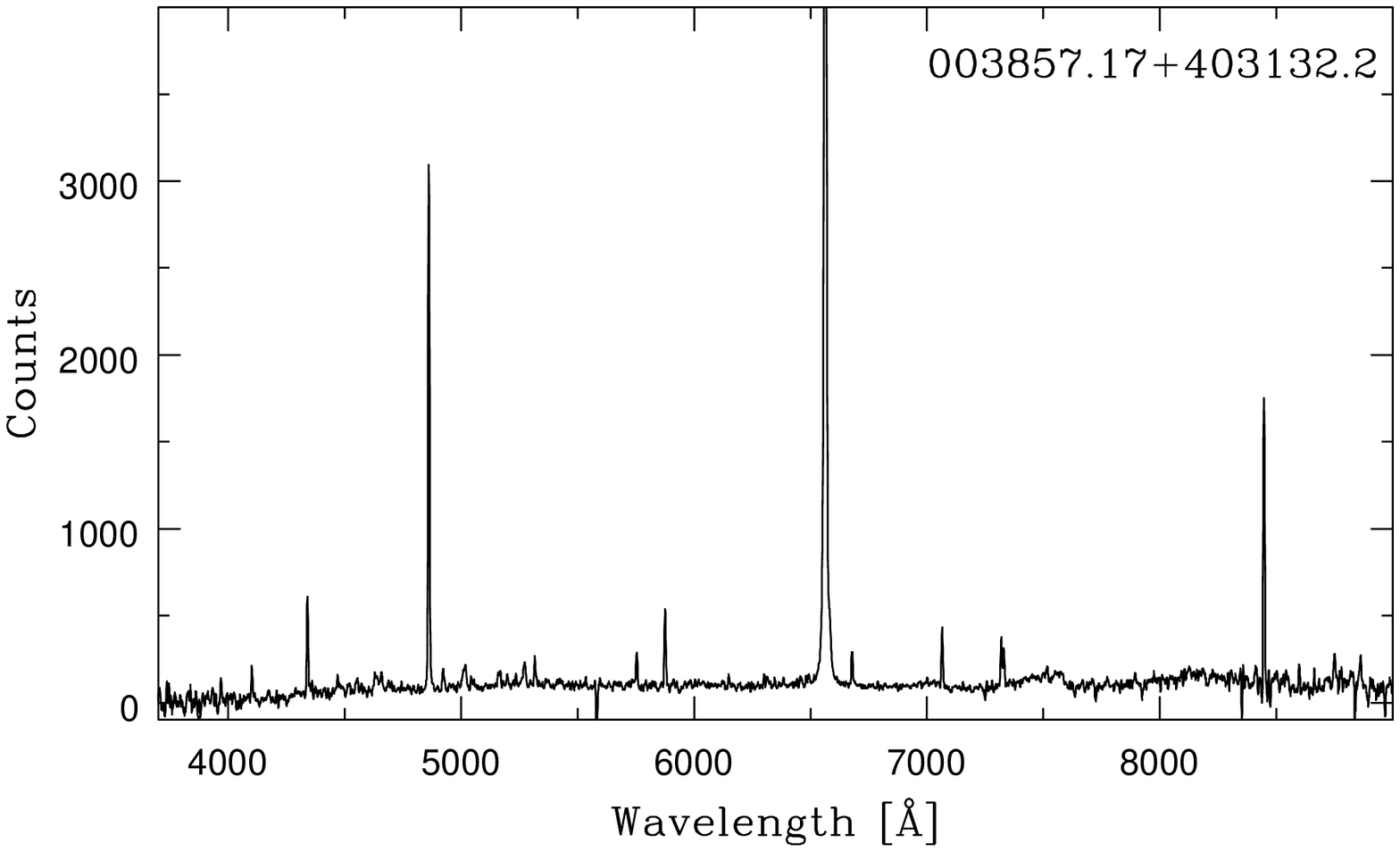}}
\centerline{\includegraphics[width=\columnwidth]{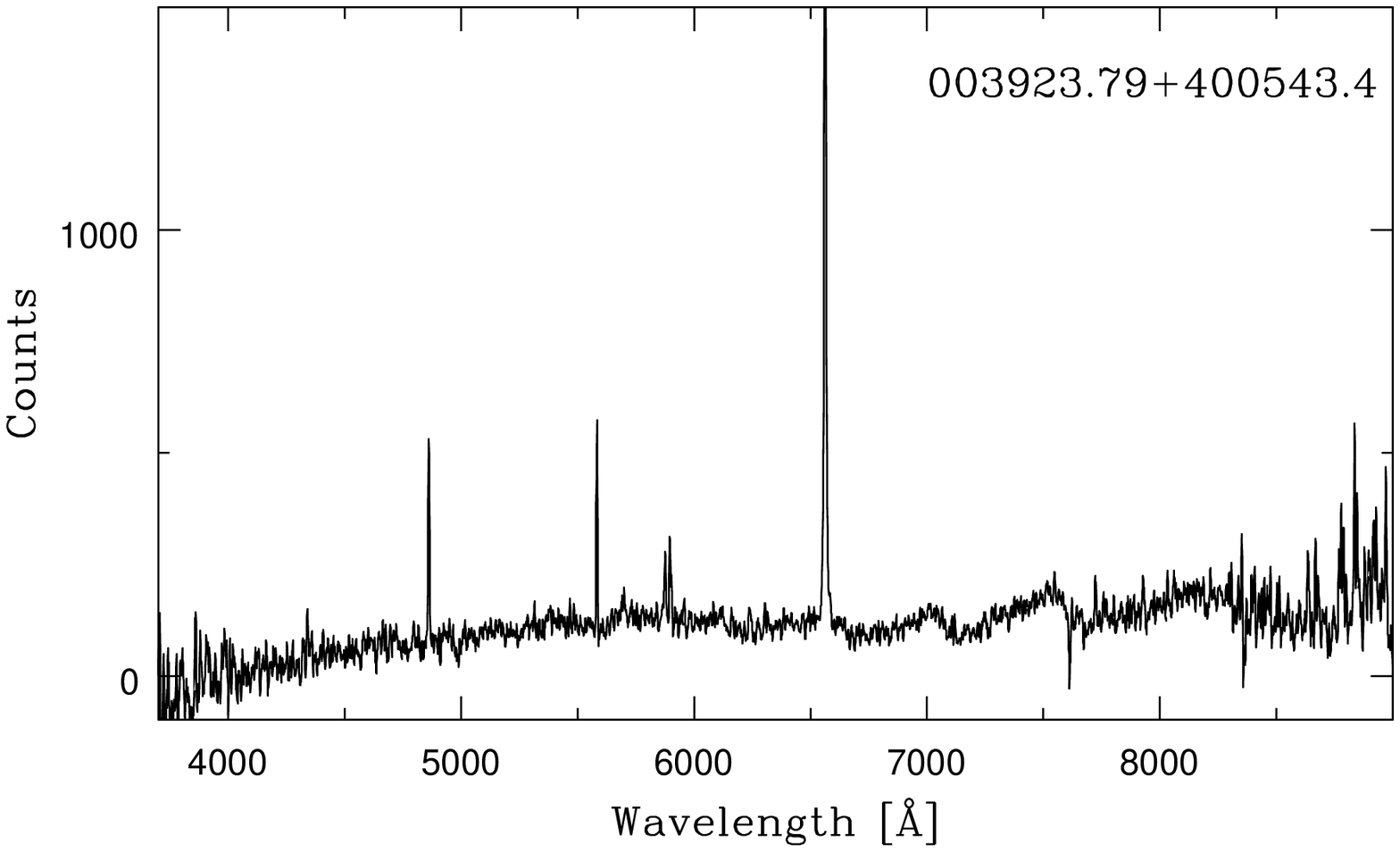}
\includegraphics[width=\columnwidth]{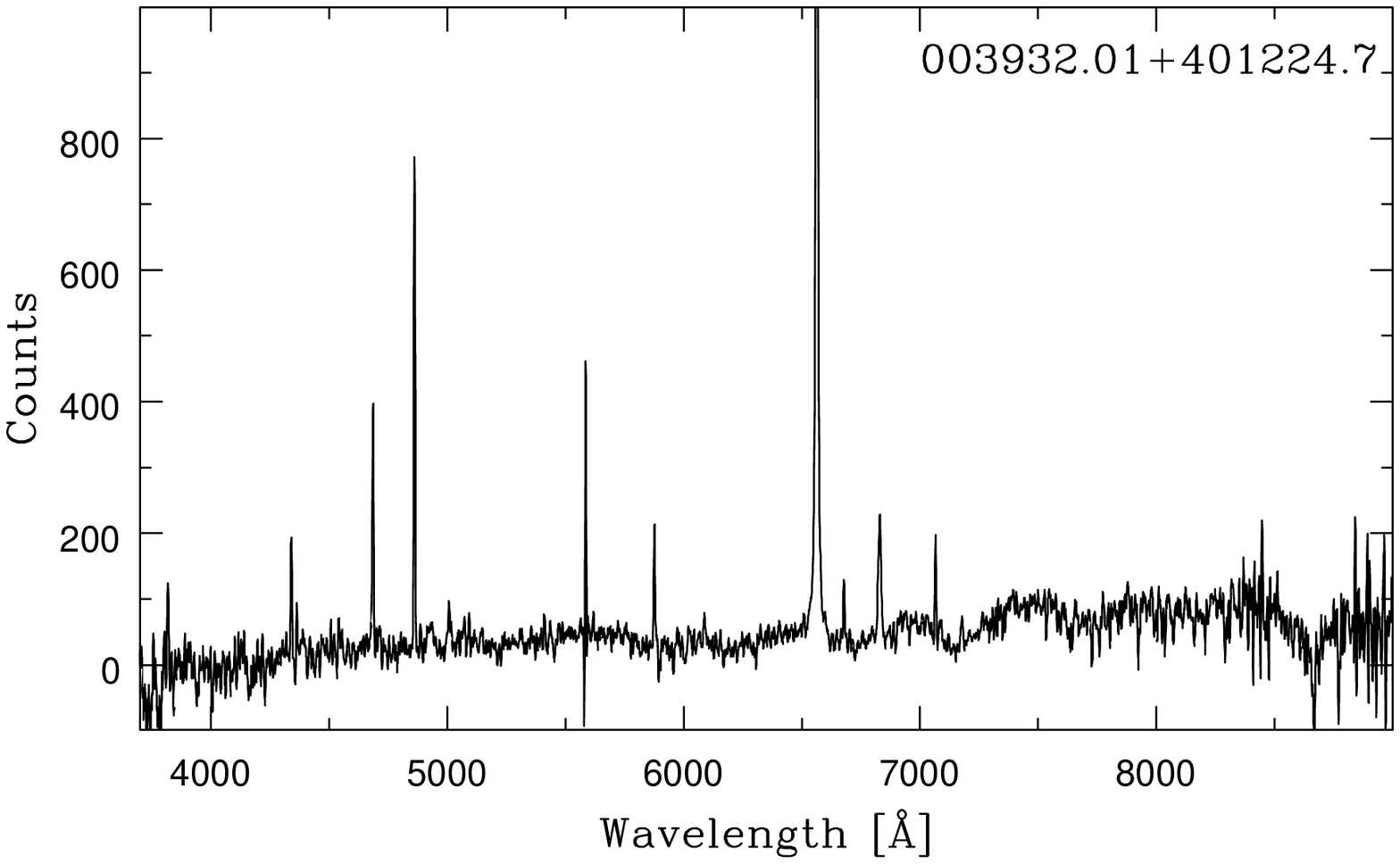}}
\centerline{\includegraphics[width=\columnwidth]{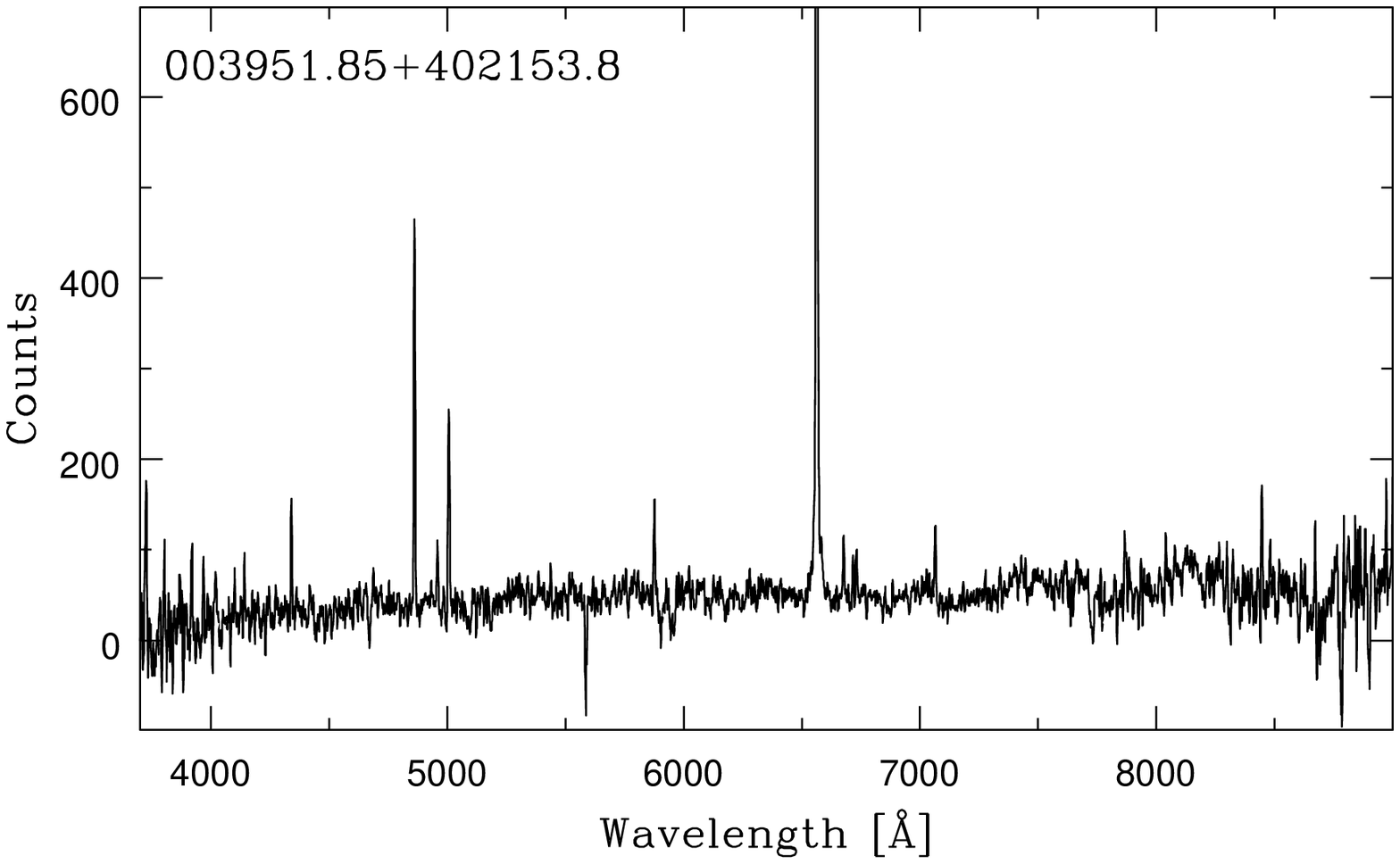}
\includegraphics[width=\columnwidth]{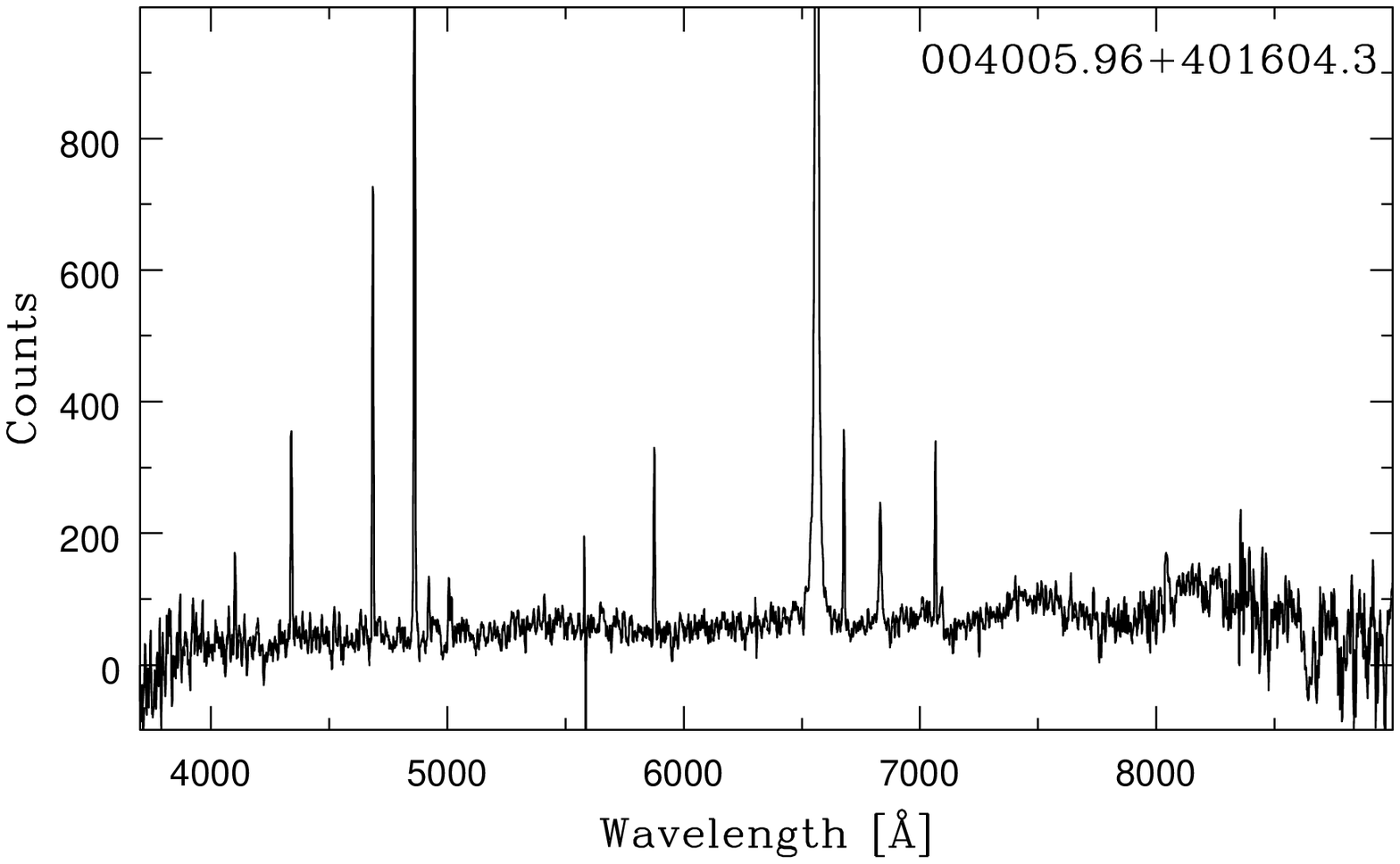}}
\centerline{\includegraphics[width=\columnwidth]{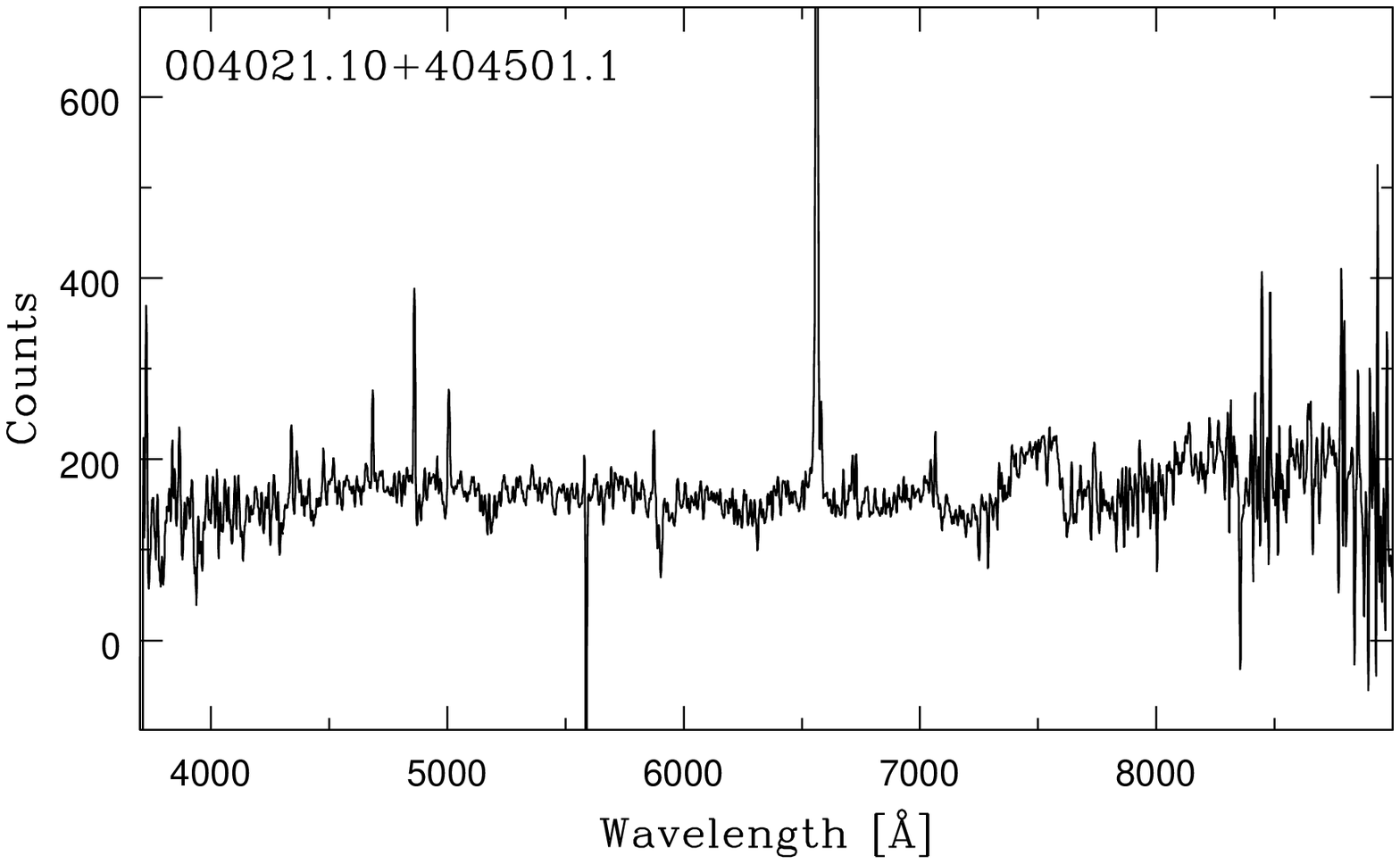}
\includegraphics[width=\columnwidth]{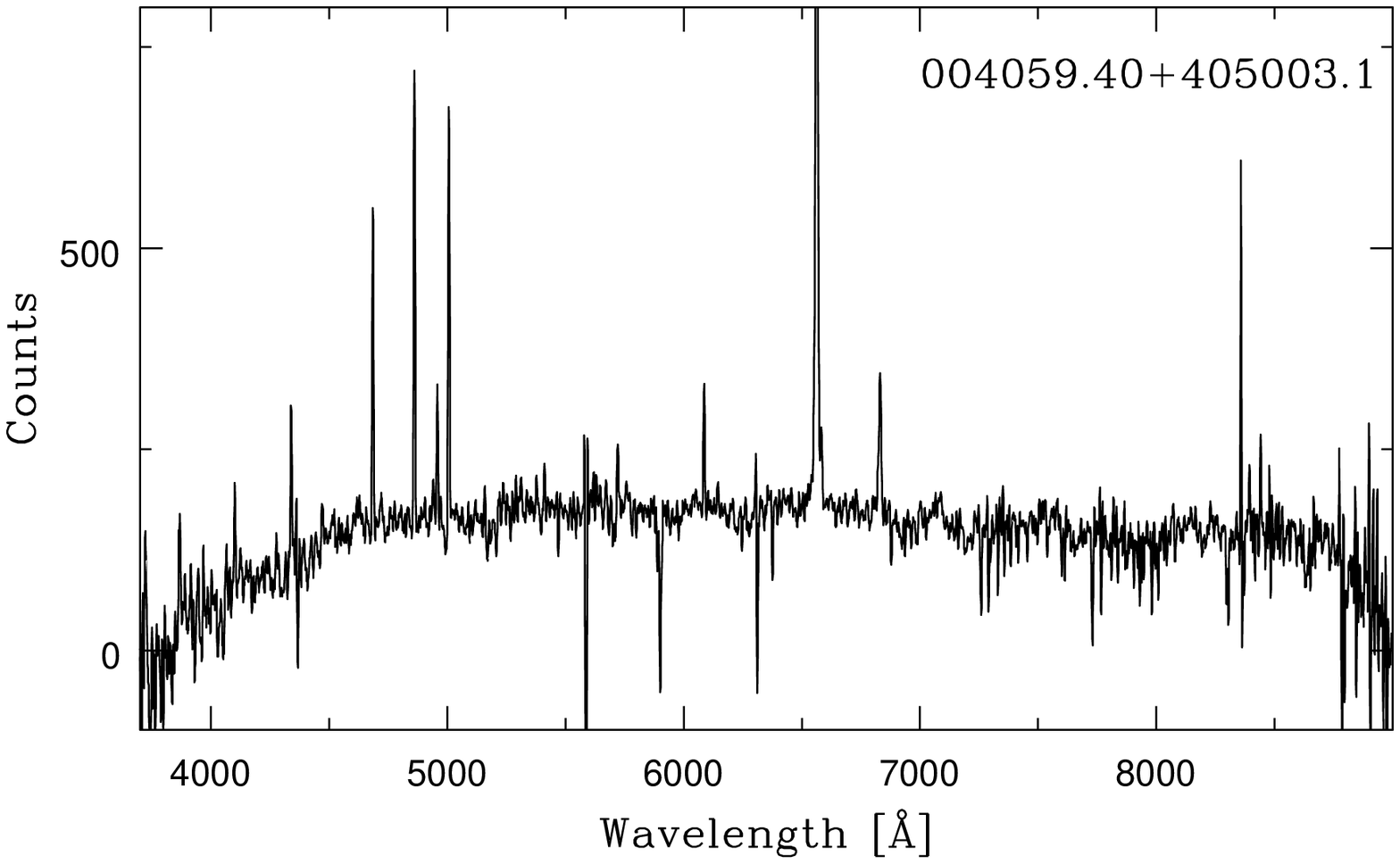}}
\caption{Spectra of SySt in M31}\label{sp1}
\end{figure*}

\begin{figure*}
\centerline{\includegraphics[width=\columnwidth]{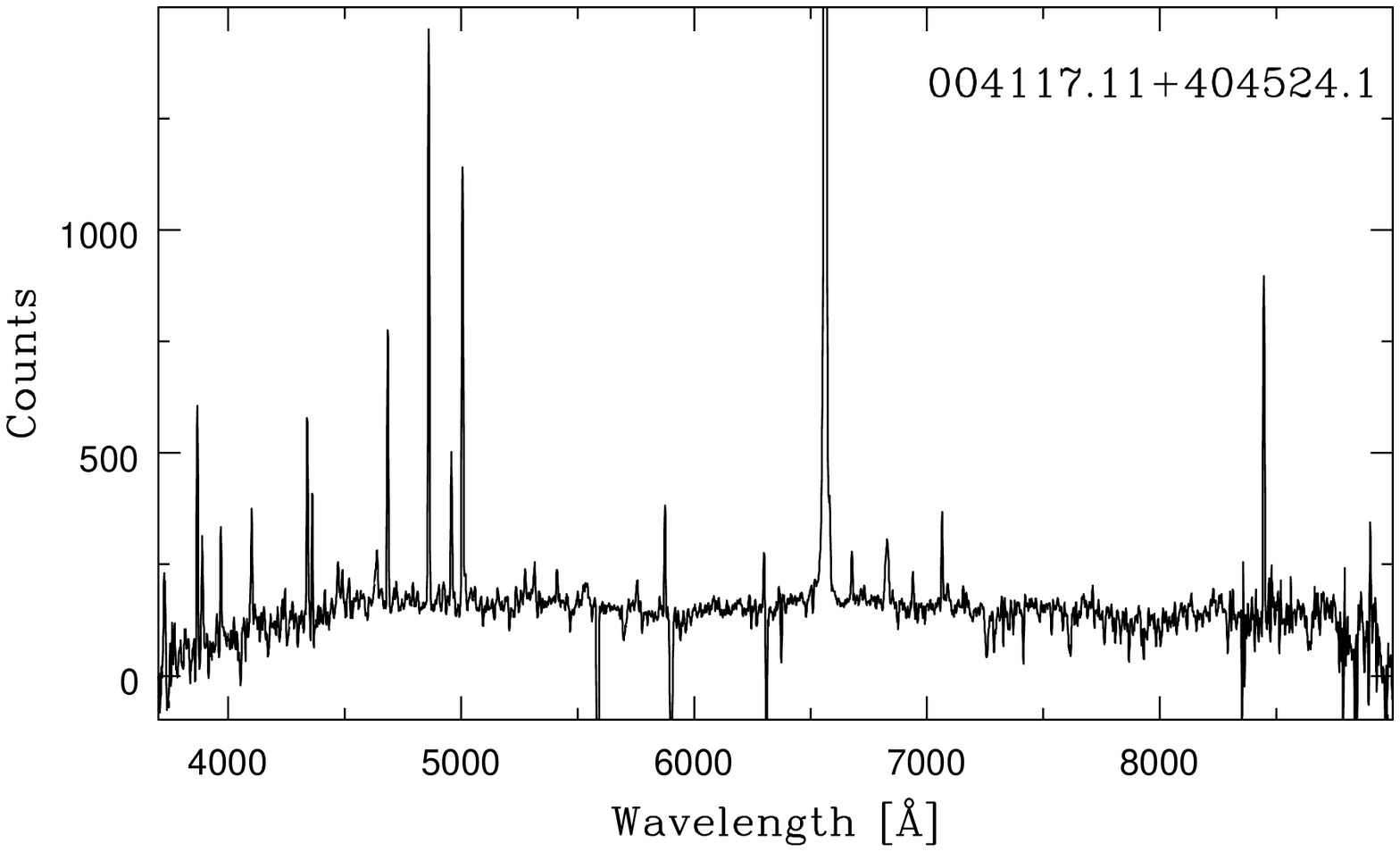}
\includegraphics[width=\columnwidth]{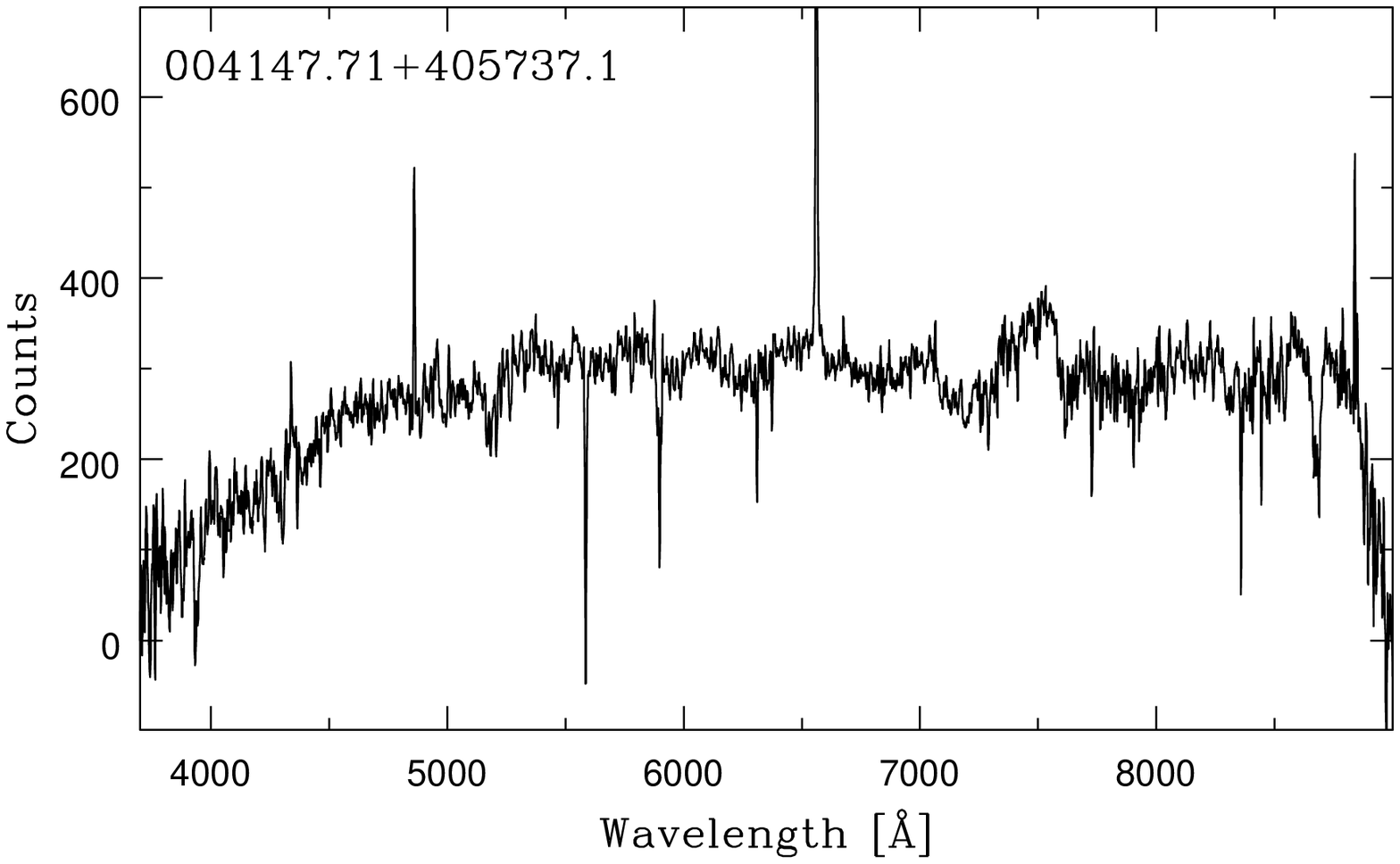}}
\centerline{\includegraphics[width=\columnwidth]{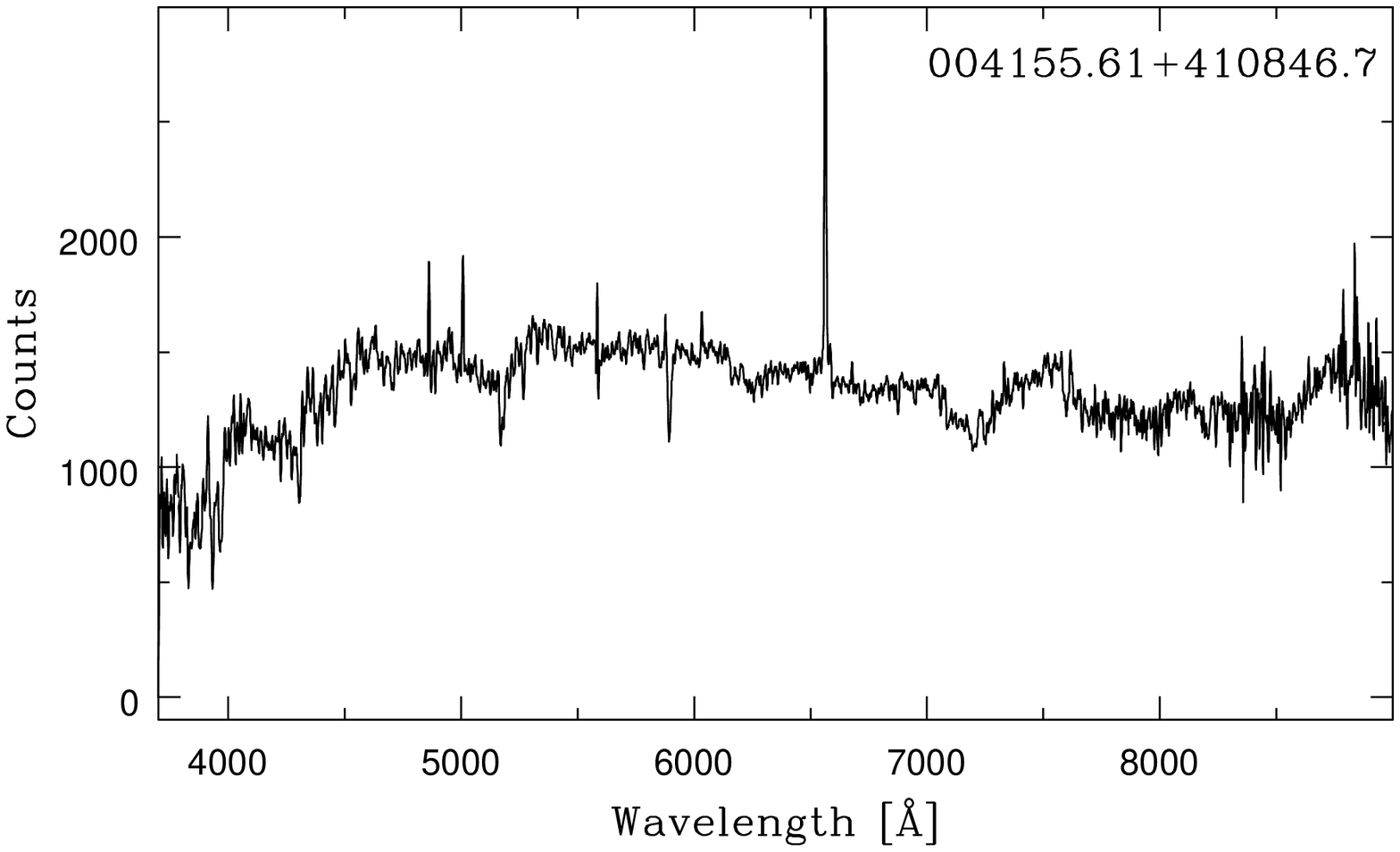}
\includegraphics[width=\columnwidth]{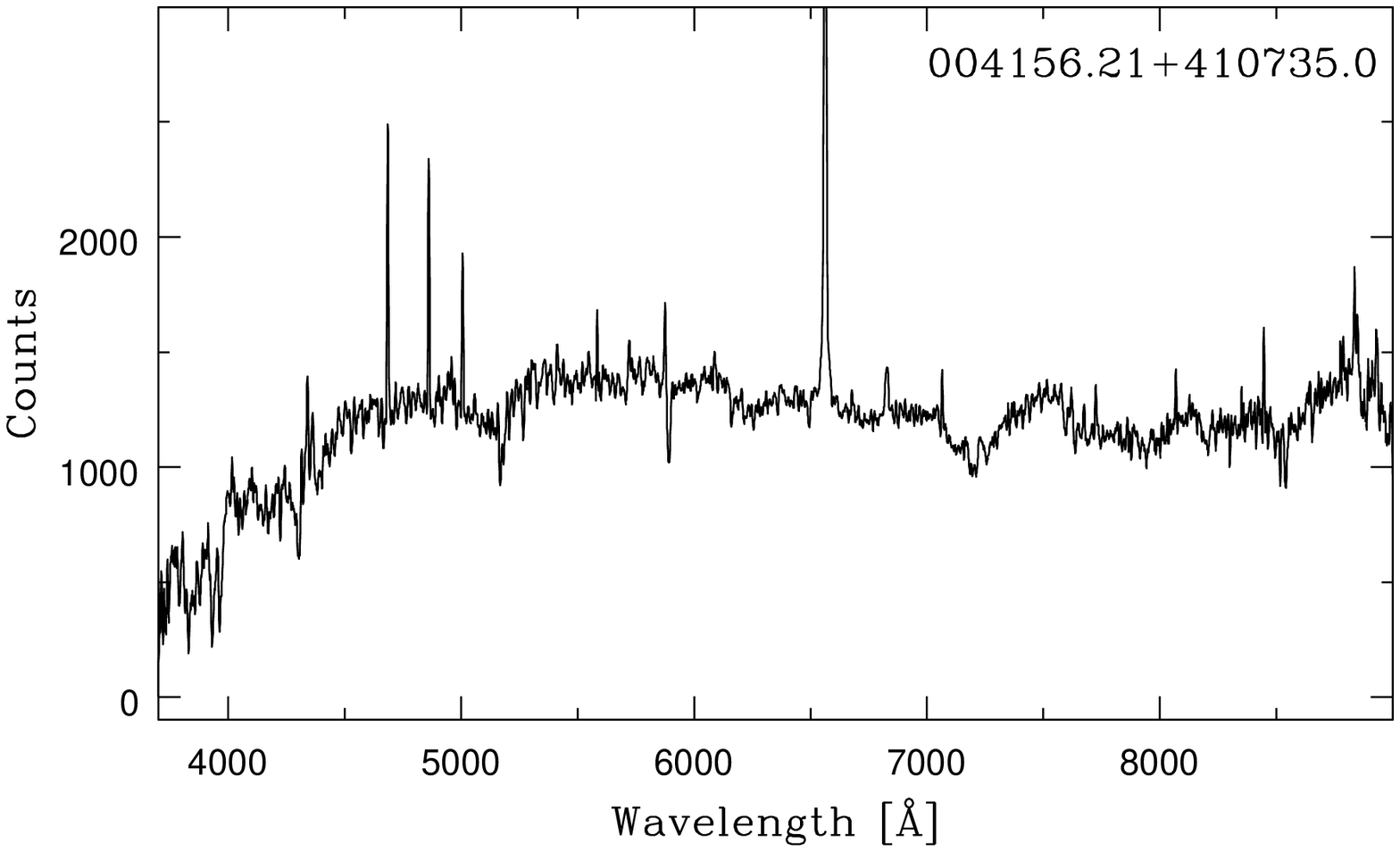}}
\centerline{\includegraphics[width=\columnwidth]{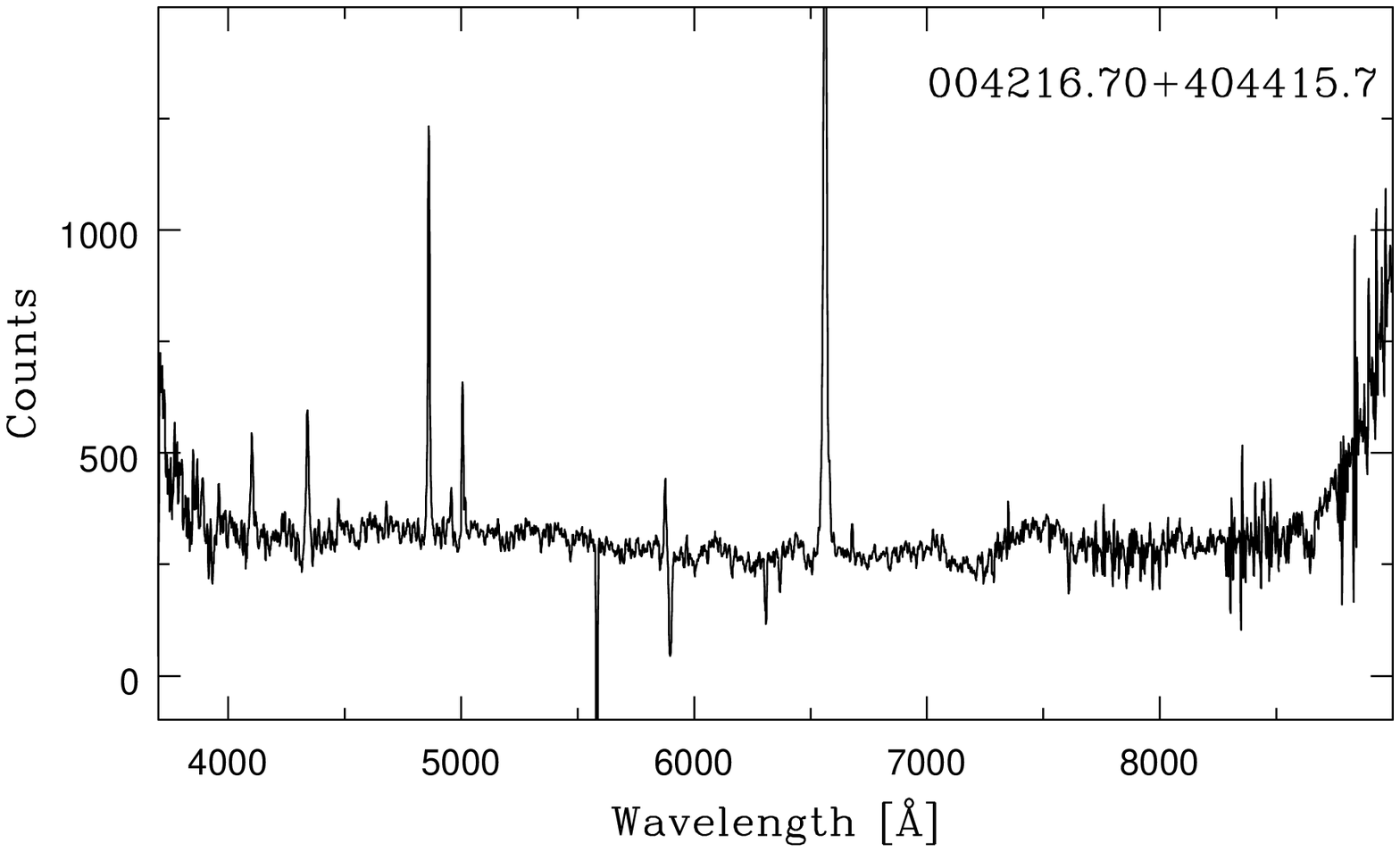}
\includegraphics[width=\columnwidth]{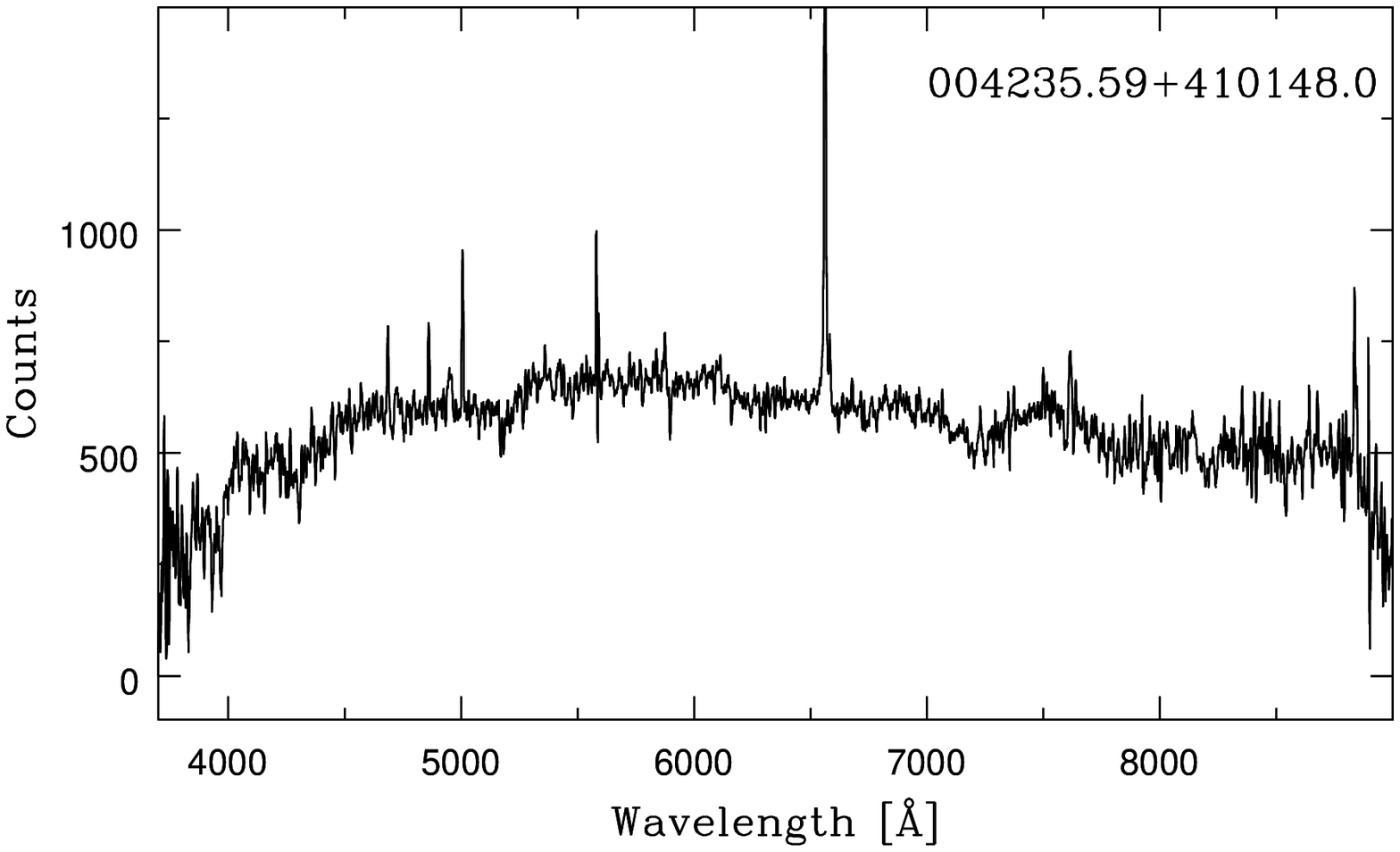}}
\centerline{\includegraphics[width=\columnwidth]{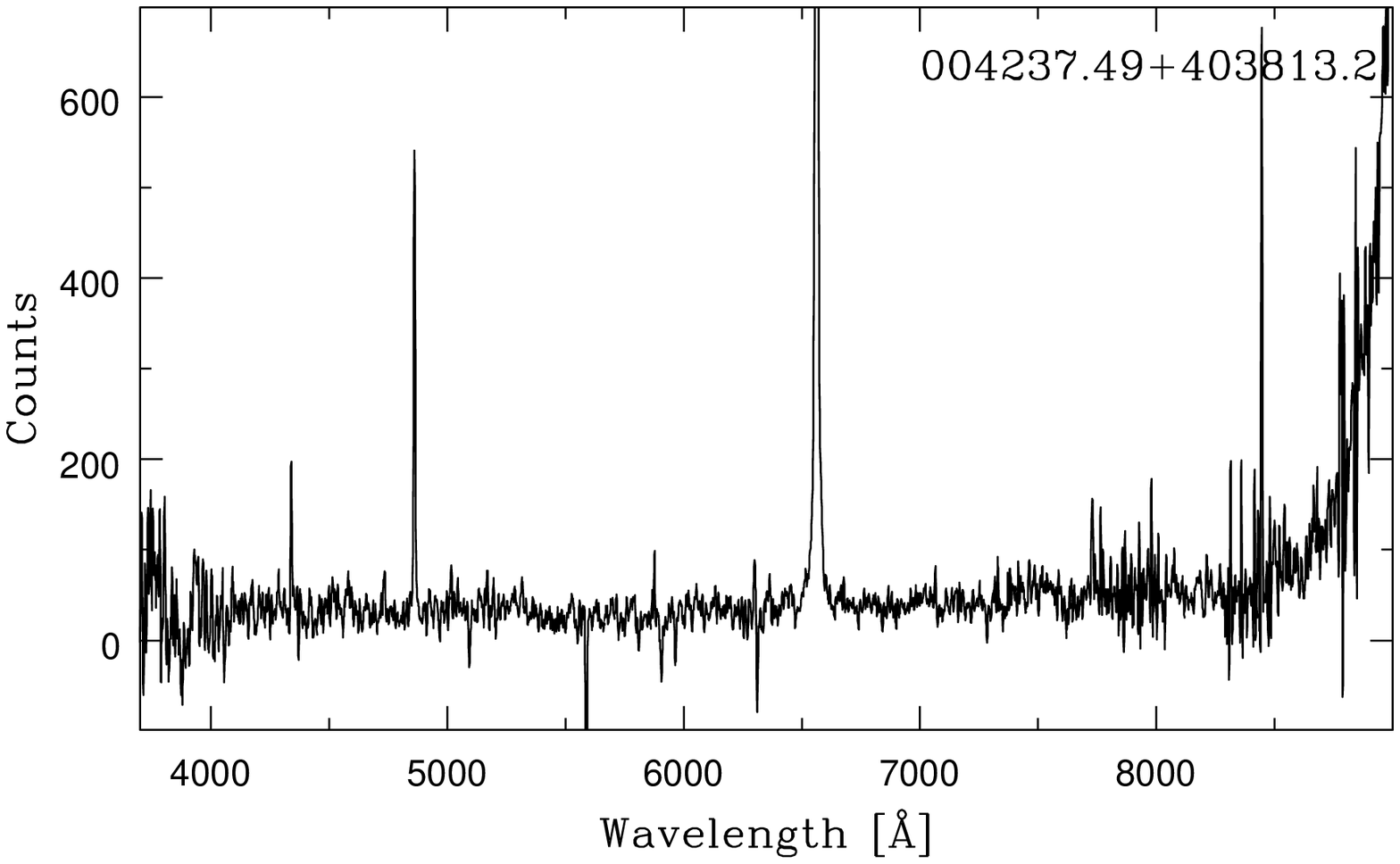}
\includegraphics[width=\columnwidth]{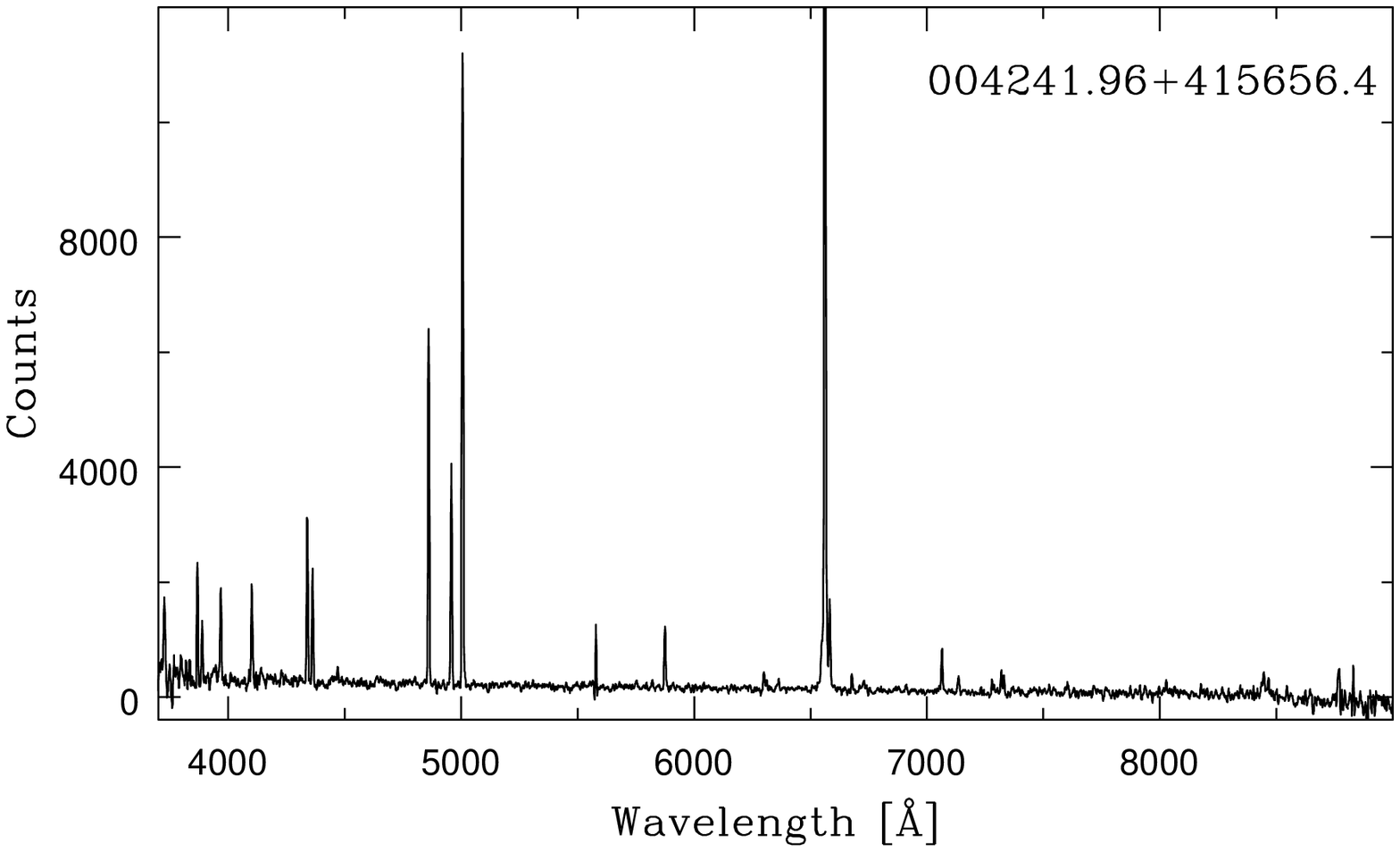}}
\caption{Spectra of SySt in M31}\label{sp2}
\end{figure*}

\begin{figure*}
\centerline{\includegraphics[width=\columnwidth]{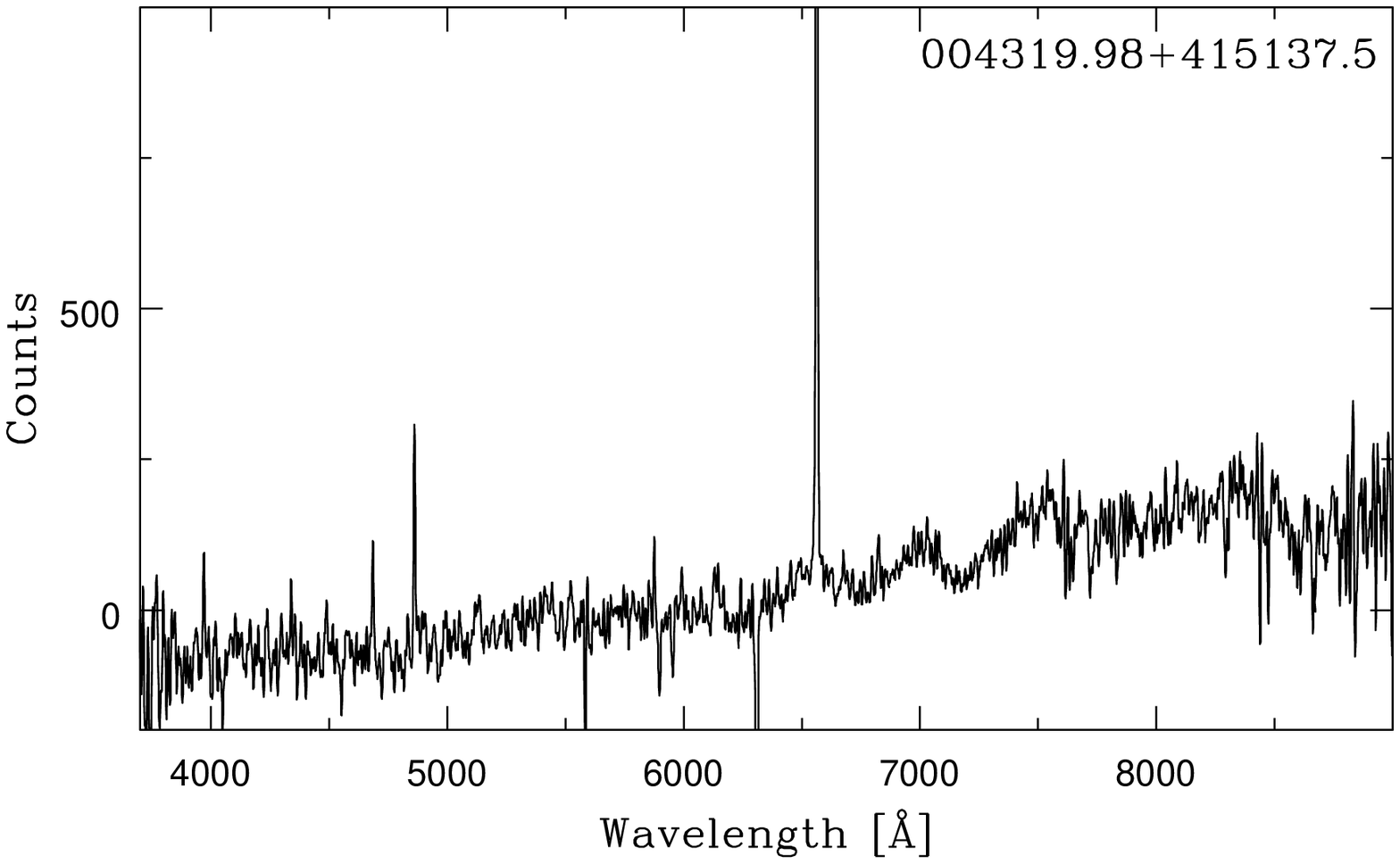}
\includegraphics[width=\columnwidth]{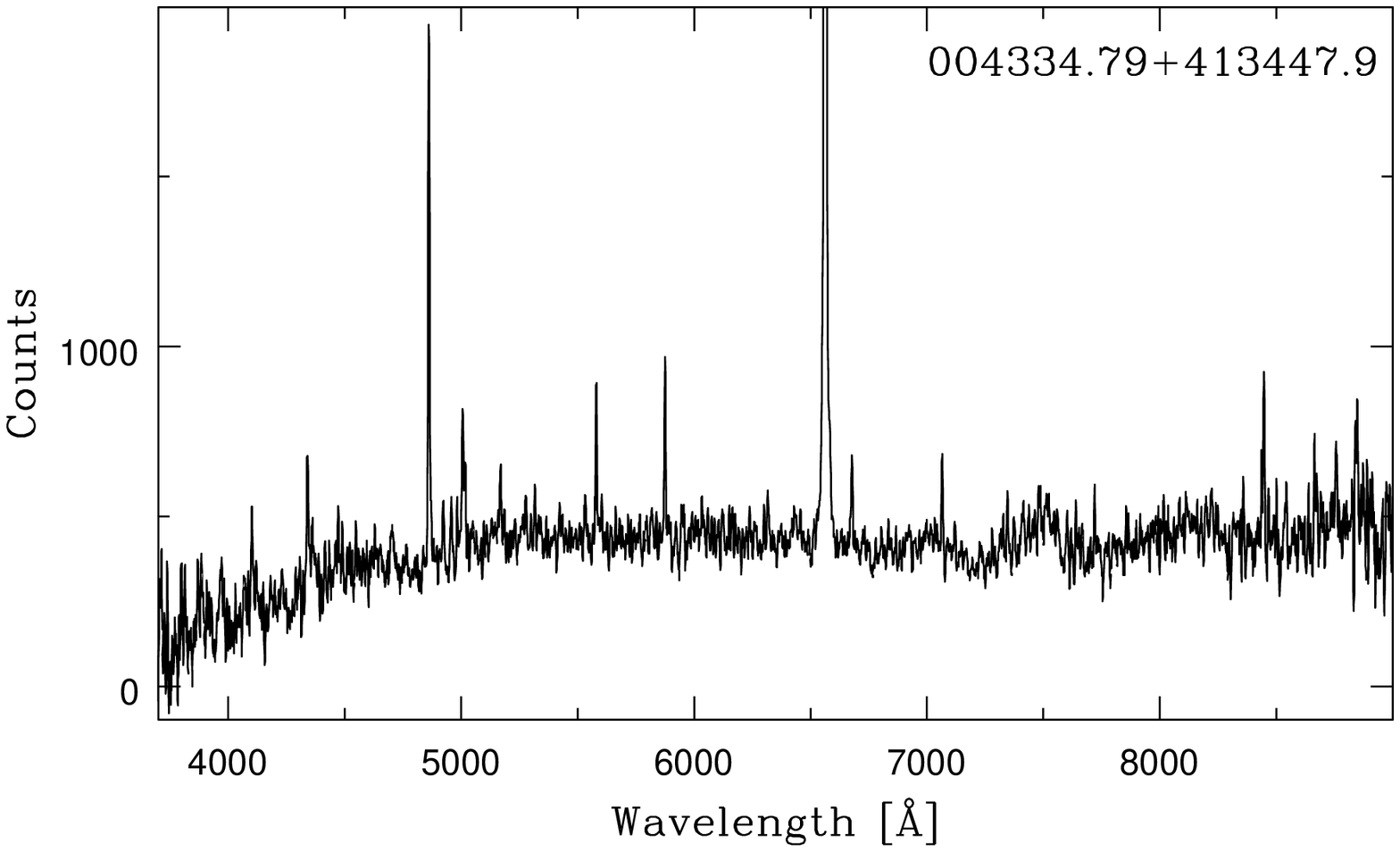}}
\centerline{\includegraphics[width=\columnwidth]{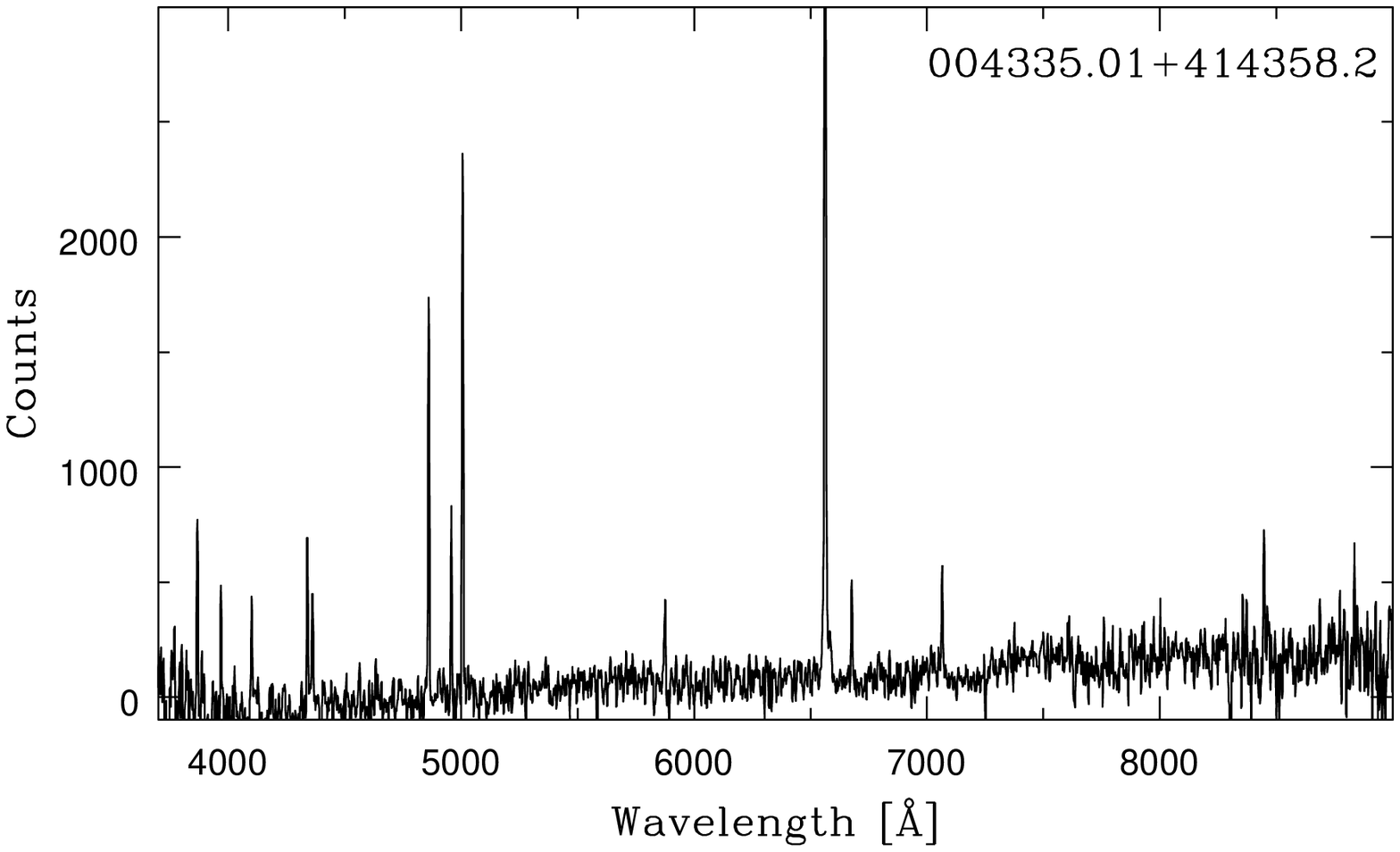}
\includegraphics[width=\columnwidth]{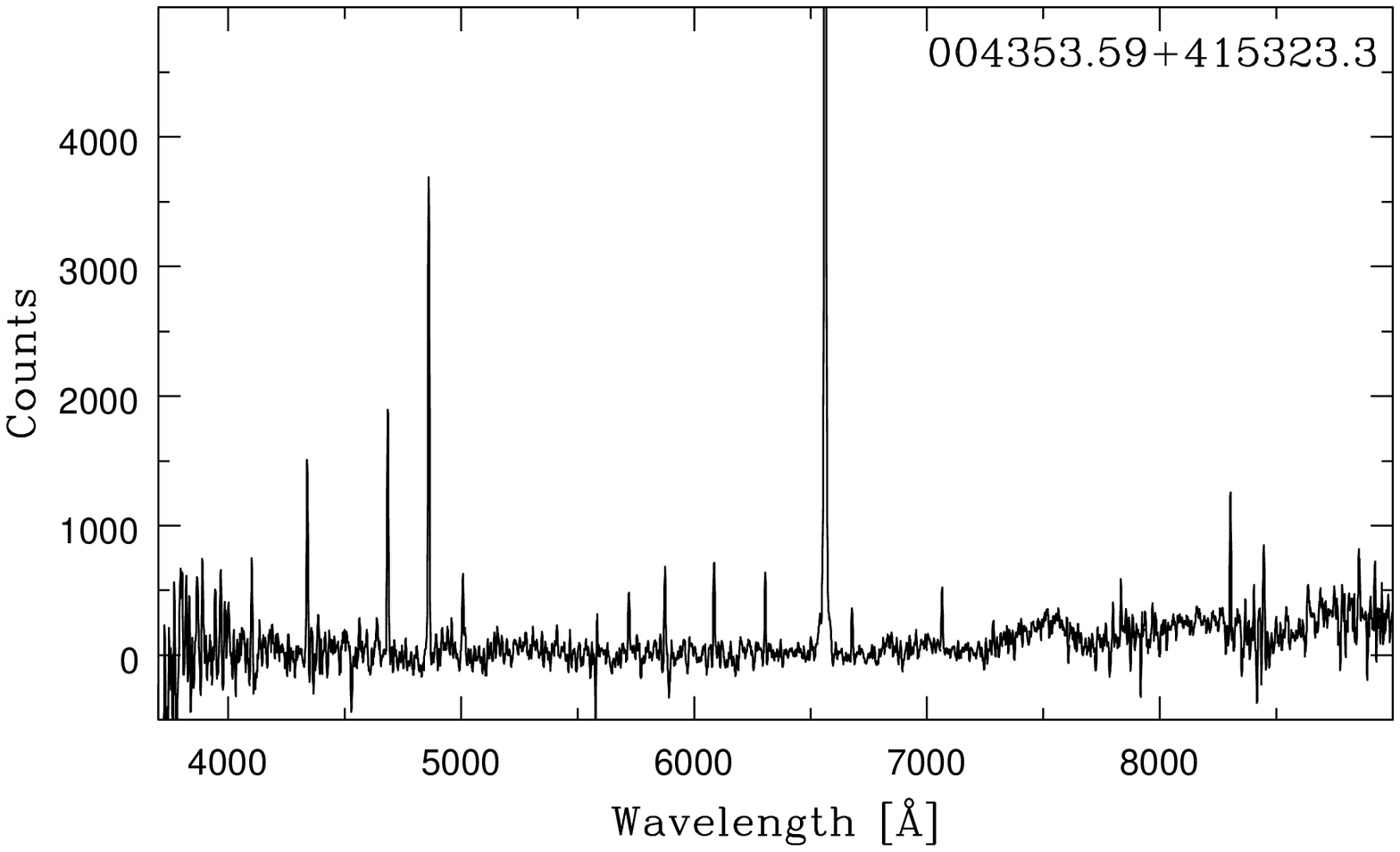}}
\centerline{\includegraphics[width=\columnwidth]{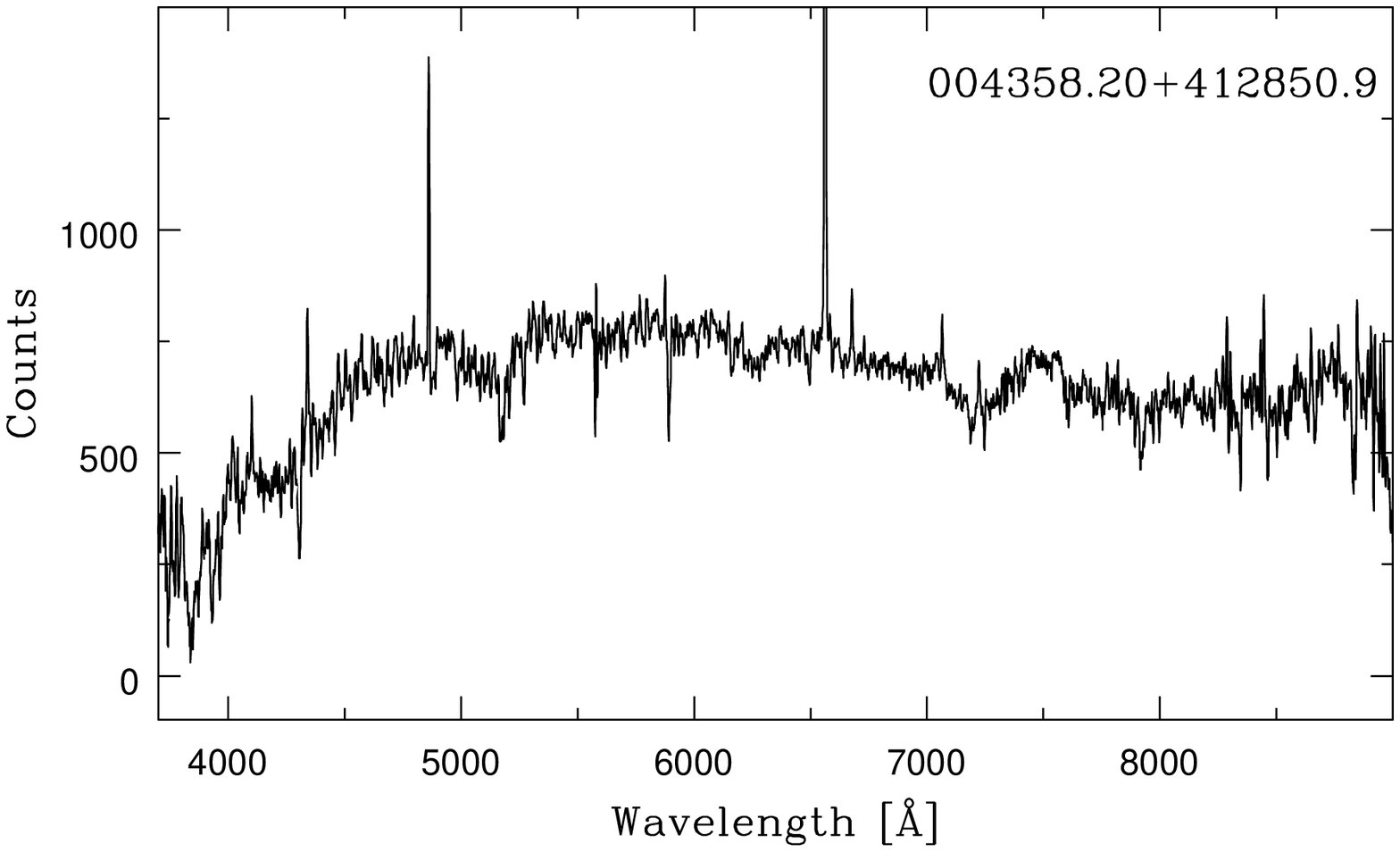}
\includegraphics[width=\columnwidth]{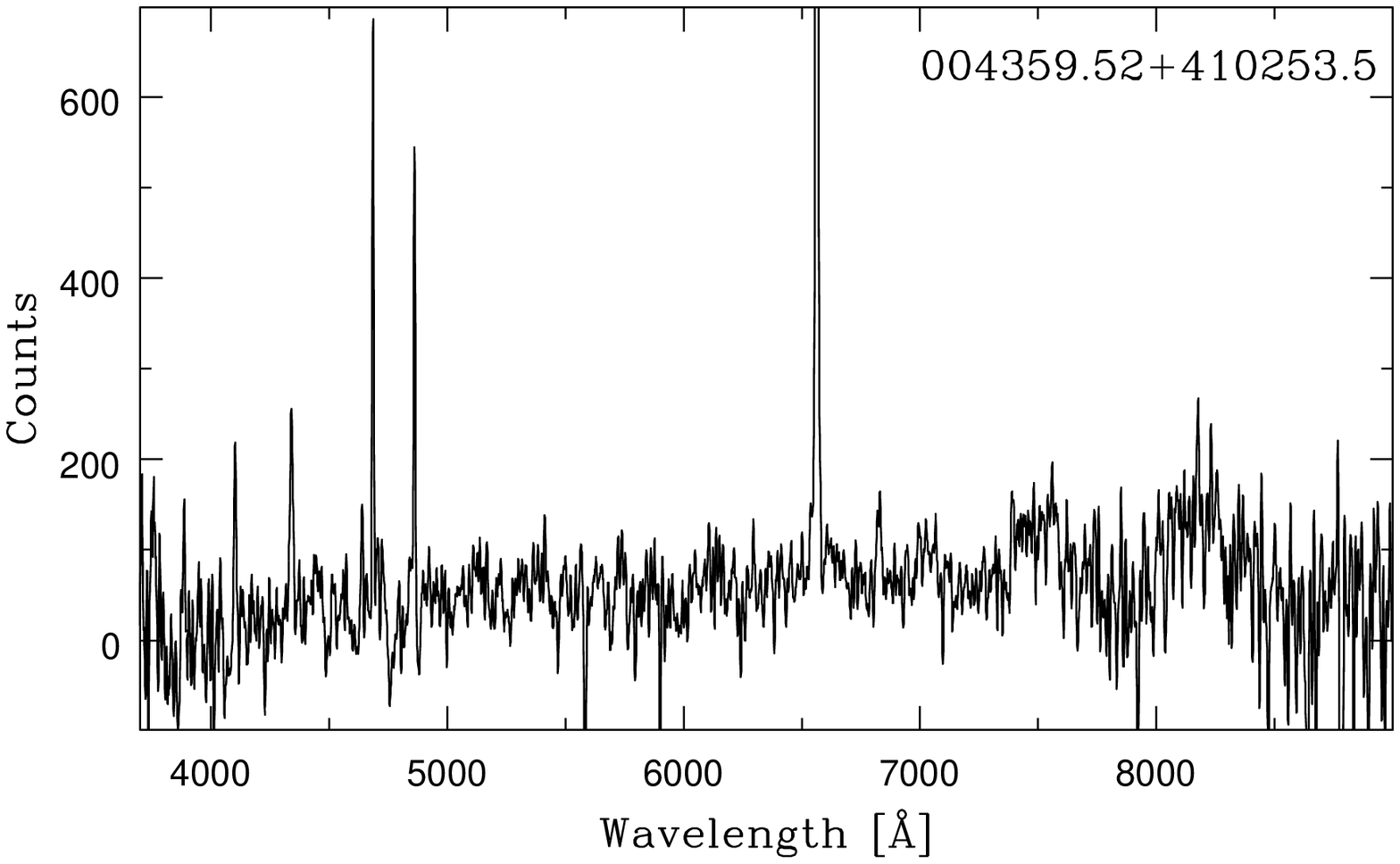}}
\centerline{\includegraphics[width=\columnwidth]{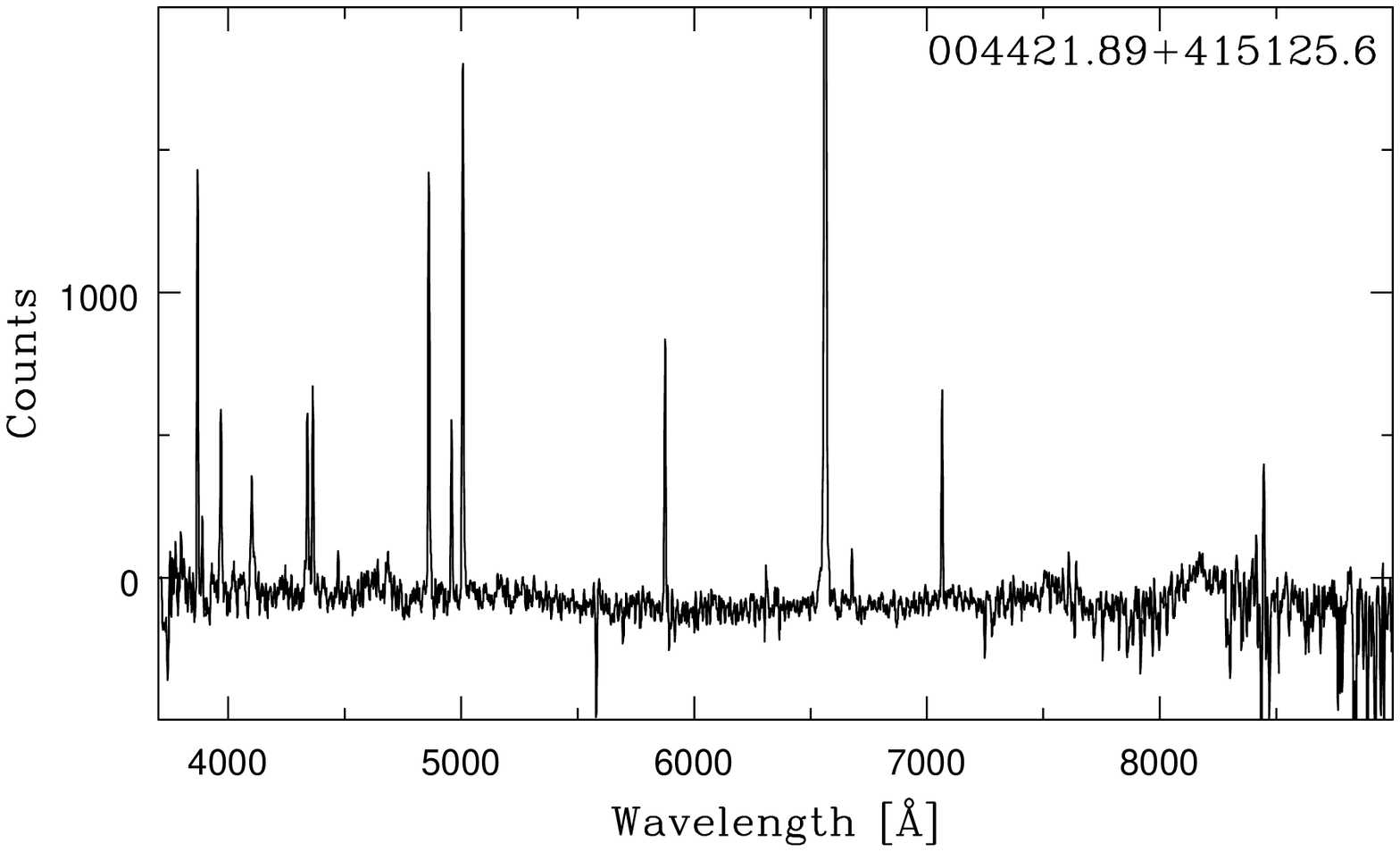}
\includegraphics[width=\columnwidth]{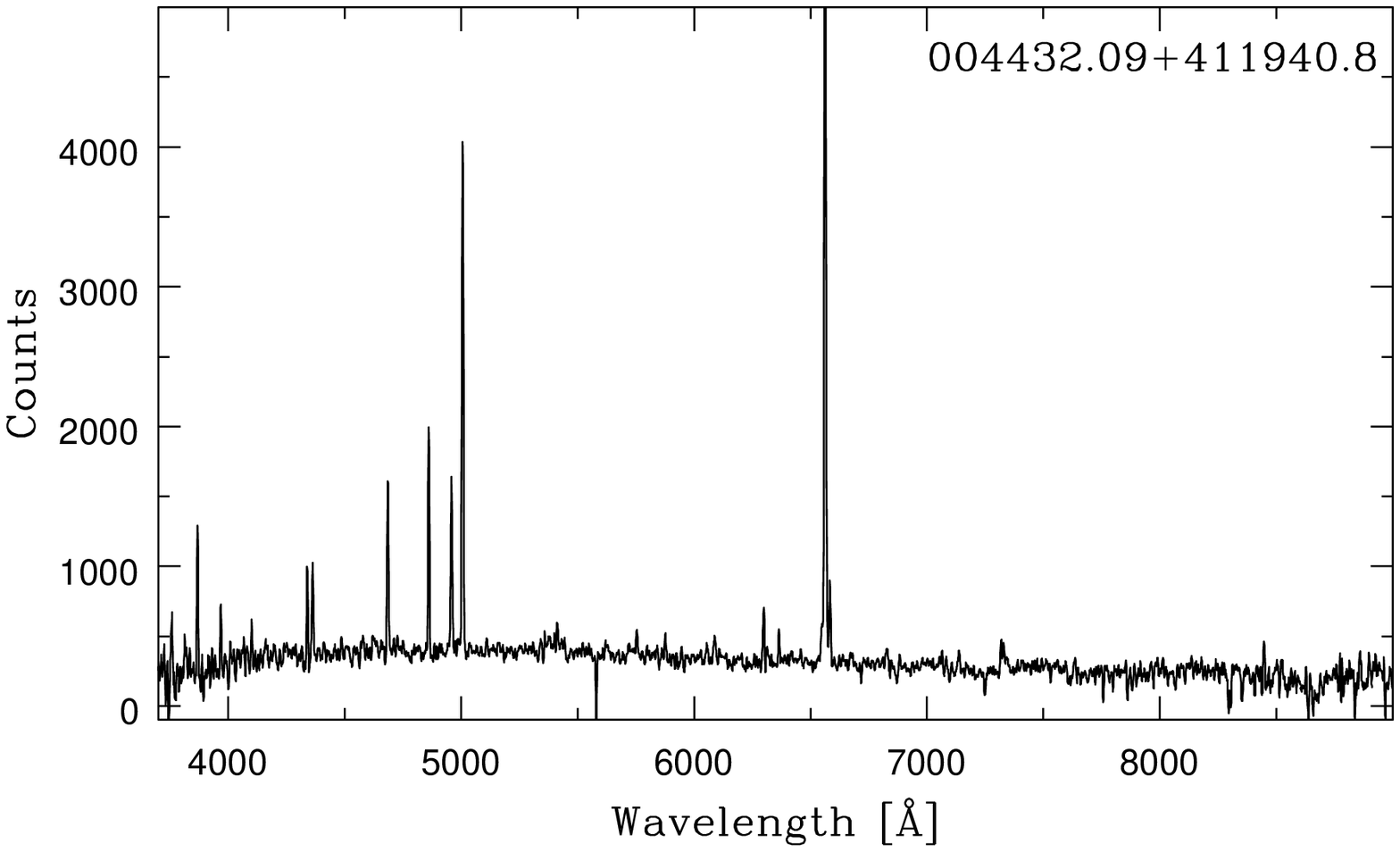}}
\caption{Spectra of SySt in M31}\label{sp3}
\end{figure*}

\begin{figure*}
\centerline{\includegraphics[width=\columnwidth]{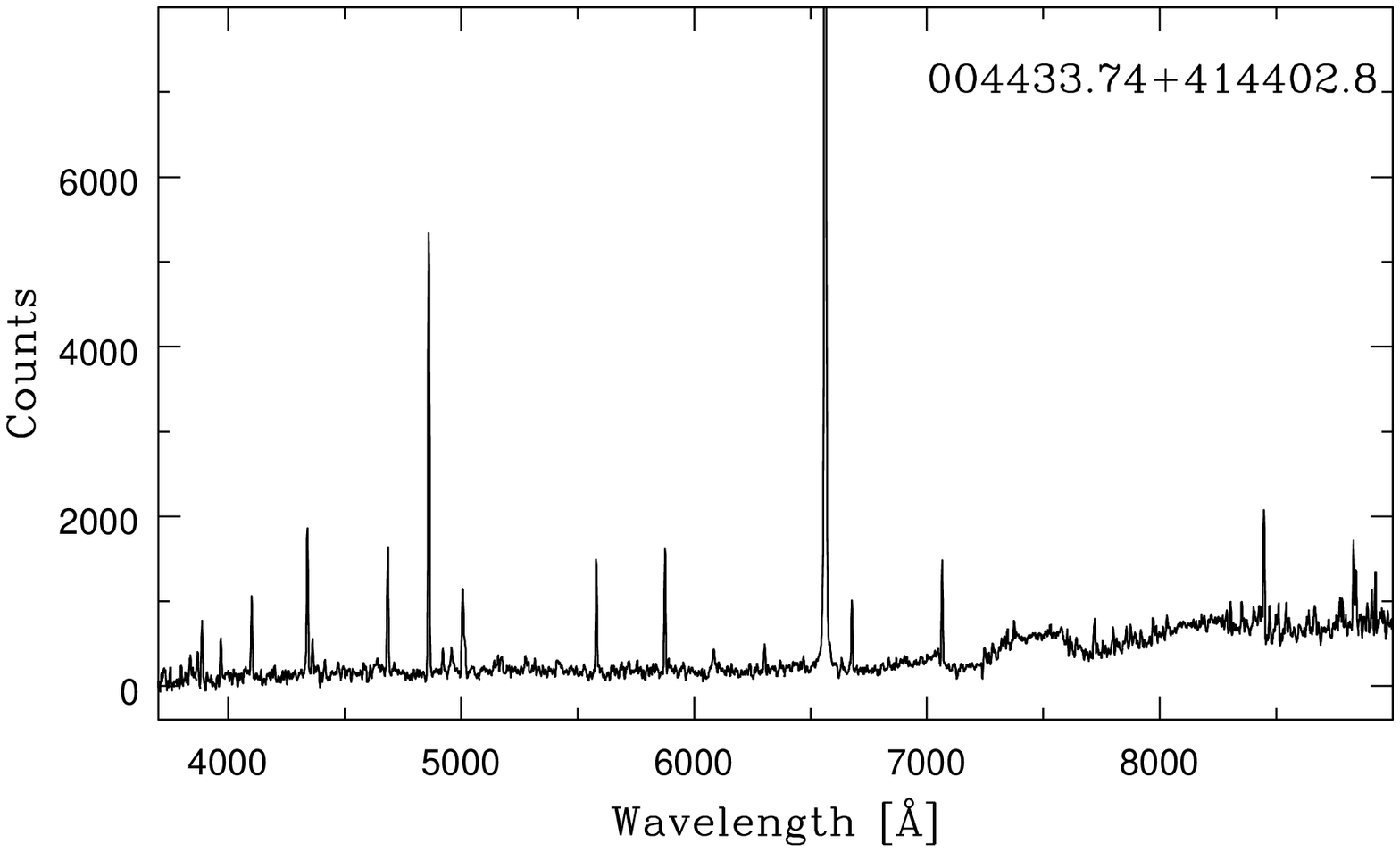}
\includegraphics[width=\columnwidth]{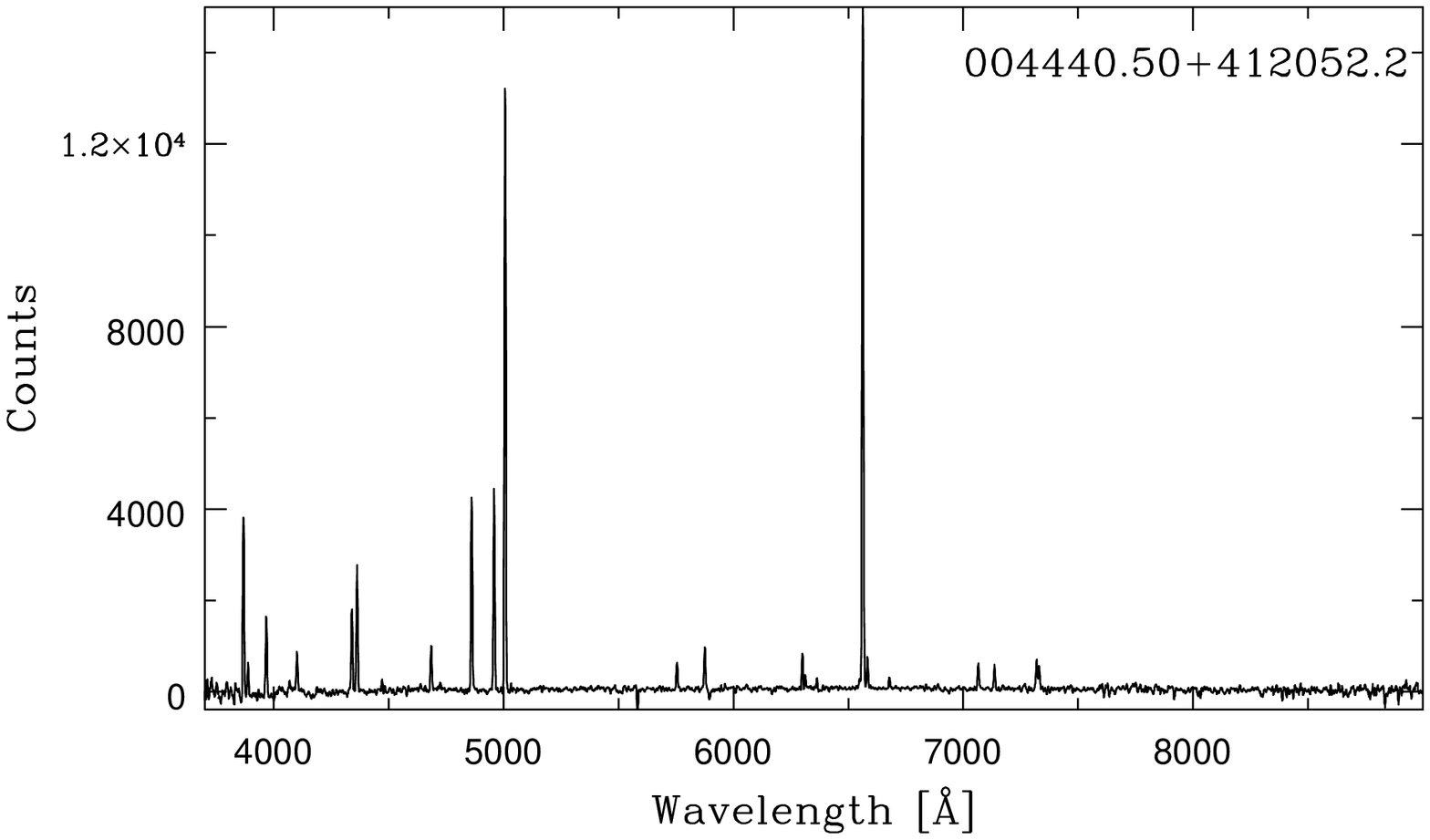}}
\centerline{\includegraphics[width=\columnwidth]{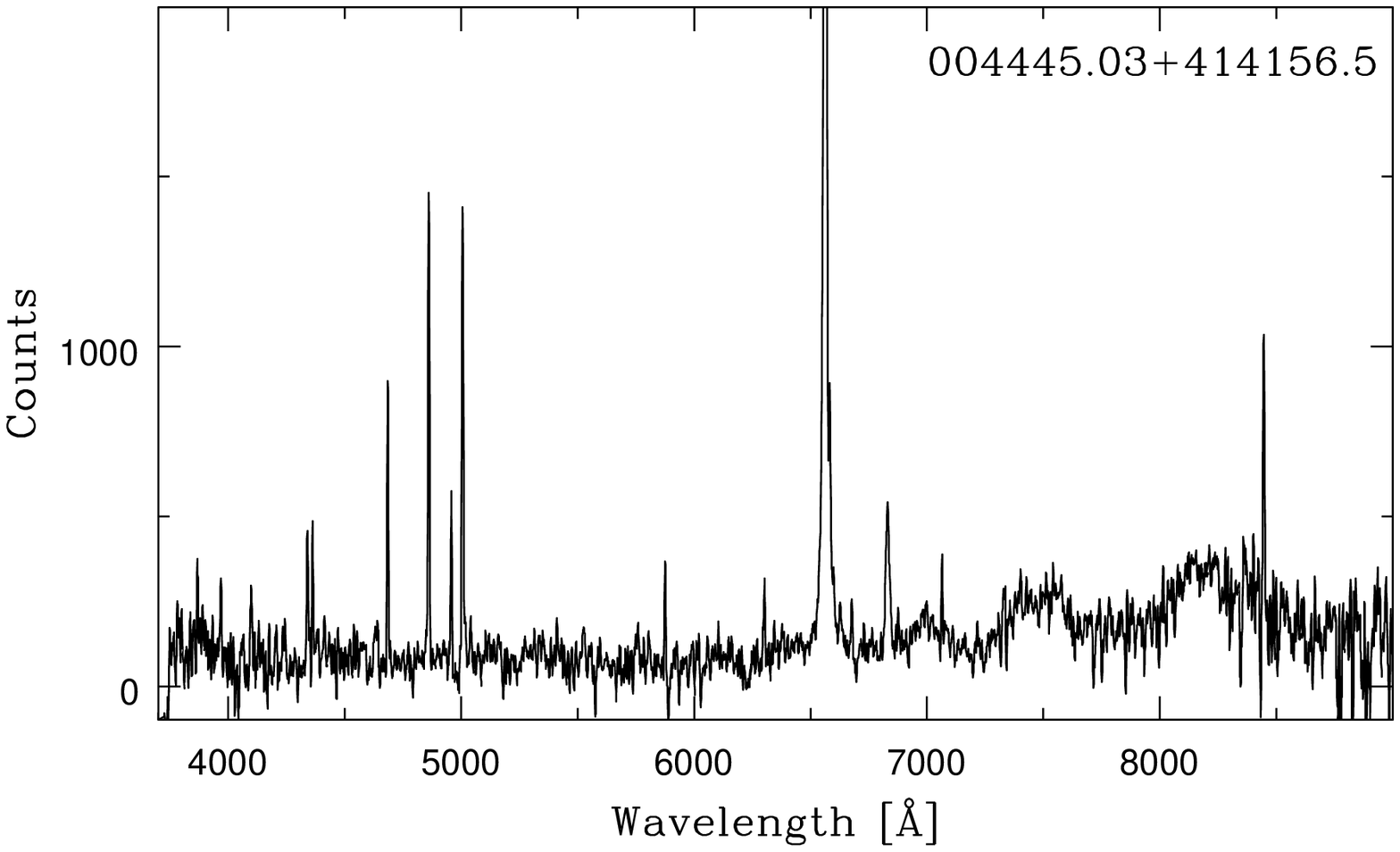}
\includegraphics[width=\columnwidth]{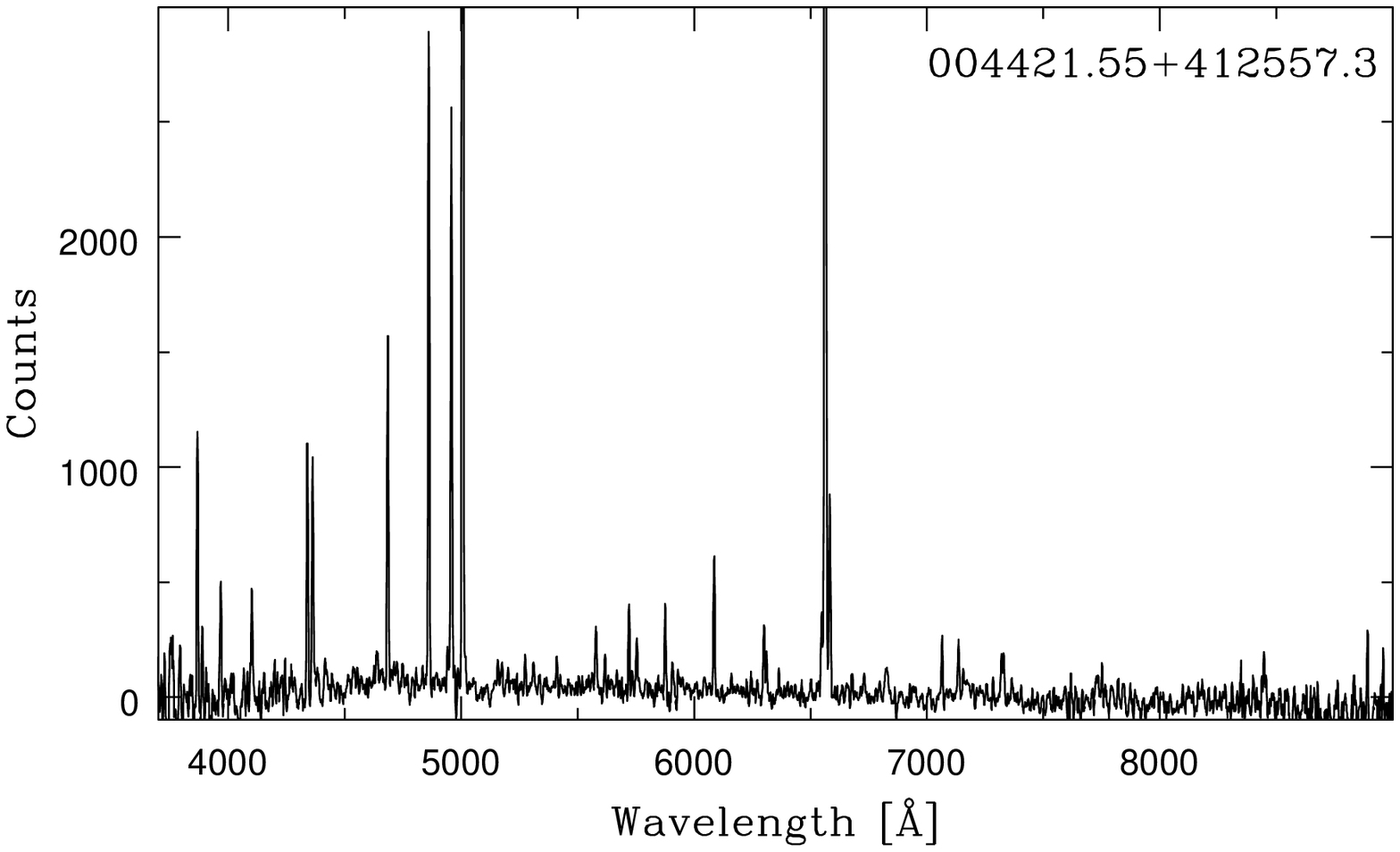}}
\centerline{\includegraphics[width=\columnwidth]{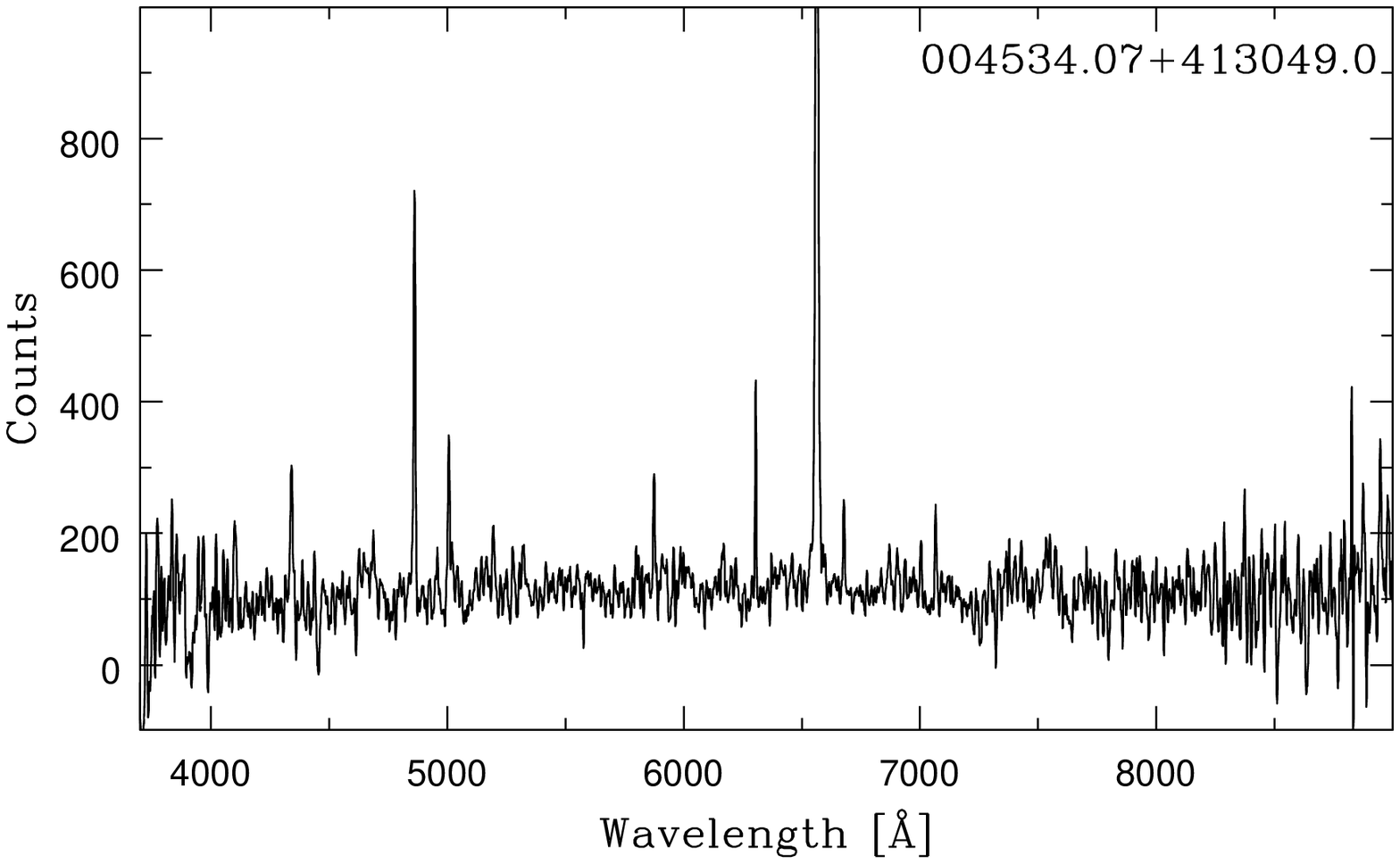}
\includegraphics[width=\columnwidth]{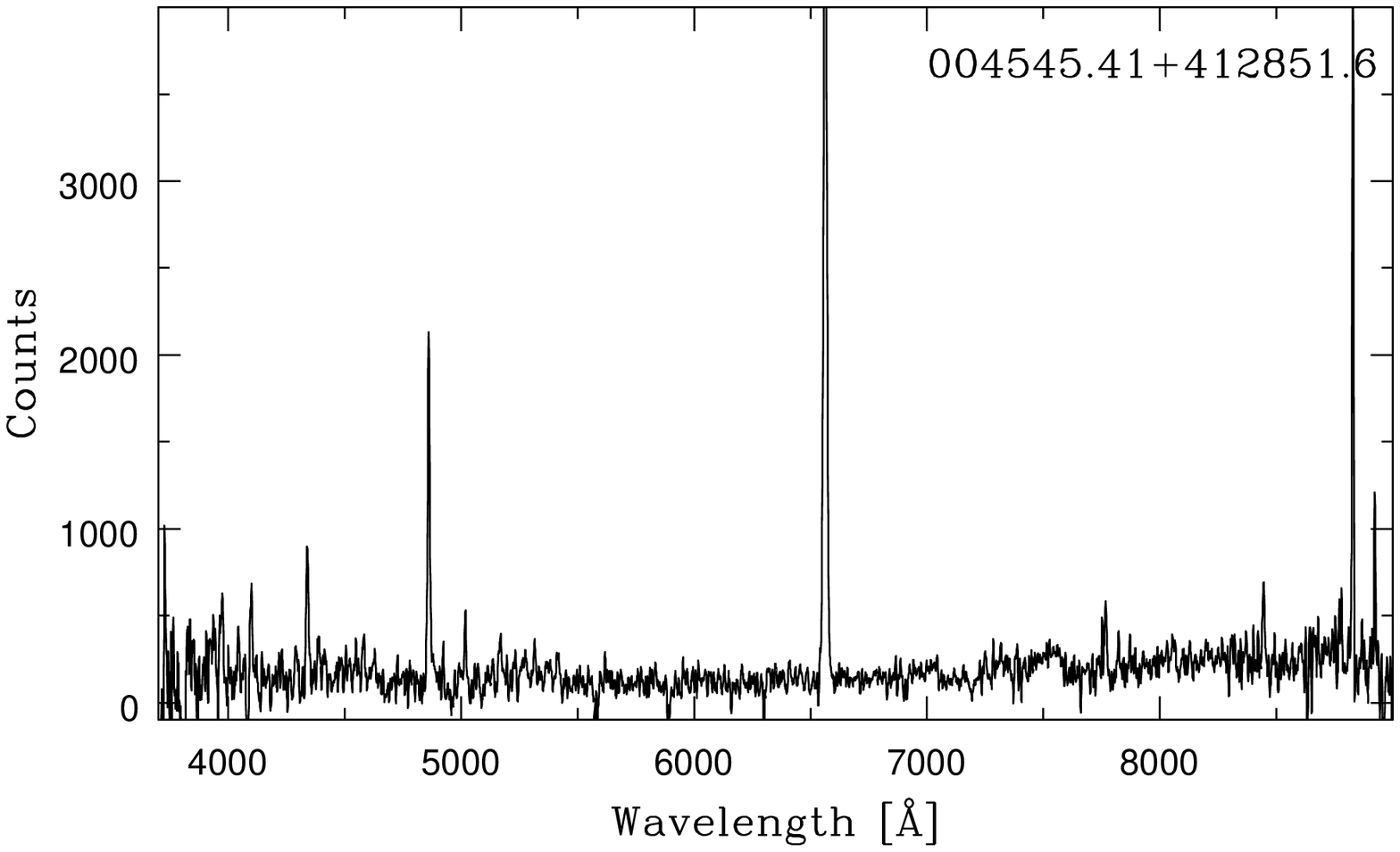}}
\caption{Spectra of SySt in M31}\label{sp4}
\end{figure*}
\bsp
\label{lastpage}

\end{document}